\documentclass[11pt]{article}
\pdfoutput=1
\usepackage[centertags]{amsmath}
\usepackage[square, comma, sort&compress,numbers]{natbib}
\usepackage{array,multirow}
\numberwithin{equation}{section}
\usepackage{amssymb}
\usepackage{graphicx}
\usepackage{color}
\usepackage{relsize}
\usepackage[utf8]{inputenc}

\usepackage{epsfig}

\def\tr{{\rm Tr}}

\def\Or[#1]{{\text{O}}\left({#1}\right)} 
\def\dotl[#1,#2]{\left\langle #1, #2 \right\rangle}
\def\dotlb[#1,#2]{[ #1, #2 ]}
\def\dotp[#1,#2]{(#1) \cdot (#2)}
\def\aff[#1,#2]{\hat{#1}(#2)}
\def\n4sym{{\cal N}=4 SYM}
\def\>{\rangle}
\def\<{\langle}
\def\weight[#1,#2,#3]{\{(#1),#2,#3\}}
\def\ads[#1]{$\text{AdS}_{#1}$}

\newcommand{\ba}{\begin{eqnarray}}
\newcommand{\ea}{\end{eqnarray}}
\newcommand{\be}{\begin{eqnarray}}
\newcommand{\ee}{\end{eqnarray}}
\newcommand{\bq}{\begin{equation}}
\newcommand{\eq}{\end{equation}}
\newcommand{\benn}{\begin{equation*}}
\newcommand{\eenn}{\end{equation*}}
\newcommand{\bi}{\begin{itemize}}  
\newcommand{\ei}{\end{itemize}}

\newcommand{\CO}{{\cal O}}

\newcommand{\CV}{{\cal V}}
\newcommand{\nn}{\nonumber}

\newcommand{\blue}{\bf\color{blue}}

\newcommand\oo\infty
\newcommand\s\sigma
\newcommand\de\delta
\newcommand\De\Delta

\newcommand\f\phi
\newcommand\g\gamma
\newcommand\x\times

\usepackage{jheppub}

\makeatletter
\def\@fpheader{\vspace{-.1cm}}
\makeatother

\title{On the Late-Time Behavior of Virasoro Blocks \\ and a Classification of Semiclassical  Saddles}

\author[a]{A.\ Liam Fitzpatrick,}
\author[b]{Jared Kaplan,}
\affiliation[a]{Department of Physics, Boston University, \\
Commonwealth Avenue, Boston, MA 02215, U.S.A.}
\affiliation[b]{Department of Physics and Astronomy,  Johns Hopkins University, \\
Charles Street, Baltimore, MD 21218, U.S.A.}

\abstract{  
Recent work has demonstrated that black hole thermodynamics and information loss/restoration in AdS$_3$/CFT$_2$ can be derived almost entirely from the behavior of the Virasoro conformal blocks at large central charge, with relatively little dependence on the precise details of the CFT spectrum or OPE coefficients.    Here, we elaborate on the non-perturbative behavior of Virasoro blocks by classifying all `saddles' that can contribute for arbitrary values of external and internal operator dimensions in the semiclassical large central charge limit.  The leading saddles, which determine the naive semiclassical behavior of the Virasoro blocks, all decay exponentially at late times, and at a rate that is independent of internal operator dimensions.  Consequently, the semiclassical contribution of high-energy states does not resolve a well-known version of the information loss problem in AdS$_3$.  However, we identify two infinite classes of sub-leading saddles, and one of these classes does not decay at late times.  
} 

\arxivnumber{} 

\begin{document}   
 
\maketitle
\flushbottom

\section{Introduction and Summary}  
 
Black hole thermodynamics can be derived \cite{Fitzpatrick:2014vua, Fitzpatrick:2015zha, Fitzpatrick:2015foa, Alkalaev:2015wia, Alkalaev:2015lca, KrausBlocks, Beccaria, Anous:2016kss, Besken:2016ooo,Fitzpatrick:2015dlt} as a universal consequence of the Virasoro symmetry algebra of AdS$_3$/CFT$_2$.  Unitarity violation or `information loss' occurs in the semiclassical limit where  Newton's constant\footnote{Throughout this paper, we use units where the AdS$_3$ radius of curvature $\ell_{\rm AdS}=1$.} $G_N \to 0$ with $G_N E$ fixed for  bulk masses and energies $E$.  Information loss persists order-by-order in a perturbation expansion in $G_N = \frac{3}{2c}$,  where $c$ is the central charge of the CFT$_2$, but it is explicitly resolved \cite{Fitzpatrick:2016ive, Chen:2016cms} by non-perturbative `$e^{-c}$' effects.  These phenomena occur independently within each irreducible representation of the Virasoro algebra.  

These observations suggest a simple and unorthodox viewpoint:  that in $2+1$ dimensions, black hole thermodynamics, information loss, and its restoration essentially depend on only the Virasoro algebra (i.e. on the gravitational sector of AdS$_3$), and more specifically on the behavior of the Virasoro conformal blocks.  This means that some of the most fascinating features of quantum gravity in AdS$_3$ appear to be largely independent of the spectrum and OPE coefficients of the CFT$_2$ dual.  This prospect opens up a powerful line of attack for understanding non-perturbative physics in  quantum gravity through the determination of the non-perturbative behavior of the Virasoro conformal blocks.

The present work will bolster these claims in two ways.  First, we will explicitly identify all of the semiclassical `instantons' or `saddles' associated with the non-perturbative or `$e^{-c}$' effects that can appear in the large $c$ expansion of the Virasoro conformal blocks.  Second, we will compute the late-time behavior of all semiclassical Virasoro blocks, with arbitrary internal and external states.  

We find that in the heavy-light  limit, which corresponds to a sub-Planck-mass object probing a BTZ black hole, the leading semiclassical Virasoro blocks all decay exponentially at late times, at a rate independent of the dimension of the exchanged state.  Roughly speaking, this means that semiclassical high energy states do not play a privileged role in resolving information loss.  However, in addition to the leading saddle, we will also identify two infinite classes of sub-leading semiclassical saddles, and one of these two classes of saddles does not decay at very late times.  

Before explaining these results we will review some background and motivation. 

\begin{figure}[t!]
\begin{center}
\includegraphics[width=0.45\textwidth]{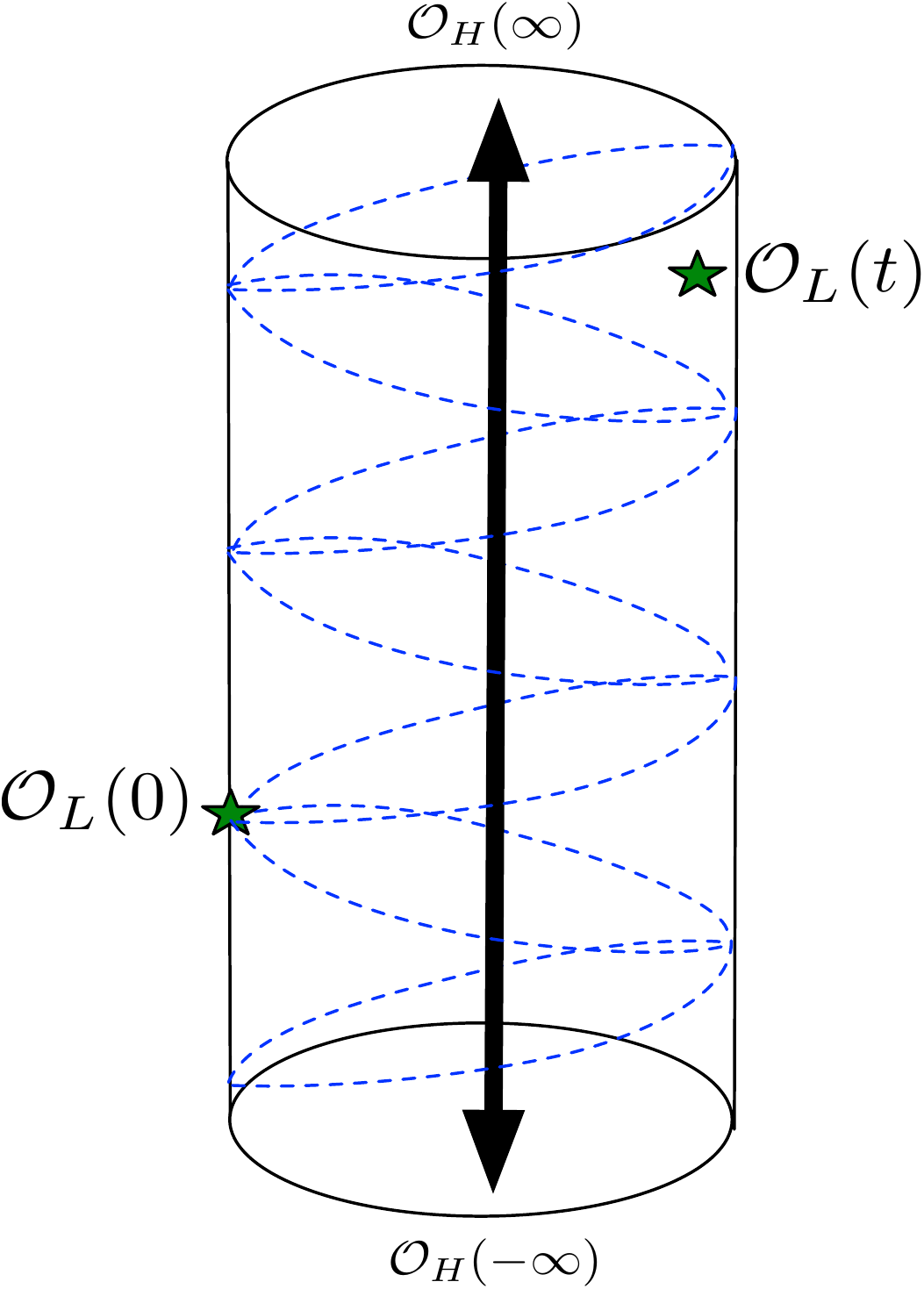}
\caption{This figure depicts a generic configuration of the Lorentzian heavy-light correlator, with dashed lines drawn in to indicate past and future lightcones emanating from the operator $\CO_L(0)$.  The lightcones appear as branch cuts in CFT correlators.}
\label{fig:HeavyLightwithLightcones}
\end{center}
\end{figure}

\subsection{Information Loss and the Late-Time Behavior of AdS Correlators}

Although any violation of unitarity in quantum gravity might be construed as `information loss', we would like to be more specific and zero in on information loss associated with black holes.  In this work we will focus on physical processes where initial data disappears because a high-energy microstate behaves too much like the thermal ensemble, i.e. as if the heavy state is a perfect thermal bath with an infinite number of degrees of freedom.  This seems to be a good description of why  encyclopedias are destroyed when they fall into semiclassical black holes.   

The information loss problem(s) we will discuss can be studied by observers who stay far away from a black hole, meaning that in the  AdS/CFT context, the problem can be diagnosed using only CFT correlators \cite{Maldacena:2001kr}, without reconstructing the AdS spacetime. Thus we are studying the `easier' information loss problem \cite{Fitzpatrick:2016ive}, rather than the `hard' problem of how to simultaneously maintain unitarity and effective field theory across the horizon of a black hole.  From the point of view of CFT the problem has two parts:  first we must understand why the CFT \emph{appears} non-unitary at large $c$, and then, from this vantage point, we need to identify the effects that restore the information.

We will be focusing on 4-point correlators in CFT$_2$, which we will write as 
\be
\< \CO_H(\infty) \CO_H(1) \CO_L(z) \CO_L(0) \>
\label{eq:HHLL4pt}
\ee 
to emphasize that we are most interested in the physics of the heavy-light limit where a relatively light probe interacts with a black hole in AdS$_3$.  The dual AdS$_3$ setup is pictured in figure \ref{fig:HeavyLightwithLightcones}.  The ratios $\frac{h_i}{c} \sim G_N M_i$ are fixed in the semiclassical large $c$ limit.  In this paper, unlike in many of our previous works, we will obtain results that hold for arbitrary values of both $\frac{h_L}{c}$ and $\frac{h_H}{c}$.  Our methods will pertain to the entire semiclassical regime.

We recently discussed two sharp signatures of information loss  \cite{Fitzpatrick:2016ive} encoded in the heavy-light CFT correlator, `forbidden singularities' and the late time behavior \cite{Maldacena:2001kr} of the correlator.  Here we will be focusing on the latter.  

If we consider a schematic decomposition of the heavy-light correlator in the $\CO_H \CO_L \to \CO_H \CO_L$ OPE channel, which corresponds to conventional time slices of the cylinder pictured in figure \ref{fig:HeavyLightwithLightcones}, we find
\be
\< \CO_H(\infty) \CO_L(t) \CO_L(0)  \CO_H(-\infty) \> = \sum_i \lambda_i e^{i E_i t}
\ee 
for some coefficients $\lambda_i$ and energy levels $E_i$.  In the presence of a large number of terms, well-known statistical considerations apply (see \cite{Barbon:2014rma} for a relevant, detailed discussion).  In holographic CFTs, these suggest that the correlator will decay exponentially until it is of order $e^{-\frac{1}{2} S_{BH}}$, where  $S_{BH}$ is the heavy-state black hole entropy.  After this point it enters  a chaotic phase where it is typically very small, but occassionally returns to an order one value due to Poincar\'e recurrences on timescales of order $e^{S_{BH}}$.  This channel seems to describe healthy unitary evolution as long as the $\lambda_i$ are discrete and finite, and in particular it is hard to see how information loss can occur.  Individual unitary conformal blocks in this channel bear little resemblance to the physics of semiclassical gravity, which produces a continuous spectrum in the limit $c\rightarrow \infty$ \cite{Fitzpatrick:2014vua}.

We will be focusing instead on the $\CO_L \CO_L \to \CO_H \CO_H$ OPE channel
\be
\label{eq:VirasoroBlockDecomposition}
\< \CO_H(\infty) \CO_L(t) \CO_L(0)  \CO_H(-\infty) \> = \sum_{h_I, \bar h_I} P_{h_I, \bar h_I} \CV_{h_I} \left(1 - e^{-it} \right) \CV_{\bar h_I} \left(1 - e^{-i \bar t} \right)
\ee
where $\CV_h$ are Virasoro conformal blocks and $P_{h, \bar h}$ are products of OPE coefficients.  We emphasize that this expansion always  converges \cite{Maldacena:2015iua} away from OPE limits.   In this channel  the familiar features of semiclassical quantum gravity, including the physics of BTZ black holes, emerge in the large $c$ limit \cite{HartmanLargeC, Fitzpatrick:2014vua, HartmanExcitedStates, TakayanagiExcitedStates,  Fitzpatrick:2015zha, Fitzpatrick:2015foa, Alkalaev:2015wia, Alkalaev:2015lca, KrausBlocks, Beccaria, Anous:2016kss, Besken:2016ooo,Fitzpatrick:2015dlt, Fitzpatrick:2016ive, Lewkowycz:2016ukf,Chen:2016dfb}.  In particular, information loss appears \cite{Fitzpatrick:2016ive, Maldacena:2001kr} as the exponential decay of correlators at arbitrarily late times.   

Previously, this behavior has been demonstrated \cite{Fitzpatrick:2016ive, Anous:2016kss} in the limit of small $\frac{h_L}{c}$ and  $\frac{h_I}{c}$, where $h_I$ is the dimension or AdS$_3$ energy of the intermediate state  $\CO_I \subset \CO_L(z) \CO_L(0)$ appearing in the light-light and heavy-heavy OPEs.   These analyses left open the question of whether the information loss problem might be ameliorated by heavy semiclassical states with large finite $\frac{h_I}{c}$, or perhaps by effects suppressed by $\frac{h_L}{c}$.  We will demonstrate that this is not the case.  All semiclassical Virasoro blocks decay exponentially at sufficiently late times, and the asymptotic rate is independent of the intermediate operator dimension $h_I$.

\subsection{The Semiclassical Saddles of the Virasoro Conformal Blocks}

We will be investigating the Virasoro conformal blocks in the limit of large central charge.  We can define these blocks very explicitly by inserting a sum over all the intermediate states that are related to each other by the Virasoro algebra.  The sum over an irreducible representation of Virasoro can be written using an intermediate state projection operator 
\be
\label{eq:SchematicVacuumBlock}
\CV_{h_I}(z) = \left\< \CO_H(\infty) \CO_H(1) \left( \sum_{\{a_k \},\{ b_l\}}  \frac{  L_{-a_1} \cdots L_{-a_k} | h_I \>  \< h_I | L_{b_l} \cdots L_{b_1}  }{\mathcal{N}_{\{b_l \},\{ a_k\}}} \right) \CO_L(z) \CO_L(0) \right\>
\ee
where $L_{-m}$ are Virasoro generators, $\cal N$ is a Gram matrix of normalizations, and $h_I$ is the dimension of the primary operator/state labeling the irreducible representation.  This sum has a natural interpretation as an OPE expansion in small $z$, with intermediate states of dimension $h_I + n$ contribution as $z^{h_I + n}$.  At finite values of $c$ and external and internal operator dimensions $h_i$ and $h_I$, the OPE has radius of convergence $1$, and this formula provides a non-perturbative definition of the blocks.

Although one can perform some interesting calculations \cite{Fitzpatrick:2015foa} using this definition, the present work will be based on a very different approach to the Virasoro blocks.  

Remarkably, Virasoro blocks have a large $c$ expansion \cite{Monodromy,Zamolodchikovq} reminiscent of the semiclassical expansion of a path integral.  This means that at large $c$ with fixed ratios $\eta \equiv h / c$,  the perturbative large $c$ expansion has the structure 
\be
\CV(h_i, c; z) &\approx& e^{ -\frac{c}{6} f(\eta_i, \eta_I,  z)} \sum_{n=0}^\infty \frac{1}{c^n} g_n(\eta_i, \eta_I,  z),
\label{eq:pertlargec}
\ee
where $f$ is finite as $c \to \infty$.  We have also separated out the internal primary dimension $\eta_I$ from the external operator dimensions.  One justification for this expansion is that Virasoro blocks should be computable using Chern-Simons theory \cite{Witten:1988hf, Witten:1988hc, Verlinde:1989ua, Elitzur:1989nr, Witten:2010cx, Gaiotto:2011nm}, a fact we hope to return to in the future. More practically, (\ref{eq:pertlargec}) has been checked against the non-perturbative definition (\ref{eq:SchematicVacuumBlock}) to high orders in a small $z$ expansion, and in other ways in various specific limits.  

Our main goal in this paper will be to compute and classify the $f(\eta_i,  z)$ and all other semiclassical `instantons' or `saddles' in certain kinematic limits.  In section \ref{sec:BlockExpansions} we will discuss the meaning of both the perturbation series in $\frac{1}{c}$ and the semiclassical saddles; then in section \ref{sec:ClassificationSaddles} we will review how these saddles can be identified and classified near $z=0$.  In the remainder of section \ref{sec:LateTimeMonodromy} we compute and classify all the saddles in the vicinity of $z=1$; as we explain in that section, this will allow us to determine their late time behavior.

\subsection{Transseries Expansion of the Virasoro Blocks in $1/c$}
\label{sec:BlockExpansions}

Equation (\ref{eq:pertlargec}) is an asymptotic series with zero radius of convergence in the small parameter $\frac{1}{c}$.  This leads us to expect that there must be sub-leading semiclassical `instantons' or `saddles' that contribute to $\CV$ at a non-perturbative level.  The sharpest way to see that (\ref{eq:pertlargec}) is missing subleading saddles is to study cases where some of the external operators are degenerate, i.e. have null Virasoro descendants.  In this case, the blocks can be computed explicitly and one can see that additional saddles appear  \cite{Fitzpatrick:2016ive} after analytically continuing $z$ along certain paths in the complex plane.  The sub-leading saddles eventually dominate over the leading saddle in certain regimes \cite{Chen:2016cms}.

However, to place the subleading saddles in a more general context where the exact form of the blocks may not be known, it is useful to motivate them in terms of Borel resummation of asymptotic series and resurgence phenomena \cite{Dorigoni:2014hea, Dunne:2012ae, Basar:2013eka, Cherman:2014ofa, Dunne:2015eoa,Dunne:2016jsr,Kozcaz:2016wvy,Aniceto:2011nu,Schiappa:2013opa,Aniceto:2013fka,Couso-Santamaria:2015wga}. To summarize the logic, we can define a Borel series $B(s)$  by sending $g_n \to \frac{1}{n!} g_n$ in equation (\ref{eq:pertlargec}).
Then  we can try to define a function
\be
\CV = c \int_0^\infty ds  \, e^{-s c}  B (s)
\ee
as the Borel transform, which reproduces the series expansion if we expand $B$ order-by-order in $s$.  If the Borel integral converges and has no singularities on the real axis, then it can be viewed as another definition of the Virasoro block $\CV$, and could be explicitly verified by comparison with equation (\ref{eq:SchematicVacuumBlock}).  

Singularities of $B(s)$ in the $s$-plane  lead to branch cuts when $\CV$ is analytically continued \cite{Basar:2013eka} in its various parameters, which include $\eta_i$, $c$, and the kinematic variable $z$.  We expect that we can identify these singularities in the Borel plane with corresponding semiclassical saddles \cite{Fitzpatrick:2016ive, 'tHooft:1977am}.  If we were able to obtain $\CV$ from a path-integral computation, then we could go further (see \cite{Witten:2010cx} for a relevant review), identifying the semiclassical saddles as solutions to specific equations of motion, and interpreting singularities in the Borel plane as due to path-integral Stokes phenomena \cite{Witten:2010cx}.

Either way, there is a picture where equation (\ref{eq:pertlargec}), supplemented with a prescription for the proper Borel or path integration contour, may be sufficient to define $\CV$ \cite{Dorigoni:2014hea, Basar:2013eka, Dunne:2015eoa,Dunne:2016jsr,Kozcaz:2016wvy,Aniceto:2011nu,Schiappa:2013opa,Aniceto:2013fka,Couso-Santamaria:2015wga,Cherman:2014ofa,Argyres:2012ka,Dunne:2012ae,Dunne:2014bca, 'tHooft:1977am}. 
  As we analytically continue away from that limit, we will eventually cross Stokes lines (or equivalently, encircle singularities in the Borel plane), and so we will be forced to add contributions from sub-leading saddles.
Bringing the non-perturbative effects into view, one can write an improved expansion for $\CV$ as a double-sum, or ``transseries'' \cite{Dorigoni:2014hea}, 
\be
\CV(h_i, \eta_I,  c; z) &=& \sum_{p=0}^\infty\sum_{n=0}^\infty e^{ - \frac{c}{6} f_p(\eta_i, \eta_I,  z)} \frac{g_{p,n}(\eta_i,  z)}{c^n} .
\label{eq:translargec}
\ee
We can think of the subleading terms $p>0$ as instanton contributions that fill in the interior of AdS$_3$ with particular geometries. 

As we continue further  in $z$, the coefficients of the various saddles will  change as we cross Stokes lines.  At any fixed finite value of $z$, we may find that $\CV$ is dominated by the saddle with minimum Re$[f_p]$ that appears\footnote{This innocuous-seeming statement may be subtle in practice \cite{Chen:2016cms} if there are an infinite number of saddles.} with a non-zero coefficient in equation (\ref{eq:translargec}). However, if we take a limit such as large $t$ with $z = e^{-it}$ (the large time limit pictured in figure \ref{fig:HeavyLightLightconesandPlane}), then we cross more and more branch cuts.  In this case we  do not expect that any particular saddle will dominate $\CV$, particularly because  the coefficients of these saddles will  depend directly on $t$.  This means that knowing the large time behavior of the individual saddles $f_p$ will not immediately tell us the large time behavior of the Virasoro blocks.\footnote{At least not in the $\CO_H \CO_H \to \CO_L \CO_L$ OPE channel, or any channel where we must cross more and more Stokes lines as $t$ increases. }  

We emphasize that these statements are not just hypothetical -- they have already been demonstrated \cite{Fitzpatrick:2016ive} in the case of simple degenerate operators.  In those special cases, the Coulomb gas formalism \cite{Dotsenko:1984nm, Dotsenko:1984ad, DiFrancesco:1997nk} plays a role analogous to that conjectured for Chern-Simons theory in the general case.  This strongly suggests that the Chern-Simons description of Virasoro blocks must reduce to the Coulomb gas in the case of degenerate operators \cite{Gaiotto:2011nm}.  We have also shown that the semiclassical saddles can be used  \cite{Chen:2016cms} to obtain a useful and explicit result for a CFT$_2$ correlator.  In that example we found that there were an infinite number of saddles, and they themselves needed to be Borel resummed in order to obtain a finite result.\footnote{Somewhat mysteriously, the result matches an AdS$_2$ computation (compare 6.57 of \cite{Maldacena:2016upp} with 4.14  of \cite{Chen:2016cms}).}  

Classifying the semiclassical saddles provides important information about the behavior of the blocks away from Stokes lines. 
In particular, it tells us the behavior of each individual instanton, including the leading saddle which governs the large time behavior of the semiclassical Virasoro blocks.  This makes it possible to prove that the leading semiclassical Virasoro blocks all decay at the same exponential rate in the heavy-light limit.  A crucial consequence of this result is that the information loss problem must persist after including semiclassical conformal blocks for heavy states with $h_I \sim \CO(c)$.

\begin{figure}[t!]
\begin{center}
\includegraphics[width=0.7\textwidth]{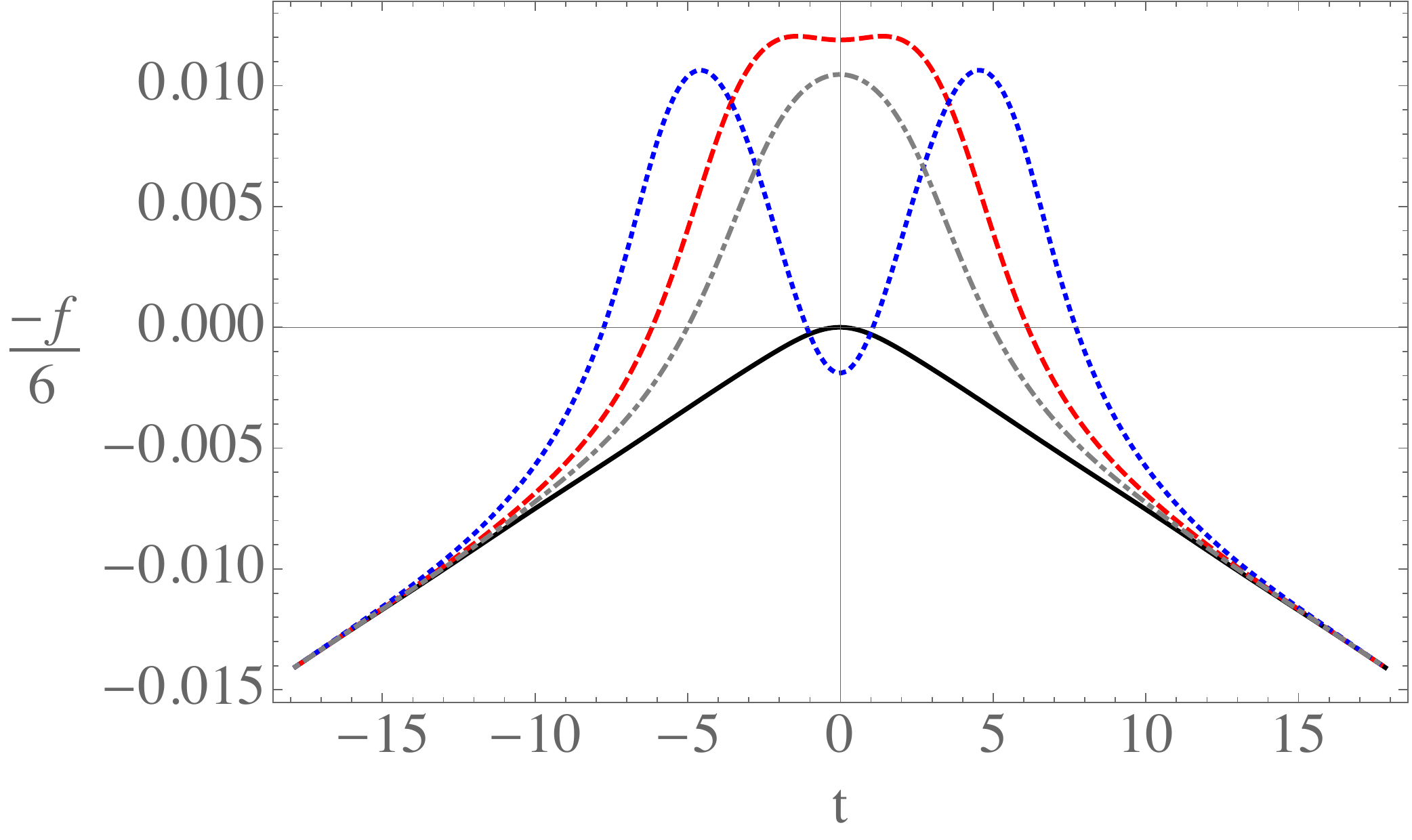}
\caption{This figure shows the time-dependence of leading semiclassical saddles contributing to $\frac{1}{c} \log \CV$, with different $\alpha_I = 1, 3/5, i/2, 5i/4$ (black, solid; gray, dot-dashed; red, dashed; and blue, dotted, respectively) and fixed $\alpha_L =0.99$  and $T_H = 2 \pi$.  The solid black line corresopnds to $\alpha_I = 1$, which is the vacuum Virasoro block. For ease of comparison we have made an overall constant shift in each $f$ to emphasize that the late-time exponential  decay is completely independent of the intermediate operator dimension. See fig. \ref{fig:FollowLeadingNonVac} for more details. }
\label{fig:MoneyPlot}
\end{center}
\end{figure}

\subsection{Summary of Results}

Writing semiclassical contributions to the Virasoro blocks as $e^{ - \frac{c}{6} f(\eta_i, \eta_I,  z)}$ with $z \equiv 1 - e^{-it}$, we find two discrete infinite classes of $\kappa = z(1-z) \partial_z f$ in the limit $t \to \pm \infty$.  All of the sub-leading saddles may be interpreted as `additional angles' in AdS$_3$, as depicted in figure \ref{fig:surplus}.  

We define $\alpha_X \equiv \sqrt{1 - \frac{24 h_X}{c}}$; we will take $\alpha_L$ to be real and $\alpha_H = 2 \pi i T_H$ to be purely imaginary, as this is the case of interest for correlators probing BTZ black holes.  
The first infinite class are the decaying saddles, with asymptotic $\kappa$ of the form
\be
\kappa_{\mathrm{dec}}(n) &=& n(1-n) - \frac{1}{2} + \left(\frac{1}{2} - n \right) (\alpha_L \pm \alpha_H) \mp \frac{\alpha_L \alpha_H}{2},
\ee
where $n$ must be an integer, as discussed near equation (\ref{eq:KappaSolns}).  The leading semiclassical contribution to the Virasoro blocks is the case $n=0$.  For all values of $n$, the $\pm$ signs are always dynamically chosen (by following the solutions from early to late times) so that 
\be
\CV(t) &\stackrel{|t| \rightarrow \infty} \sim& 
e^{i \theta(t)} \exp \left[- \frac{\pi}{6} \Big( |2n-1|\pm \alpha_L \Big)  c T_H |t|  \right]
\ee
decyas as $|t| \to \infty$ for real $\alpha_L$ and real $T_H$.  

The other infinite class are the oscillating saddles which approach
\be
\kappa_{\mathrm{osc}}(m) = \frac{1}{4} \left( \alpha_H^2 + \alpha_L^2 - 1 \right) - \frac{1}{4} \delta^2(t)
\ee
at late times, where the function $\delta(t) \approx \frac{2 \pi m}{t} + \cdots$ is given in terms of an arbitrary integer $m$, and is specified more precisely in and around equation (\ref{eq:newsolutions}).    For physical values of the external operator dimensions, $\kappa_{\mathrm{osc}}$  approaches a real number at late times; this is what indicates that these saddles oscillate rather than decay.  

Near the OPE limit $z \sim 0$ classification of saddles becomes very easy, as it depends solely on the power-law behavior of $\CV(z)$ as $z \to 0$.   However, the connection between the classification of saddles near $z \sim 0$ and $z \sim 1$ appears to be rather complicated, and we have not fully mapped it out.

All semiclassical saddles have a leading large time behavior that is independent of the intermediate operator dimension $h_I$.   This somewhat surprising fact accords with all prior calculations.  The dependence on $h_I$ appears only at sub-leading order in the late time limit, controlling the rate at which $\kappa(t)$ approaches its asymptotic value.  In figure \ref{fig:MoneyPlot}  we show the time-dependence of $-\frac{1}{6} f$ for a variety of leading and sub-leading saddles to illustrate the range of possible behaviors.  Some saddles can grow before they ultimately decay; it would be interesting to understand the parametric details of this phenomenon.

The outline of this paper is as follows.  Section \ref{sec:LateTimeMonodromy} is devoted to an analytic classification of the saddles near $z \sim 0$ and $z \sim 1$.  We review the monodromy method and the classification near $z \sim 0$ in section \ref{sec:ClassificationSaddles}.  Then in section \ref{sec:HowtoGetLateTime} we explain why the behavior of the saddles near $z \sim 1$ is sufficient to understand their late time behavior.  Finally in section \ref{sec:MonodromyComputation} we solve the monodromy problem analytically near $z \sim 1$ and classify the possible late time behaviors of the saddles with section \ref{sec:SemiclassicalSaddlesLateTime}.  In section \ref{sec:BehaviorIntermediateTime} we connect the saddles near $z \sim 0$ to those $z \sim 1$, and study their time-dependence and the way they approach their asymptotic behavior at late times.  We match our analytic solutions to numerics, and in the process obtain many consistency checks.

In section \ref{sec:DegenerateStates} we use a very different `algebraic' method to compute the semiclassical saddles associated with correlators of degenerate external operators.   We also provide a partial derivation of the monodromy method, based on analytic continuation from the algebraic method, in section \ref{sec:MonodromyMethodfromAlgebraicMethod}.  We conclude with a discussion in section \ref{sec:Discussion}.  In appendix \ref{app:Generality} we make some more detailed comments about the generality of our results.

\section{Saddles and Their Late-Time Behavior from the Monodromy Method}
\label{sec:LateTimeMonodromy}

In this section we will classify all semiclassical saddles contributing to the Virasoro conformal blocks, and we will calculate their late-time behavior.  We show that the leading semiclassical Virasoro blocks all decay exponentially, and at a rate that is independent of intermediate operator dimensions.  We also identify two much larger classes of semiclassical solutions; one class decays even more rapidly at late times, while the other approaches a constant magnitude.

These computations are tractable because the late-time behavior of the semiclassical Virasoro blocks can be determined  by focusing on the region of small $|z - 1|$, as illustrated in figure \ref{fig:HeavyLightLightconesandPlane}.  So our strategy will be to solve the monodromy method directly in this kinematic region, keeping the phase of $z-1$ arbitrary.  

First, in section \ref{sec:ClassificationSaddles}, we will review the monodromy method and classify all saddles near the OPE limit $z \to 0$.  In section \ref{sec:HowtoGetLateTime} we explain the kinematics of the late time limit, and then in section \ref{sec:MonodromyComputation} we perform the relevant monodromy method computations in this limit.  
In section \ref{sec:SemiclassicalSaddlesLateTime} we classify the solutions analytically and determine their late time behavior, while in section \ref{sec:BehaviorIntermediateTime} we analyze the behavior of the solutions at intermediate times.

\subsection{The Monodromy Method and the Classification of Saddles Near $z=0$}
\label{sec:ClassificationSaddles}

\begin{figure}[t!] 
\begin{center}
\includegraphics[width=0.45\textwidth]{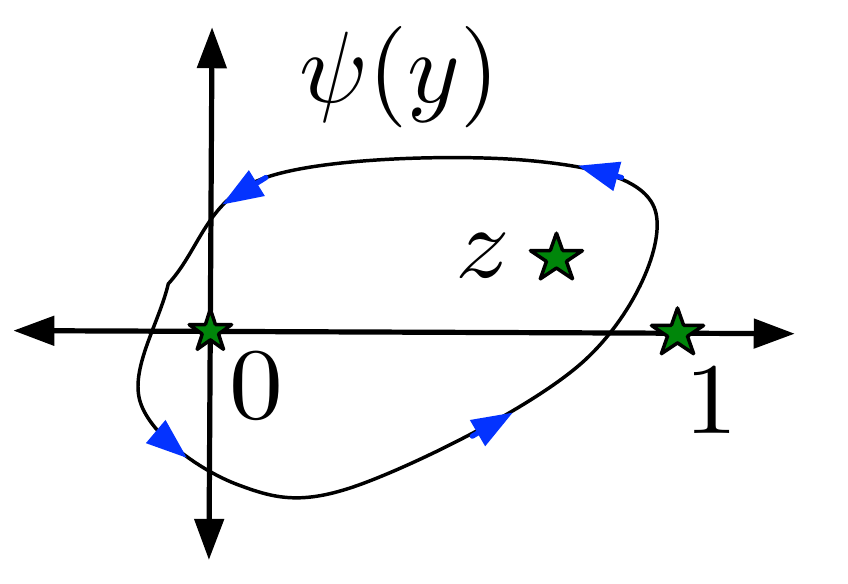}
\caption{ The path along which we must translate $\psi(y)$, the solutions to equation (\ref{eq:MonodromyEquation}), in order to define a $2 \times 2$ monodromy matrix.   The path must encircle $y=0$ and $y=z$, but not $y=1$. }
\label{fig:MonodromyPath} 
\end{center} 
\end{figure}

A remarkably effective method for investigating the semiclassical functions $f_p(\eta_i, \eta_I, z)$ is the ``monodromy method'' developed\footnote{For detailed reviews see e.g.  appendix C of \cite{Fitzpatrick:2014vua} or appendix D of \cite{HarlowLiouville}; sometimes the method is known as ``the method of auxiliary parameters''.}   in  \cite{Monodromy,Zamolodchikovq}.    It involves solving the following math problem, which is formulated in terms of the auxiliary parameter $\kappa(z)$ defined by
\be
 \kappa(z) \equiv z (1-z) \partial_z f_p(\eta_i, \eta_I, z).
 \label{eq:KappaVsFp}
 \ee
  By inserting into the conformal correlator an additional operator $\psi_{2,1}$ that has a degenerate level 2 Virasoro descendant, $(L_{-2} - \frac{3}{2(2h_\psi+1)} L_{-1}^2)| \psi_{2,1}\>=0$, one obtains the second order differential equation
\be
\label{eq:MonodromyEquation}
\psi''(y) + T(y;z) \psi(y) = 0,
\ee
where $T(y; z)$ is the value the classical stress tensor takes at position $y$, while $z$ is the holomorphic cross-ratio of equation (\ref{eq:HHLL4pt}).  In this paper, we will be focused on the case $\eta_1 = \eta_2 \equiv \eta_L, \eta_3 = \eta_4 \equiv \eta_H$, mainly for simplicity but also because this is the case where the identity block can contribute.  Then,
\be
\label{eq:MonodromyStressTensor}
T(y;z) = -\frac{\kappa(z)}{(1-y) y (y-z)}+\frac{6 \eta _H}{(1-y)^2}+6 \eta _L
   \left(\frac{1}{y^2}+\frac{1}{(y-z)^2}+\frac{2}{y}+\frac{2}{1-y}\right).
\ee
There will be two solutions, $\psi_1$ and $\psi_2$, and if we track their behavior as $y$ follows a closed path encircling $0$ and $z$, as pictured in figure \ref{fig:MonodromyPath}, they must transform into a new linear combination according to
\be
\left( \begin{array}{c}
\psi_{1} \\
\psi_{2}  \end{array} \right) \to M \left( \begin{array}{c}
\psi_{1} \\
\psi_{2}  \end{array} \right) 
\ee
for some $2 \times 2$ matrix $M$. The auxiliary parameter $\kappa(x)$ is fixed by the condition that $M$ have eigenvalues 
\be
{\rm eigenval}(M) = -e^{\pm i \pi \alpha_I}, \qquad  \alpha_I \equiv \sqrt{1 - \frac{24 h_I}{c} },
\label{eq:MonodromyInternal}
\ee
where  $h_I$ is the intermediate operator dimension.  Since the product of the eigenvalues is $1$, we can simplify this to a constraint on the trace of $M$.  The main challenge then is to determine $M$ as a function of $\kappa$.  

This problem is difficult because solutions to equation (\ref{eq:MonodromyEquation}) do not have a simple integral representation for general values of the parameters.  It was solved by Zamolodchikov in the limit $\kappa \to \infty$, and more recently by us \cite{Fitzpatrick:2014vua} and others \cite{Anous:2016kss,Hijano:2015rla,deBoer:2014sna} in a perturbative expansion in $\eta_L$ and $\eta_I$.  The general problem is equivalent to the question of the monodromy of Heun's functions  \cite{ronveaux1995heun}, and to the connection problem of Painlev\'e IV \cite{ZamolodchikovPainleve}, and the solution is not known in analytic form.

Now suppose that one has obtained $M$ as a function of $\kappa$.  At this point, the reader may  be wondering which of the distinct saddles $f_p$ (with fixed values of all $\eta$s) is related to $\kappa$.  The answer is all of them! 

For generic values of parameters, solving the monodromy condition will produce an infinite number of solutions for $\kappa$. 
One of these solutions will be the $\kappa$ corresponding to the leading saddle, but in fact the other solutions correspond to subleading saddles.  In this way, the function $M[\kappa]$ contains an enormous amount of information about the Virasoro block $\CV$, holding the key not only to the leading semiclassical behavior but also to other non-perturbative effects as well, as we discussed in section \ref{sec:BlockExpansions}.

One can argue that the sub-leading saddles should all have the same monodromy as the leading saddle by applying the original logic of the monodromy method to the full non-perturbative series (\ref{eq:translargec}). Adding $\psi_{2,1}(y)$ in the correlator produces the sequence
\be
\Psi_{\rm tot}(\eta_i, \eta_I, c; z, y) &=& \sum_{p=0}^\infty \psi_p(z,y) e^{ - \frac{c}{6} f_p(\eta_i, \eta_H, z)} ,
\label{eq:translargecPsi}
\ee
the main point being that because $\psi_{2,1}$ is a light (i.e. $h_{2,1} \sim \CO(1)$) operator, its presence can shift the perturbative parts $g_{p,n}$ but not the non-perturbative pieces $f_p$.  Acting with the degenerate combination $L_{-2} + \frac{3}{2(2h_\psi+1)} L_{-1}$ produces the differential equation (\ref{eq:MonodromyEquation}) for each $\psi_p$, but with $\kappa$ given by the derivative of the corresponding $f_p$.  
To leading order at large central charge, the monodromy of the total combination $\Psi_{\rm tot}$ must still be given by the matrix $M$ as $y$ encircles $0$ and $z$.  However, it is manifest that the saddles do not mix under this monodromy, since they are completely independent of $y$.  So each $\psi_p$ should individually have a monodromy matrix $M$, and thus $M[\kappa= z(1-z)\partial_z f_p]$ is the same for all $p$.\footnote{ This argument is not completely rigorous since in fact it has never been proven directly from the definition of the blocks even that the leading exponential in the large $c$ limit (with $\eta_i, \eta_I$ fixed) grows like $\CO(c)$, though there are by now a large number of highly non-trivial consistency checks of this behavior; we are assuming this $\CO(c)$ scaling holds for all subleading saddles as well.  The second unproven assumption is that the addition of the $\psi_{2,1}$ light degenerate operator does not affect the $\CO(c)$ part of the exponentials.  Part of the motivation for this assumption comes from Liouville theory where the action is $\CO(c)$ and one can see explicitly that adding light operators creates an $\CO(1)$ shift rather than an $\CO(c)$ shift; however, this last statement concerns the full correlator and it is not clear how to turn it into a proof for the individual blocks. One might also be more skeptical of this assumption for the subleading saddles than for the leading ones, since for any given value of parameters, and to leading order at large $c$, only one saddle will dominate, while the others will be negligible; it is not clear if one can exhibit a (perhaps unphysical) region in kinematic or parameter space where each saddle dominates. } 
 
\begin{figure}[t!]
\begin{center}
\includegraphics[width=0.9\textwidth]{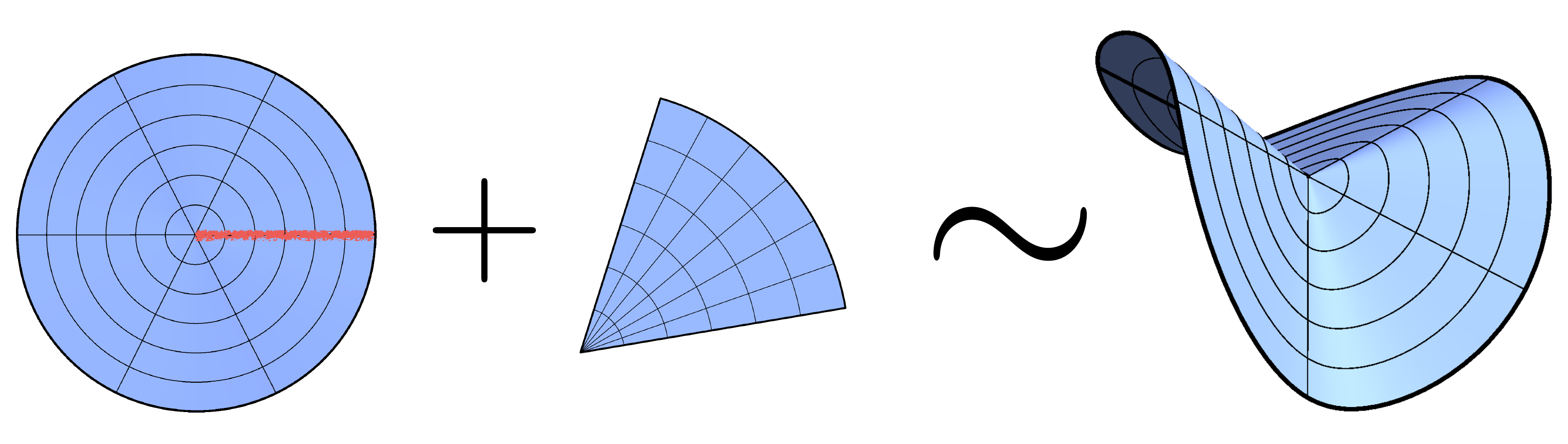}
\caption{A surplus angle geometry from the insertion of a negative weight state in AdS$_3$.}
\label{fig:surplus}
\end{center}
\end{figure} 

Although we cannot solve for $M[\kappa]$ in complete generality, we can immediately solve for it in the limit $z \sim 0$.  This limit includes the OPE limit, though it is more general since it also includes small $z$ on every sheet after analytic continuation.  To obtain $M[\kappa]$ in this limit, we can first set $z=0$ and study the behavior of $\psi(y)$ at small $y$.  To obtain the monodromy as $y$ encircles $0$ and $z$ (now equal), it is sufficient to keep just the leading power, 
\be
\psi(y) \stackrel{y \sim 0 }{\approx} y^\nu . 
\ee
The monodromy equation at small $y$ immediately implies $ \nu = \frac{1\pm \sqrt{1+4 \kappa - 48 \eta_L}}{2}$, i.e. the eigenvalues of $M[\kappa]$ at $z=0$ are
\be
z=0&:& {\rm eigenval}(M)[\kappa] = -\exp \left( \pm i \pi  \sqrt{1+4 \kappa - 48 \eta_L} \right).
\label{eq:SmallYMono}
\ee
Comparing with (\ref{eq:MonodromyInternal}), we see that to have the correct monodromy, the square root term in (\ref{eq:SmallYMono}) must be $\alpha_I + 2p$ for an integer $p$. So,
\be
\label{eq:smallzkappa}
\kappa - 12 \eta_L &\stackrel{z \rightarrow 0}{=}& \frac{1}{4} \Big((\alpha_I+2p)^2-1\Big).
\ee
  This relation in turn gives us the small $z$ behavior of the ``saddles'',
\be
e^{-\frac{c}{6} f} &\stackrel{z \sim 0}{\approx}& z^{h_{\rm inst}  - 2h_L}, \\ h_{\rm inst} &=& \frac{c}{24} (1- (2p+\alpha_I)^2).
\ee
This small $z$ behavior looks naively like an OPE singularity for an exchanged operator of weight $h_{\rm inst}$, which we can identify as the weight of the instanton.  Note that for $p=0$, this weight is just the weight of the block itself,  
\be
p=0 &:& h_{\rm inst} = h_I. 
\ee
Of particular interest is the vacuum block, $h_I=0, \alpha_I=1$, for which we have
\be
h_{\rm inst} = - \frac{c}{24} (1 - (2p+1)^2) .
\ee
This is exactly the large $c$ weight of the $\CO_{2p+1,1}$ degenerate operators \cite{Fitzpatrick:2016ive, Chen:2016cms}. 

 The appearance of negative weights may be surprising, in particular because this means that the saddles will produce stronger singularities at $z\sim 0$ than the identity block OPE singularity.  On further reflection, however, these negative weights and their corresponding singularities are in fact a necessary consequence of the structure of the large $c$ expansion of the blocks.  The point is that the subleading saddles are not present on the first sheet in the complex $z$ plane (we will define this region more precisely in later sections), but rather are generated upon analytic continuation. Passing to higher sheets, one does indeed find stronger singularities in a large $c$ expansion than the OPE singularity \cite{Roberts:2014ifa,Chen:2016cms,Fitzpatrick:2016thx,Hartman:2015lfa}.  In fact, we have already used the behavior of these saddles on the second sheet \cite{Chen:2016cms} to successfully compute the behavior Lorentzian correlators associated with chaos \cite{Roberts:2014ifa} in 2d CFTs \cite{Chen:2016cms}.

Heavy states in AdS back-react on the geometry, and static eigenstates create geometries that at long distances look like deficit angles or BTZ black holes \cite{Fitzpatrick:2014vua}.  In the case of saddles contributing to the vacuum Virasoro block, the conformal weight $h_{\rm inst}$ are negative, so the expression for the deficit angle $\Delta \phi$ is also negative \cite{Deser:1983nh,Fitzpatrick:2014vua}:
\be
\Delta \phi = 2 \pi \left( 1 - \sqrt{1-\frac{24 h_{\rm inst}}{c}} \right) = -4\pi p.
\ee 
Therefore we expect that saddles to look like {\it surplus} angles (depicted in fig. \ref{fig:surplus}) on intermediate slices of AdS$_3$ . 

Going beyond the leading behavior of each saddle at small $z$ is straightforward and can be determined directly from the small $z$ expansion of (\ref{eq:SchematicVacuumBlock}).  This is because all dependence on the index $p$ for the saddle enters in the combination $(\alpha_I+2p)^2$, so the corresponding subleading saddle is the same as the leading saddle for a conformal block with $h_I =c \Big( 1-(\alpha_I+2p)^2 \Big)/24$.  The first few correction terms to (\ref{eq:smallzkappa}) can easily be determined this way, e.g.
\be
\kappa \approx 12 \eta_L (1-z) + \frac{2-z}{8} \left(  (\alpha_I + 2p)^2 -1 \right) + \CO(z^2) .
\ee
In all examples where we can compute $\kappa(z)$, it has a finite radius of convergence around $z \sim 0$.  However, branch cuts develop at larger $z$, and we now turn to methods that will allow us to determine its behavior far from the OPE sheet.

\begin{figure}[t!]
\begin{center}
\includegraphics[width=0.95\textwidth]{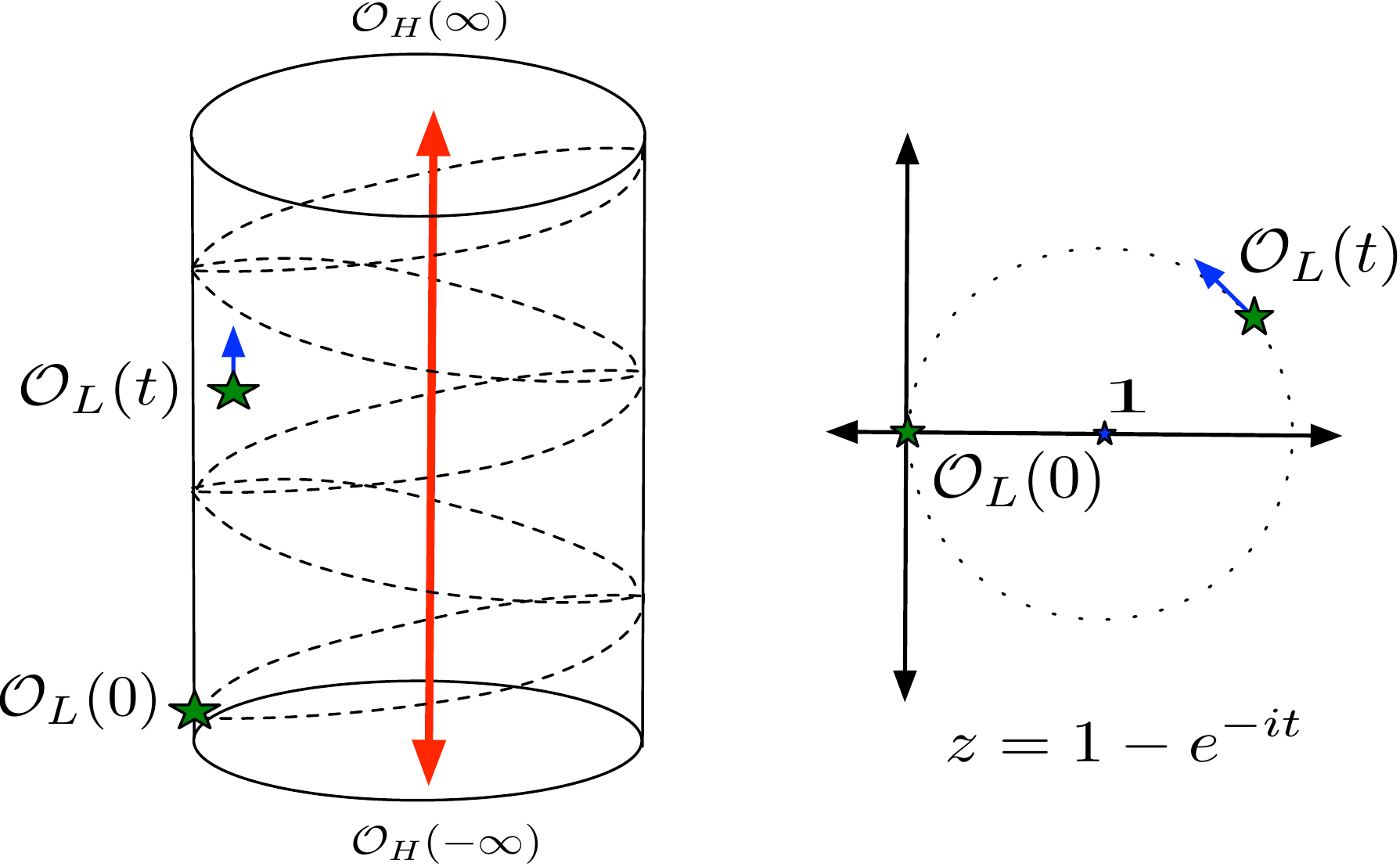}
\caption{This figure depicts a generic configuration of the Lorentzian heavy-light correlator, with dashed lines drawn in to indicate past and future lightcones emanating from the operator $\CO_L(0)$.  Due to the cylindrical geometry, as $t$ increases the operator $\CO_L(t)$ must pass through the future lightcone of $\CO_L(0)$ at regular intervals.  From the point of view of the conventional $z$-plane, depicted on the right, the multi-sheeted CFT correlator transitions to a different sheet for each $2\pi$ increment of $t$. }
\label{fig:HeavyLightLightconesandPlane}
\end{center}
\end{figure}

\subsection{The Kinematic Limit Associated with Late Time Behavior}
\label{sec:HowtoGetLateTime}
 
We want to study the configuration of CFT operators depicted in figure \ref{fig:HeavyLightLightconesandPlane}, which can be interpreted as the 2-pt function of  $\CO_L$ in the background of an energy (dilatation) eigenstate:
\be
\< \CO_H(\infty) |  \CO_L(t_1) \CO_L(t_2)  | \CO_H(-\infty) \> .
\ee  
Here the CFT lives on a cylinder, and $t = t_1 - t_2$ is a Lorentzian time separation.  The $\CO_H$ operators act in the infinite past and future and create a primary state, and so the correlator will be independent of the average time $t_1 + t_2$.  Thus $t$ and a relative angular coordinate on the cylinder will be the only physical variables.

\begin{figure}[t!]
\begin{center}
\includegraphics[width=0.95\textwidth]{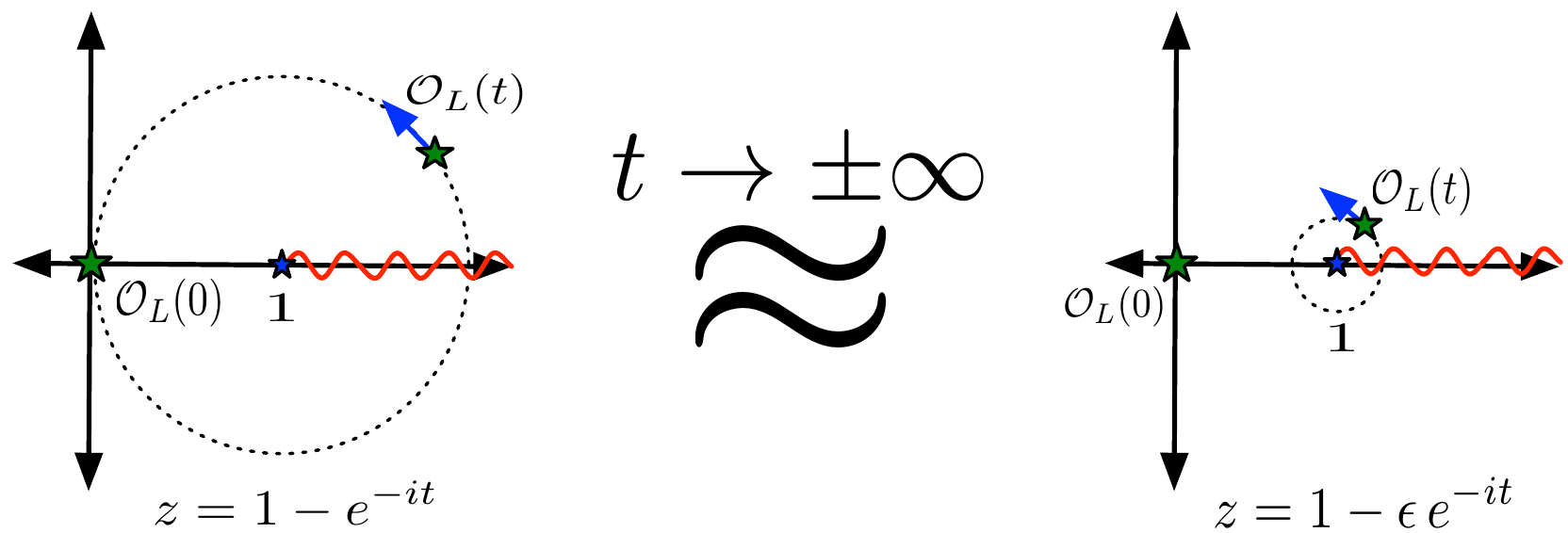}
\caption{This figure suggests how the large time behavior of the heavy-light correlator can be obtained by taking $|1-z|$ small. The semiclassical Virasoro blocks depend on $|t| \gg 1$ through the winding of $z$ roughly $\frac{|t|}{2 \pi}$ times around $z=1$.  This means the late time behavior can be determined by expanding at small $|z-1|$ while keeping the phase of this quantity arbitrary. See figure \ref{fig:HeavyLightLightconesandPlane} for the interpretation of this time-dependence in AdS.}
\label{fig:EquivalenceSmallvsLargeZ}
\end{center}
\end{figure}

This Lorentzian correlator can be obtained from the usual Euclidean 4-pt function, 
\be
\< \CO_H(\infty)   \CO_H(1) \CO_L(z, \bar z)  \CO_L(0) \>,
\ee  
by the substitution $z \to 1 - e^{-it + i\phi}$ and $\bar z \to 1 - e^{-it - i \phi}$ and a simple overall rescaling, which is necessary to pass from the plane to the cylinder.  Perhaps it is surprising that $\CO_H(\infty)$ and $\CO_H(0)$ in the Euclidean plane immediately produce the desired eigenstates in the infinite \emph{Lorentzian} future and past.  This follows from the standard $i \epsilon$ prescription: to pick out the lowest-energy state created by $\CO_H$,  we must give time a small imaginary part as we push the operator insertion into the infinite future or past, and we are left with $\lim_{T \to \pm \infty} \CO_H(e^{\epsilon T})$ which becomes $\CO_H(0)$ or $\CO_H(\infty)$.  The analytic continuations of $\CO_L(0)$ and $\CO_L(z)$ are conventional; for a nice  review in this context see \cite{Hartman:2015lfa}.

Thus as $t \to t + 2\pi$, the $z$ coordinate encircles $1$.  Because both correlators and conformal blocks typically have branch cuts from $1$ to $\infty$,  we will pass onto a new sheet in the complex $z$-plane, as depicted in figure \ref{fig:HeavyLightLightconesandPlane}.  The late time behavior will therefore be governed by the change in the value of the correlator or Virasoro block between sheets.  

In section \ref{sec:ClassificationSaddles}, we saw that it is useful to compute the semiclassical Virasoro blocks $\CV = e^{-\frac{c}{6} f(z)}$ by writing $f$ in terms of a function $\kappa$:
\be
f(z) = \int^z \frac{dx}{(1-x)x} \kappa(x) .
\ee
At late times, $z$ will encircle $1$ again and again, so that the leading late-time behavior will be given by the residue $2\pi i \kappa(1)$.  This dramatically simplifies our task to the computation of a single number!\footnote{Strictly speaking, we will see that $\kappa(z)$ for $z \sim 1$ has non-trivial dependence on the phase of $1-z$, so it is a bit of an abuse of notation at this point to write $\kappa(1)$.  However, we will see that in the limit $t \rightarrow \pm \infty$, $\kappa(1-\epsilon e^{-i t})$ approaches a constant, which is all that is needed in the present discussion.  Astute readers might also worry about encountering forbidden singularities, but these are only present on the Euclidean sheet, and do not interfere with the large time analysis.}

As an example of how this works, let us re-interpret the results of \cite{Fitzpatrick:2014vua} in this light; we will also note two potential pitfalls.  In \cite{Fitzpatrick:2014vua} we found the leading saddle has
\be
\label{eq:OldHeavyLightKappa}
\kappa(z) \approx 6z \frac{ \eta_L \left(\alpha_H - 1 + (1-z)^{\alpha_H} (1 + \alpha_H) \right) -  \eta_I \alpha_H (1-z)^{\frac{\alpha_H}{2}}   }{1 - (1-z)^{\alpha_H}}
\ee
to leading order in $\eta_L$ and $\eta_I$, where $\alpha_H \equiv \sqrt{1 - 24 \eta_H}$.  When $\alpha_H$ is positive and real, we can take the limit $z \to 1$ unambiguously to find
\be
\kappa(1) \approx 6 \eta_L (\alpha_H - 1) .
\ee
The dependence on $\eta_I$ has dropped out entirely.  Continuing to $\eta_H > \frac{c}{24}$, we find the late-time decay of the semiclassical Virasoro block
\be
\CV \sim e^{ -2 \pi T_H h_L  t },
\ee
in agreement with previous results.  However, this analysis neglected two important questions:  how can we control potentially non-analytic behavior of $\kappa(z)$ near $z =1$, and how do we know that the semiclassical blocks are exponentially decaying, rather than exponentially growing? 

 We will deal with the first question by explicitly computing the leading non-analytic pieces of $\kappa(z)$ as a function of $z-1$.  In our example here, it is easy to see that when $z = 1 - e^{-it}$ and $\alpha_H$ is imaginary, at late times the non-analytic pieces such as $(1-z)^{\alpha_H}$ vanish at late times.\footnote{Or for the opposite choice of sign of $\alpha_H$, these terms grow in such a way that there is a cancellation between numerator and denominator in equation (\ref{eq:OldHeavyLightKappa}), so we obtain identical late-time behavior. }  We will obtain similar results for the general semiclassical blocks.  The resolution to the second issue relates closely to that of the first.  Just as the perturbative (in $\frac{h_L}{c}$) solution for $\kappa$ flipped sign at $t \to -\infty$ as compared to $t \to +\infty$, the general solutions will also be dominated by different terms in these two regimes, so that we find decay rather than growth whenever $|t| \to \infty$.  We also demonstrate the transition between these two behaviors numerically in section \ref{sec:BehaviorIntermediateTime}.

\subsection{Evaluation of the Monodromy Near $z=1$}
\label{sec:MonodromyComputation}

In this section we will compute $\kappa(z) \equiv z(1-z) \partial_z f(z)$, where $\CV = e^{-\frac{c}{6} f(z)}$,  to leading order in an expansion in small $|1-z|$.  As we have discussed in section \ref{sec:HowtoGetLateTime}, this is sufficient to determine the late time behavior of the Virasoro conformal blocks in the semiclassical limit of $c \to \infty$ with all $h_i / c$ fixed. Note in particular that the phase of $1-z$ can be  large, as it is only the absolute value of this quantity which is presumed small.

\begin{figure}[t!]
\begin{center}
\includegraphics[width=0.45\textwidth]{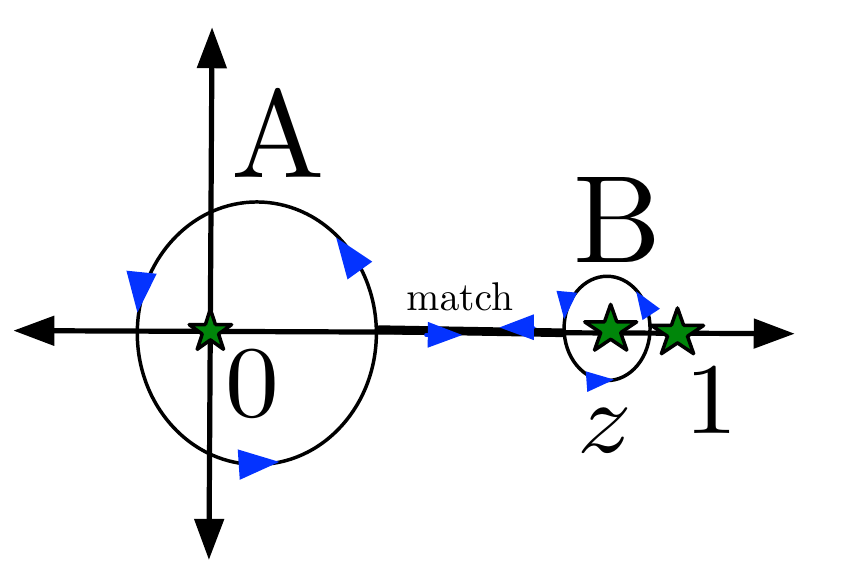}
\caption{ 
To compute the monodromy in the desired limit $z \to 1$, we consider separately (A) generic values of $y$ and (B) the region where $|y - 1| \propto |z - 1| \ll 1$, and then we construct the full monodromy matrix by matching the two solutions in an intermediate regime.  }
\label{fig:MonodromyPath2}
\end{center} 
\end{figure}

Our strategy will be to divide the monodromy path pictured in figure \ref{fig:MonodromyPath2} into two regions where (A)  $y$ is far from $z$ and $1$ and (B) where $y$ and $z$ both approach $1$.\footnote{More precisely, region A is defined as the limit $z \rightarrow 1$ with $y$ fixed, and region B as the limit $z\rightarrow 1$ with the ratio $\frac{1-y}{1-z}$ fixed.}  Remarkably, it is possible to solve equation (\ref{eq:MonodromyEquation}) exactly in both limits.  Then we will compute the full monodromy matrix $M$ as the product of four matrices: the monodromy from circling $0$ using solutions in region (A), a matching matrix between the two regions, the monodromy from region (B) and encircling $z$, and a final matching matrix.

We begin by studying equation (\ref{eq:MonodromyEquation}) in region A, and so we take the limit $z \to 1$ immediately.  From figure \ref{fig:MonodromyPath2}, this should be a good approximation when $y$ is far from both $z$ and $1$, but it must break down as $y$ passes between these two points because $T(y;z)$ has singularities when $y = z$ or $1$.  Equation (\ref{eq:MonodromyEquation}) at $z=1$ simply becomes a hypergeometric differential equation after a re-definition of parameters\footnote{Here and throughout this section, $C$ is a hypergeometric parameter unrelated to the central charge.}
\be
\label{eq:RedefinitionofParameters}
\eta _H &=& \frac{1}{24}  \left(1-(A-B)^2\right),
\nn \\
\eta _L&=& \frac{1}{24} (2-C) C,
\nn   \\
\kappa &=&\frac{1}{2} C (A+B-1)-A B .
\ee
We will write the solutions as
\be
\psi_{11} &=& y^{C/2} (1-y)^{\frac{1}{2} (A+B-C+1)} \,_2F_1(A,B;C;y),
\\
\psi_{21} &=&    y^{1-\frac{C}{2}} (1-y)^{\frac{1}{2} (A+B-C+1)} \,
   _2F_1(A-C+1,B-C+1;2-C;y).
\ee
These solutions have the simple monodromy matrix
\be
\left( \begin{array}{cc}
e^{i \pi C} & 0 \\
0 & e^{-i \pi C}  \end{array} \right) 
\ee
as $y$ encircles $0$, which is labeled as region A in figure \ref{fig:MonodromyPath}.  

 To solve in region B, we will change coordinates to 
 \be
 Z \equiv 1-z, \qquad W \equiv \frac{y-z}{1-z},
 \ee
 so that region B is the limit $|Z| \ll 1 $ with $W$ fixed, and the path B shown in fig \ref{fig:MonodromyPath2} is just $W\rightarrow e^{2\pi i}W$.
Expanding  equation (\ref{eq:MonodromyEquation}) to leading order at small $Z$ leads to another hypergeometric differential equation, with solutions
\be
\psi_{12} &=& W^{C/2} (1-W)^{\frac{1}{2} (A-B+1)} \, _2F_1(A,C-B;C;W),
\\
\psi_{22} &=&  W^{1-\frac{C}{2}} (1-W)^{\frac{1}{2} (A-B+1)} \, _2F_1(1-B,A-C+1;2-C;W),
\ee
which are completely independent of $Z$.
Once again, it is very easy to determine these solutions' relevant monodromy because we need only expand at small $W$, where the solutions behave as simple power-laws.  In fact, as a $2 \times 2$ matrix the monodromy from encircling $z$ in region B is identical to that from circling $0$ in region A. 

The only remaining challenge is to match the solutions between region $A$ and $B$ in the regime where they are both valid, namely $|W| \gg 1$ and $|1-y| \ll 1$.  This can be accomplished with some standard hypergeometric identities,\footnote{For reference, it is useful to note that
\be
\, _2F_1(A,B;C;y) &=& \frac{\Gamma (C) (1-y)^{-A-B+C} \Gamma (A+B-C) \,
   _2F_1(C-A,C-B;-A-B+C+1;1-y)}{\Gamma (A) \Gamma (B)}
\nn   \\
   &&+ \frac{\Gamma (C)
   \Gamma (-A-B+C) \, _2F_1(A,B;A+B-C+1;1-y)}{\Gamma (C-A) \Gamma (C-B)}
\ee
in order to transform from solutions near $y=0$ to those near $y=1$ and vice versa.  Also
\be
\, _2F_1(A,B;C;y)=(1-y)^{-A-B+C} \, _2F_1(C-A,C-B;C;y)
\ee
can be useful when matching.} and can be summarized by a $2 \times 2$ matrix $M_{match}$ defined by
\be
\left( \begin{array}{c} \psi_{11} \\ \psi_{12} \end{array} \right) &=& M_{match} \left( \begin{array}{c} \psi_{21} \\ \psi_{22} \end{array} \right), 
\ee
in the regime of overlapping validity of the two solutions.  The result $M_{match}$ is fairly complicated algebraically, but the end result is given in terms of it by the product. 
\be
M = \left( \begin{array}{cc}
e^{i \pi C} & 0 \\
0 & e^{-i \pi C}  \end{array} \right) 
 \cdot M_{match} \cdot \left( \begin{array}{cc}
e^{i \pi C} & 0 \\
0 & e^{-i \pi C}  \end{array} \right) 
 \cdot M_{match}^{-1}
\ee
Note that $\det M = 1$ automatically, and so the eigenvalues of $M$ will be entirely determined by its trace. To leading order at small $Z$, the trace can be written
\be
\boxed{\tr (M) = 2- 2m_+ m_- - m_+^2 Z^{A+B-C} - m_-^2 Z^{C-A-B} }\ , 
\label{eq:TraceM}
\ee
where it is convenient to separate out the coefficients of the different powers of $Z$ that appear at this order:
\be
\label{eq:mpmm}
m_+ &=&\frac{2 \pi  \Gamma (C-A-B) \Gamma (C-A-B+1)}{\Gamma (1-A) \Gamma (1-B) \Gamma
   (C-A) \Gamma (C-B)}  , \nn\\
    m_- &=& \frac{2 \pi  \Gamma (A+B-C) \Gamma (A+B-C+1)}{\Gamma (A) \Gamma (B) \Gamma (A-C+1)
   \Gamma (B-C+1)}.
\ee
The product $m_+ m_-$ that appears in the $Z$-independent term can be written purely in terms of   sines:
\be
 m_+ m_- =  \frac{4 \sin (\pi A) \sin (\pi B) \sin (\pi (A-C)) \sin (\pi (B-C))}{\sin^2 (\pi (A+B-C))}.
 \ee
Equation (\ref{eq:TraceM}) is a key result in this paper, and it contains the information that we will need to determine the late time behavior of all Virasoro conformal blocks. We expect that deviations from equation (\ref{eq:TraceM}) at higher order in $Z$ will take the form of a power series in $Z$ multiplying each of the terms of order $Z^0, Z^{A+B-C}$, and $Z^{C-A-B}$.  This means that the non-analyticity of the monodromy near $Z=0$ is fully captured by equation (\ref{eq:TraceM}).\footnote{The series coefficients of the higher order terms in $Z$ can have singularities at some values of $A,B,C$, and for such values the higher order terms would not be negligible.  The fact that the non-analyticity is captured by eq. (\ref{eq:TraceM}) means that the size of such correction terms can be diagnosed on the first sheet of the complex $z$ plane, where a numeric analysis of the monodromy differential equation is straightforward.}

\subsection{Classification of Semiclassical Solutions Near $z=1$ at Late Times}
\label{sec:SemiclassicalSaddlesLateTime}

To analyze the solutions for $\kappa(z)$ near $Z=0$, we must first discuss the relative size of the terms in the trace (\ref{eq:TraceM}).  As depicted in fig. \ref{fig:EquivalenceSmallvsLargeZ}, the limit $|Z| \rightarrow 0$ should not be thought of simply as $Z=0$, since for any small finite value $Z=1-z$ has a phase that depends on Lorentzian time.  Furthermore, as time evolves $Z$ repeatedly crosses branch cuts, which typically extend from $1$ to $\infty$.  These observation are crucial for understanding the relative size of the terms equation (\ref{eq:TraceM}), since even for fixed $|Z|$ their relative size will be time-dependent.  

We can parameterize the real and imaginary parts of $Z$ as in fig. \ref{fig:EquivalenceSmallvsLargeZ},
\be
Z &=& \epsilon e^{- i t},
\ee
where $\epsilon$ and $t$ are real.  The approximations leading to the formula (\ref{eq:TraceM}) for the trace assumed $\epsilon \ll 1$, but $t$ can take any value.  So the magnitude
\be
|Z^{A+B-C}| = e^{t \textrm{Im}(A+B-C) - \textrm{Re}(A+B-C) \log(1/\epsilon) }
\label{eq:ZABCmag}
\ee
depends on time $t$ and generically transitions from small values to large values around $t \sim \frac{ \textrm{Re}(A+B-C) \log \epsilon^{-1}}{\textrm{Im}(A+B-C)}$ unless the imaginary part of $A+B-C$ vanishes.  Of course,  this quantity depends on $\kappa$ itself via
\be
A+B-C &=& \pm \sqrt{ \alpha_H^2 + \alpha_L^2 -4 \kappa -1},
\ee
so it is not quite correct to treat $A+B-C$ as a constant for all $t$.  We will see, however, that all solutions have the property that  $A,B,C$ approach fixed values at large $t$.

We will find two classes of solutions:  those with $A+B - C \neq 0$ at large times, and those where $A+B-C \to 0$.   For the former, from (\ref{eq:ZABCmag}) we expect  that either $|Z^{A+B-C}|$ or its inverse will diverge at late times.   So our strategy will be to find all solutions for $\kappa$ assuming that one of $|Z^{\pm(A+B-C)}|$ is large, and then to check for self-consistency of this assumption once we have the solutions in hand.  Then we will study the case $A+B-C \to 0$, finding a separate class of solutions to the monodromy condition.

Note that at generic intermediate times the monodromy condition $\tr(M) = - 2 \cos \pi \alpha_I$ is quite complicated; we will solve it numerically in section \ref{sec:BehaviorIntermediateTime}, demonstrating the validity of  our analytic solutions, and showing how they interpolate between small and large times.

\subsubsection{Solutions with $A+B-C \neq 0$ at Late Times}
\label{sec:ABCneq0}

Let us assume that $A+B-C \neq 0$ at late times.
In the limit that one of $Z^{\pm (A+B-C)}$ can be taken to be large, the constraint equation $\tr(M) = -2\cos \pi \alpha_I$ drastically simplifies because it is  dominated by the singular term from (\ref{eq:TraceM}).  Therefore the leading order equation for $\kappa(z)$ is  either
\be
m_+ = 0 , \ Z^{ (C-A-B)} \rightarrow 0 \ \  \ \ \mathrm{or} \ \ \ \ m_- = 0 , \ Z^{- (C-A-B)} \rightarrow 0
\label{eq:ConsistentRequirement}
\ee
and both are completely independent of $\alpha_I$!   In fact, the value of $\alpha_I$ only becomes relevant when we study the rate at which either $m_\pm \to 0$.

The $m_{\pm}$ coefficients vanish only when one or more of the $\Gamma$ functions in their denominators become infinite. It is then straightforward to write down the solutions to $m_{\pm}=0$ in terms of $A,B,C$. To translate these solutions into solutions for $\kappa$ in terms of $\alpha_H, \alpha_L$, note that the inverse of the set of equations (\ref{eq:RedefinitionofParameters}) has eight solutions.  This is because (\ref{eq:RedefinitionofParameters}) is invariant under the following symmetries:
\be
U_1: (A,B,C) &\rightarrow & (B,A,C), \nn\\
 U_2: (A,B,C) &\rightarrow & (C-A, C-B, C), \nn\\
U_3:  (A,B,C) &\rightarrow& (1-A, 1-B, 2-C),
\label{eq:ABCSym}
\ee
and the full eight transformations generated by them.  The first of these leaves $m_{\pm}$ and $A+B-C$ alone, whereas the second and third interchange the signs of $m_{\pm}$ and $\pm(A+B-C)$: 
\be
U_1: && m_{\pm} \rightarrow m_{\pm} , \qquad A+B-C \rightarrow A+B-C , \nn\\
U_2 \textrm{ or } U_3: &&  m_\pm \rightarrow m_\mp , \qquad A+B-C \rightarrow C-A-B.
\ee
This observation is quite useful because it means that $m_+=0$ and $m_-=0$ have the {\it same} solutions for $\kappa$ as a function of $\alpha_H, \alpha_L$, but that the value of $A+B-C$ for each solution depends on whether it is a zero of $m_+$ or $m_-$.  

Without loss of generality, we can therefore  study all solutions to $m_+ m_-=0$, which means that
we must have
\be
\sin (\pi  A) \sin (\pi  B) \sin (\pi  (A-C)) \sin (\pi  (B-C))  = 0 ,
\label{eq:KappaConditionSimp}
\ee
and $A+B-C \notin \mathbb{Z}$.
Translating this condition into the solutions for $\kappa$ using equation (\ref{eq:RedefinitionofParameters}), we obtain the following two infinite towers of solutions, each parameterized by an integer $n$:
\be
\kappa &=& n(1-n) - \frac{1}{2} + \left(\frac{1}{2} - n \right) (\alpha_L \pm \alpha_H) \mp \frac{\alpha_L \alpha_H}{2},
\label{eq:KappaSolns}
\ee
The two towers are related to each other by $\alpha_H \rightarrow - \alpha_H$.   Note that $n \to 1-n$ is equivalent to simultaneously flipping the signs of both $\alpha_L$ and $\alpha_H$.

We are not quite done because we must now check that these solutions are self-consistent with the assumption on the magnitude of $Z^{A+B-C}$.  For this, we have to keep track of when the solutions to (\ref{eq:KappaConditionSimp}) come from $m_+=0$ or $m_-=0$, since this determines the value of $A+B-C$.  
After some straightforward book-keeping, one finds that 
\be
\left\{ \begin{array}{cc} C-A-B, & m_+=0 \\

 A+B-C, & m_-=0 \end{array} \right\} &=&{\rm sgn} \left(n-\frac{1}{2} \right) \left( 1 -2n \mp \alpha_H - \alpha_L\right),
\label{eq:CAB}
\ee
where $n$ and the sign on $\alpha_H$ correspond to those in (\ref{eq:KappaSolns}). Consistent solutions are those satisfying (\ref{eq:ConsistentRequirement}), i.e.  $|Z^{A+B-C}| \rightarrow 0$ if $m_-=0$ or $|Z^{C-A-B}|\rightarrow 0$ if $m_+=0$.  

There are a number of different cases to treat depending on the relative sizes of the imaginary and real parts of $\alpha_H$ and $\alpha_L$.  For the sake of brevity, here we will assume that $h_H > \frac{c}{24} > h_L$, which includes the physically interesting case corresponding to a probe correlator in a black hole background. 
We can then also take Im$(\alpha_H) \equiv 2\pi T_H >0$ without loss of generality.  

Now we have Im$(A+B-C) \ne 0$, so the early- and late-time magnitude of $|Z^{A+B-C}|$ is determined by the sign of Im$(A+B-C) $. Using (\ref{eq:CAB}), we can  summarize the consistency requirement (\ref{eq:ConsistentRequirement}) as 
\be
&&\pm {\rm sgn} \left( n-\frac{1}{2} \right) = +1, \qquad  \textrm{late times}, \, t \rightarrow \infty, \\
&&\pm {\rm sgn} \left( n-\frac{1}{2} \right) = -1, \qquad  \textrm{early times}, \, t\rightarrow - \infty.
\label{eq:nindexcondition}
\ee
where the $\pm$ symbol has been chosen to be consistent with the $\pm$ symbol in equation (\ref{eq:KappaSolns}).  The above condition on $n$ separates the solutions (\ref{eq:KappaSolns}) into a set of non-overlapping early-time and late-time solutions, so we expect that at intermediate times $\kappa$ should interpolate between them.  We will see this explicitly in our numeric analysis in section \ref{sec:BehaviorIntermediateTime}, and find that for generic values of parameters, each solution for  $\kappa$ at $t=-\infty$ follows a path at intermediate times to  its complex conjugate $\kappa^*$ at $t=\infty$. 

We can use these results to predict the late time behavior of the semiclassical Virasoro blocks.  
Taking $\alpha_H = 2 \pi i T_H$, we find the key result
\be
\CV(t) &\stackrel{t\rightarrow \infty} \sim& 
e^{i \theta(t)} \exp \left[- \frac{\pi}{6} \Big( |2n-1|\pm \alpha_L \Big)  c T_H t  \right]
\label{eq:LateTimeBehavior}
\ee
at late times.  We have used the fact that only $n$ satisfying (\ref{eq:nindexcondition}) are late-time solutions. The terms shown above encode the exponential decay, and we see that just like $\kappa$, they are independent of the intermediate operator dimension $h_I$, and know only about the integer $n$ parameterizing the discrete infinity of semiclassical solutions. We have also separated out a pure phase piece,
\be
\theta(t) &=&    \frac{1}{6} \left( n(1-n) - \frac{1}{2} + \left( \frac{1}{2} - n   \right) \alpha_L \right) c t ,
\ee
which does not affect the magnitude of $\CV$. If we take $n=0$ and the ``$-$'' branch of the $\pm$ solutions, and expand $\CV$ at small $\frac{h_L}{c}$, the result exactly matches the late time behavior of known heavy-light blocks \cite{Fitzpatrick:2014vua, Fitzpatrick:2015zha, Fitzpatrick:2016ive}.

Thus we have found an infinite class of solutions that all decay exponentially as $|t| \to \infty$.  The solutions with $n=0$ are the leading semiclassical Virasoro blocks, which connect continuously with known solutions \cite{Fitzpatrick:2014vua, Fitzpatrick:2015zha} at small $h_L$.

\subsubsection{Solutions with $A+B-C \to 0$ at Late Times}
\label{sec:ABCto0}

Now let us consider the case where $A+B-C \to 0$ at late times, meaning  that we cannot immediately assume that $|Z^{A+B-C}| \to 0$ or $\infty$.  In this case, and in fact more generally if $A+B-C \to n$ for some integer $n$,  the parameters $m_\pm$ appearing in the monodromy relation develop poles (rather than zeroes, as we saw in section \ref{sec:ABCneq0}).  So let us assume that
\be
A+B-C = n + \delta(t)
\ee
for some function $\delta(t) \to 0$ as $t \to \infty$.  Let us first demonstrate that there are no solutions with $n \neq 0$.  To leading order at large times, such a solution would need to behave as
\be
\frac{c_1}{ \delta^2(t) } e^{i t \left( n +  \delta(t) \right) } +  \frac{c_2}{ \delta^2(t) }  \sim 0
\ee
for some complicated but easily computed non-zero constants $c_i$.  But this equation can never be satisfied if $\delta(t) \to 0$ at large $t$, because the first term has a time-dependent phase.  

Solutions can  exist only when $n=0$, and so we will now focus on that case.  From equation (\ref{eq:mpmm}) we see that both $m_+$ and $m_-$ diverge as $\frac{1}{\delta(t)}$, so all terms in the monodromy relation of equation (\ref{eq:TraceM}) will be important.  Expanding that equation at small $\delta$ and large $t$, we find the solution 
\be
\delta(t) &=& \frac{2 \pi m}{t} \left( 1+  i \frac{\delta^{(2)}}{t} + \CO(1/t^2)\right), \nn\\
\delta^{(2)} &=&-H_{\frac{1}{2} \left(-\alpha _H-\alpha _L-1\right)}-H_{\frac{1}{2} \left(\alpha _H-\alpha _L-1\right)}-H_{\frac{1}{2} \left(-\alpha _H+\alpha
   _L-1\right)}-H_{\frac{1}{2} \left(\alpha _H+\alpha _L-1\right)} \nn\\
    && \pm \frac{2 \pi   \cos \left(\frac{\pi  \alpha _I}{2}\right)}{\cos \left(\pi  \alpha
   _H\right)+\cos \left(\pi  \alpha _L\right)}-\log (|Z|),
   \label{eq:newsolutions}
\ee
where the parameter $m$ must be a non-zero integer, and the $H_a$ are harmonic numbers.  
If instead we had taken $|Z| \to 0$ with $t$ fixed, we would have found $\delta(t) \approx \frac{2 \pi i m}{\log Z}$, which presumably interpolates between the regime of large $t$ and small $|Z|$.
 Solving for $\kappa(t)$, at large times
\be
\kappa(t) = \frac{1}{4} \left( \alpha_H^2 + \alpha_L^2 - 1 \right) - \frac{1}{4} \delta^2(t),
\ee
which means that for all values of $m$, we have the same asymptotic limit for $\kappa$.  We also see that the sign of $m$ is physically meaningless, since $\kappa$ does not depend on it.  Furthermore, as long as $h_H$ and $h_L$ are real numbers, $\kappa(\infty)$ will be real.  Thus these solutions do not decay at very late times, but oscillate with a phase independent of both $m$ and $\alpha_I$.  To better understand their behavior at intermediate times, we will need to depend on the numerical solutions of section \ref{sec:BehaviorIntermediateTime}.

\subsubsection{A Comment on Degenerate Operators}

Degenerate operators are those with null Virasoro descendants.  We can obtain much more detailed information about their correlation functions, as we will discuss in section \ref{sec:DegenerateStates}.  But for now we simply note that some of these operators have positive integer values of $\alpha$.  So it is natural to ask if they play any special role in the solutions above?

To answer this question, note that our parameters $C$ and $A-B$ take the values
\be
C &=& 1 \pm \alpha_L
\nn \\
A-B &=& \pm \alpha_H
\ee
so if either of the light or heavy operators are degenerate, then these take integer values.  If we are near a zero or pole of $m_\pm$, then some combination of $A,B,C$ must take an integer value.  Then the fact that $A-B$ or $C$ are also integers can alter the character of the solutions.  For example, in the case where both $\CO_L$ and $\CO_H$ are degenerate, the poles and zeroes of $m_\pm$ will always coincide.    More generally, if $m_+$ has a zero when $A = 0, 1, 2, \cdots$ and $\CO_H$ is degenerate, we find that $B$ will also be an integer, so that in fact $m_+$ can have a double pole.

\section{Saddles at Intermediate Times}
\label{sec:BehaviorIntermediateTime}

To get a more complete picture of the saddles, we now turn to an analysis of their behavior at intermediate values of $z$, away from the OPE and late time regimes.  In order to better understand the connection with the solutions described in the previous sections, we will first follow the solutions from $z=0$ to $z \sim 1$ along the real $z$ axis, which is $t=0$.  Then we will follow the solutions in the regime $z\sim 1$ from $t=0$ to large $|t|$, demonstrating how the early and late time solutions connect to each other.  

This section is rather technical, so readers uninterested in the details concerning the connection between saddles near $z=0$ and $z=1$ may wish to skip to the punchline -- the behavior of the saddles as a function of time $t$.  This is illustrated in some detail with figure \ref{fig:FollowLeadingNonVac}, as was anticipated in figure \ref{fig:MoneyPlot}.

\begin{figure}[t!]
\begin{center}
\includegraphics[width=0.5\textwidth]{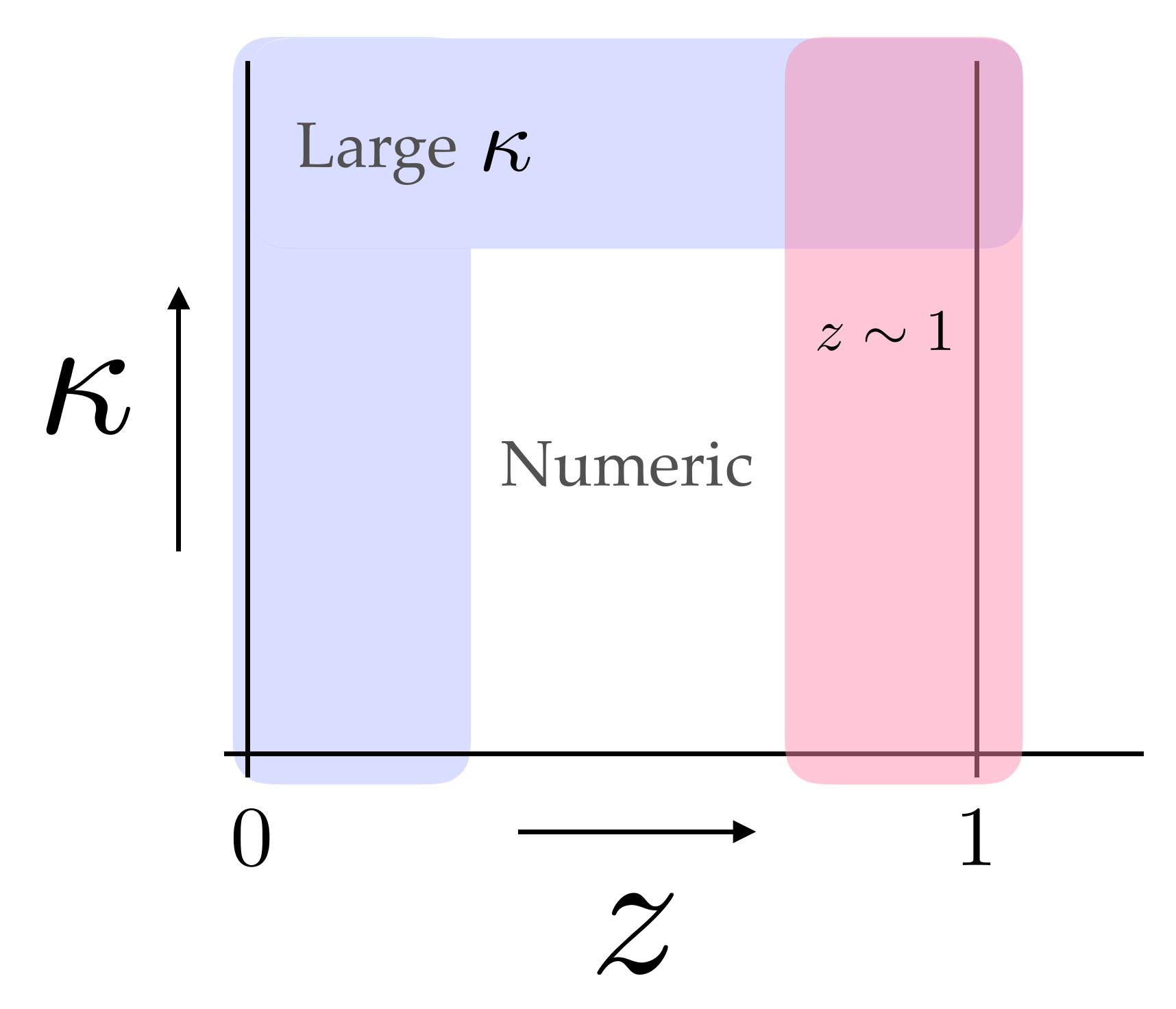}
\caption{Shaded regions indicate where the trace $\tr(M)$ of the monodromy matrix is under analytic control, whereas unshaded region indicate where a numeric analysis of solutions to (\ref{eq:MonodromyEquation}) is needed in order to compute $\tr(M)$.  When $h_L \ll c$, the heavy-light approximation (\ref{eq:OldHeavyLightKappa}) is valid in the additional region of small $\kappa$, for any $z$.}  
\label{fig:analytictraceMregions}
\end{center}
\end{figure}
 
\begin{figure}[t!]
\begin{center}
\includegraphics[width=0.95\textwidth]{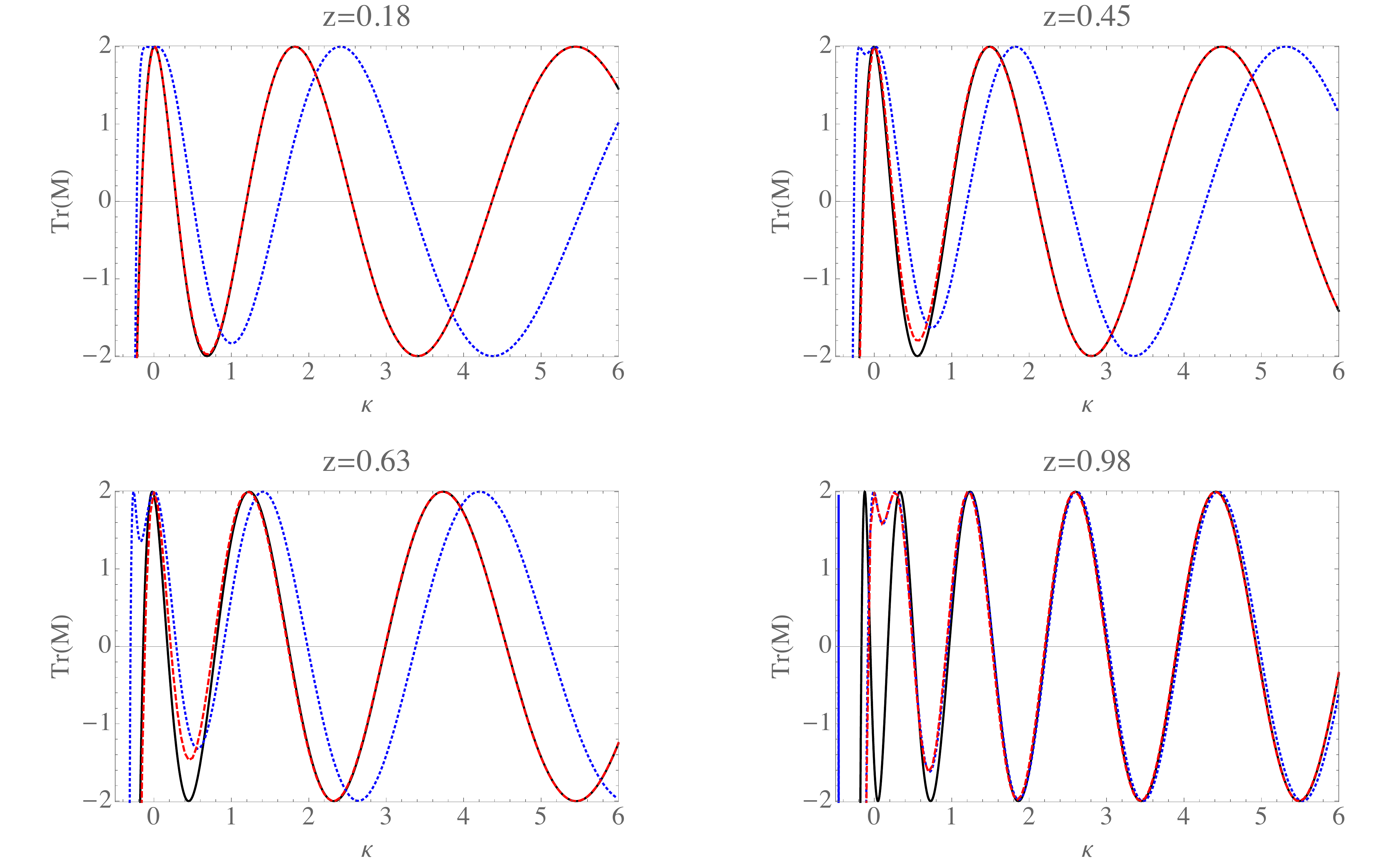}
\caption{Comparisons of analytic approximations to $\tr(M)$ with numerical results.  {\it Black, solid}: the large $\kappa$ and/or small $z$ approximation (\ref{eq:TraceMZam}).  {\it Blue, dotted}: the $z\sim 1$ approximation (\ref{eq:TraceM}).  {\it Red, dashed}: exact numeric result.   
Parameters are $\alpha_L = 0.99, \alpha_H = i$.  One can see the exact numerical solution transition between the $z \sim 0$ and $z \sim 1$ approximations in the series of plots.   }  
\label{fig:traceMcomparisons}
\end{center}
\end{figure}

\subsection{Connecting the $z \sim 0$ and $z \sim 1$ Regions}

Between $z \sim 0$ and $z \sim 1$, we do not have an analytic expression for the trace of the monodromy matrix as a function of $\kappa$.  To obtain precise results, we can resort to a numerical analysis of the solutions to the monodromy differential equation, and we will in fact do this below using code adapted from \cite{HartmanLargeC}.  However, we can also use the result of Zamolodchikov \cite{Zamolodchikovq} for $\tr(M)$ at large $\kappa$:
\begin{equation}
\tr(M) \approx - 2 \cos \left[ 2 K(z) \sqrt{  4 \kappa -2 \alpha _H^2+z \left(\alpha _H^2-\alpha _L^2+1\right)+\frac{ E(z)}{K(z)} \left(2 \alpha _H^2+2 \alpha _L^2-1\right)} \right] .
\label{eq:TraceMZam}
\end{equation}
Here, $K(z)$ and $E(z)$ are elliptic functions.  By construction, this result is exact in the limit $\kappa \rightarrow \infty$. Notably, it is also exact in the limit $z \rightarrow 0$, and deviates from the exact answer only at $\CO(z^2)$. So between (\ref{eq:TraceM}) and (\ref{eq:TraceMZam}), we have analytic expressions for $\tr(M)$ at small $z$ and at $z \sim 1$, and also at any $z$ when $\kappa$ is large.  These two analytic expressions complement each other and between the two of them, one gets a reasonably good description of $\tr(M)$ for intermediate values of $z$, as shown in figure \ref{fig:traceMcomparisons}.

 \begin{figure}[t!]
\begin{center}
\includegraphics[width=0.95\textwidth]{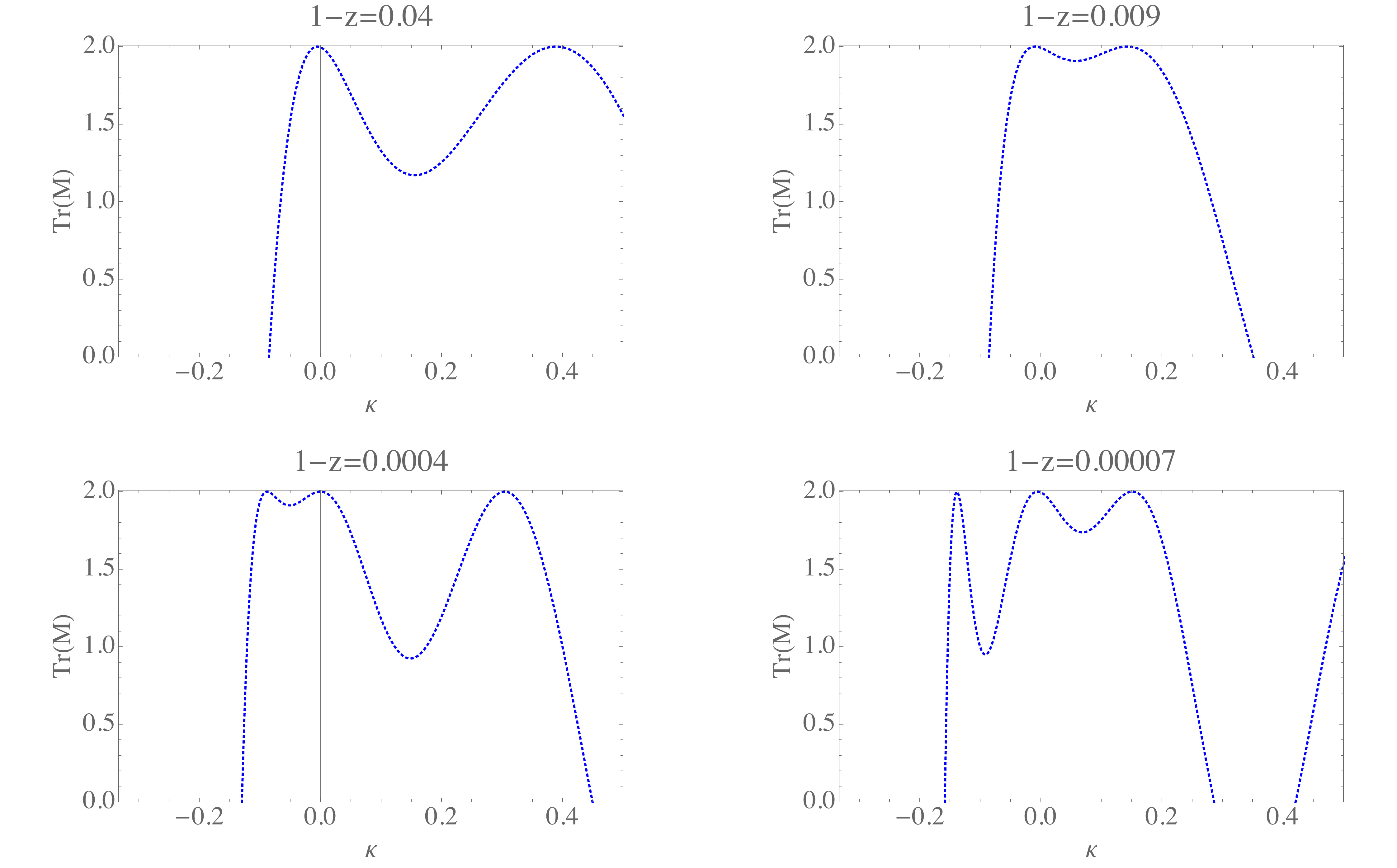}
\caption{Plot of $\tr(M)$ (using (\ref{eq:TraceM})) as $z$ comes very close to 1.  Troughs of the sinusoid become increasingly shallow as they approach $\kappa=0$, and then become deeper again after they pass to $\kappa <0$.  Therefore solutions to $\tr(M) = -2 \cos(\pi \alpha_I)$ approach $\kappa=0$, then move off into the complex plane to circle around $\kappa=0$, and then move back onto the real axis again, but at negative real $\kappa$.  Parameters $\alpha_L, \alpha_H$ are the same as in fig. \ref{fig:traceMcomparisons}.  }  
\label{fig:traceMrealsmallZ}
\end{center}
\end{figure}

A single plot such as those in figure \ref{fig:traceMcomparisons} conveniently allows one to understand solutions for any value of the intermediate dimension $h_I$.  The reason for this is that $h_I$ enters only through the condition that $\tr(M) = -2 \cos (\pi \alpha_I)$, so for different values of $h_I$ one can use the same plot of $\tr(M)$ and simply draw different horizontal lines, looking for where they intersect the curve.  Following solutions from $z \sim 0 $ to $z \sim 1 $ is as simple as watching these intersection points as the curves vary with $z$.  The qualitative behavior of the intersection points as $z$ increases toward 1 can be read off by inspection of (\ref{eq:TraceMZam}) at large $\kappa$ and $z\sim 1$:
\be
\tr(M) \sim -2 \cos \left[ 2 \kappa^{1/2} \log(1-z) \right].
\ee
That is, there are an infinite number of solutions spaced as $\kappa_n \sim n^2/\log(1-z)$, so they collapse toward $\kappa \sim 0$  as $z$ approaches 1. 

The leading saddle is of course of particular interest.  At $z=0$, the leading saddle is $\kappa = 12 \eta_L -6 \eta_I$, and by inspection of  (\ref{eq:smallzkappa}) it is the only $z=0$ solution with $\kappa<0$. This is also clearly visible in fig. (\ref{fig:traceMcomparisons}), where at $z=0$, $\tr(M)$ is monotonically decreasing at $\kappa <0$ and so has only one solution in this regime.  This also makes it particularly simple to follow the leading solution away from $z \sim 0$, since it sits on this left-most falling edge.  

 Once $z$ is sufficiently close to 1, it is more accurate to switch over to the approximation (\ref{eq:TraceM}), where we can follow the solutions to even smaller $1-z$, or from $t=0$ to early or late times.  What is already visible  from fig. \ref{fig:traceMcomparisons} is that as $z$ approaches 1, the bottom of each trough of the sinusoid starts to rise up.  In the last (lower right) panel, one can see that the first (farthest left) trough rises quite significantly. Eventually it rises above the horizontal line at $-2 \cos (\pi \alpha_I)$,\footnote{The identity block $h_I =0$ is an exceptional case since $-2 \cos (\pi \alpha_I) = 2$.} and at this point the $\kappa$ solution moves off into the complex plane.  In fig \ref{fig:traceMrealsmallZ}, we plot what happens as $z$ moves even closer to 1, in the case $\alpha_H=i, \alpha_L = 0.99$.  There, one can see that the troughs of the sinusoid become increasingly shallow as they approach $\kappa=0$.  Eventually, they pass through $\kappa=0$ and move to negative $\kappa$, and start to become increasingly deep again.  At some point, they will be deep enough to intersect the horizontal line $-2 \cos (\pi \alpha_I)$ again, at which point the corresponding solution for $\kappa$ becomes real again.  From that point on, they continue moving to the left, asymptotically approaching the value $\kappa = \frac{1}{4} (\alpha_H^2 + \alpha_L^2 -1 )$ (which is always negative in the region $h_H > \frac{c}{24} > h_L$ that we are considering).

\begin{figure}[t!]
\begin{center}
\includegraphics[width=0.48\textwidth]{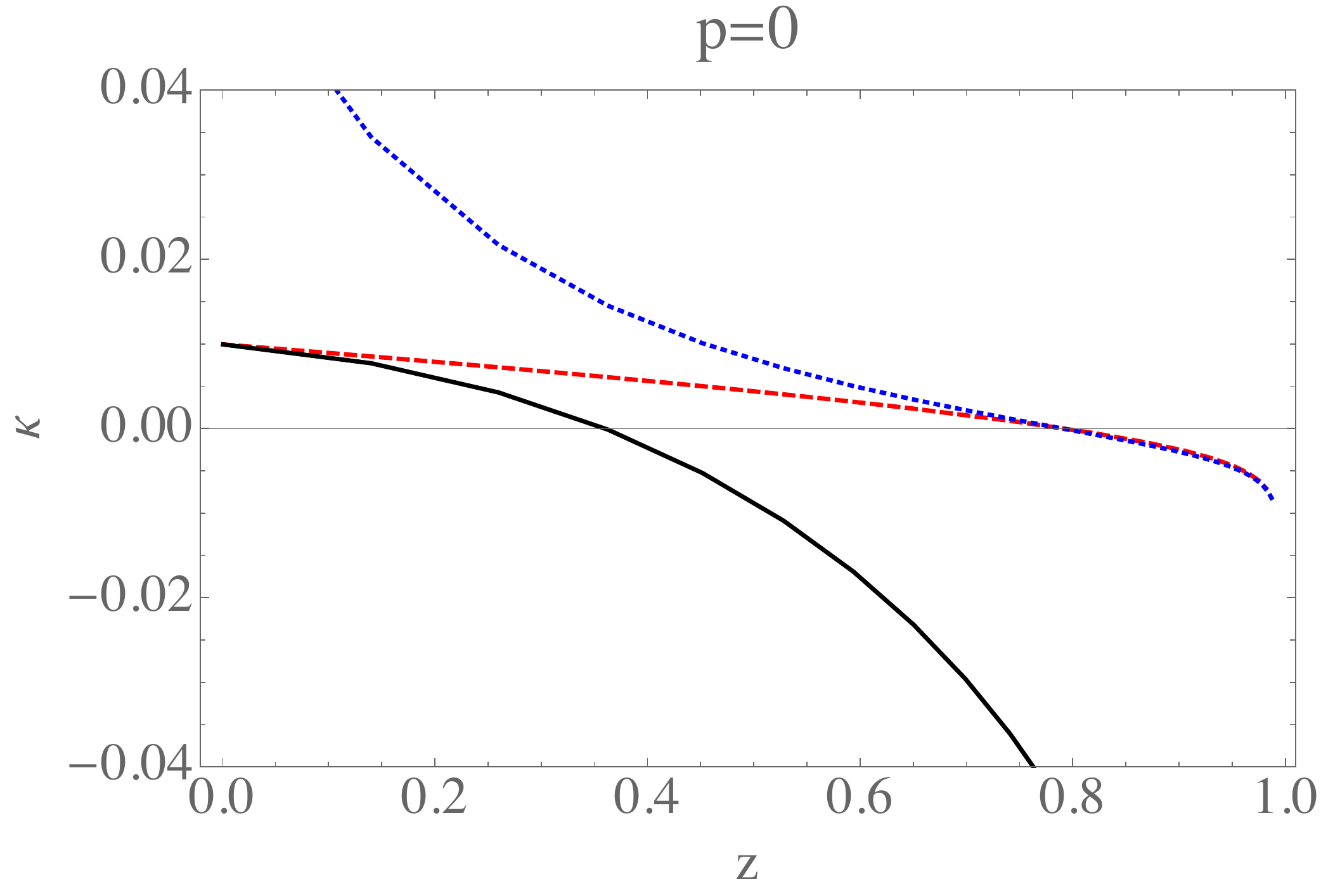}
\includegraphics[width=0.48\textwidth]{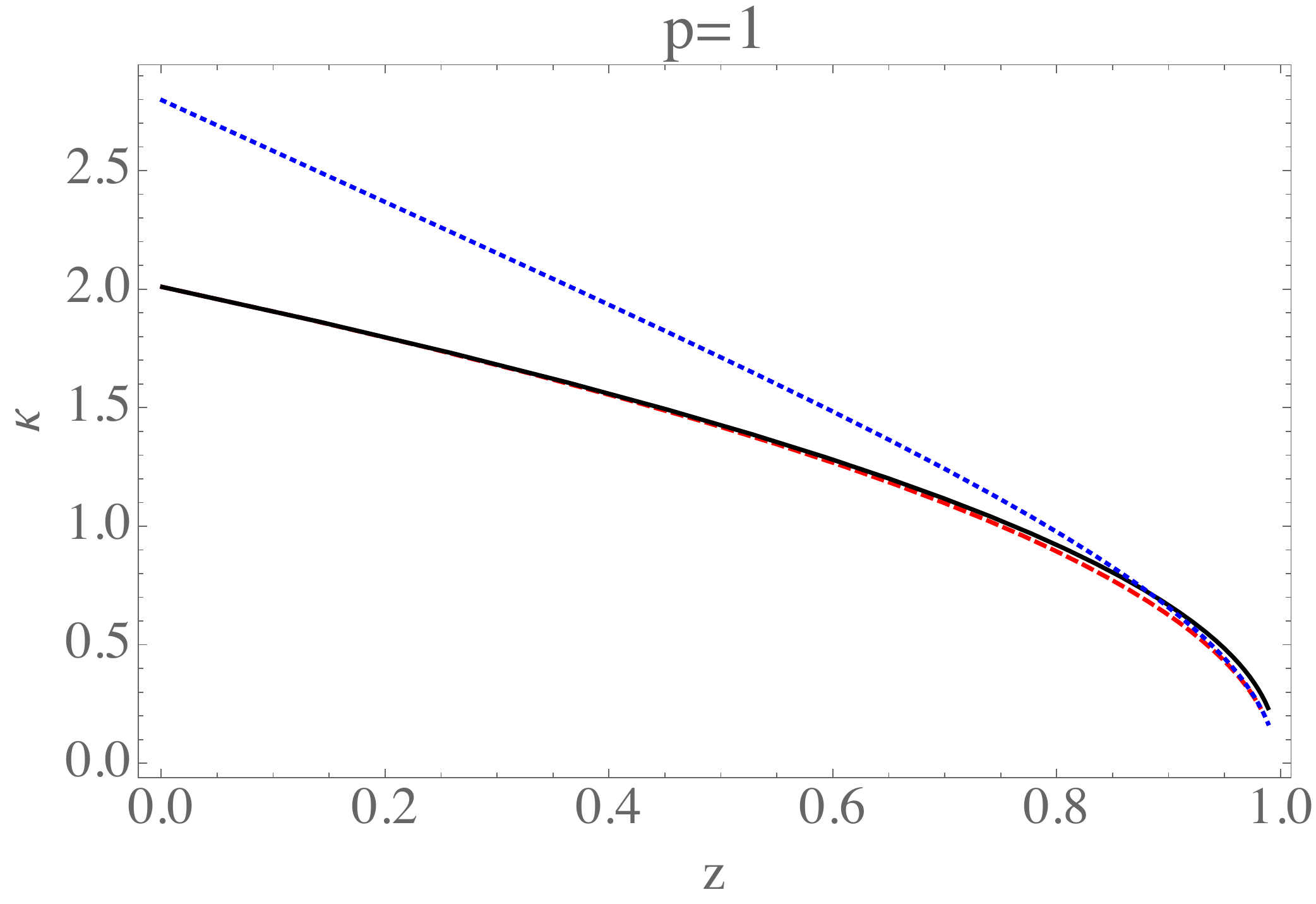}
\caption{ Interpolation of the leading $(p=0)$ and a sub-leading $(p=1)$ saddle for the vacuum block between $z\sim 0$ and $z \sim 1$.  Legend for the curves is as in fig.  \ref{fig:traceMcomparisons}: ({\it black, solid}) is the large $\kappa$/small $z$ approximation, ({\it blue, dotted}) is the small $|1-z|$ approximation, and ({\it red, dashed}) is the exact numeric result.  Parameters $\alpha_L =0.99, \alpha_H = i$ are chosen as in fig. \ref{fig:traceMcomparisons}.}
\label{fig:FollowLeadingVac}
\end{center}
\end{figure}

\subsection{Connecting Early and Late Times in the $z\sim 1$ Region}

Now, we turn to an analysis of the behavior of $\kappa(z)$ as a function of the phase  $t \equiv -\arg(1-z)$ for small but fixed $|1-z|$.  Using the formula (\ref{eq:TraceM}) for the trace in the limit $|1-z| \ll 1$, we can write the monodromy condition as
\be
\left( m_+ + m_- Z^{C-A-B} \right) \left( m_- + m_+ Z^{A+B-C} \right) &=& 4 \cos^2 \left( \frac{ \pi \alpha_I}{2} \right).
\ee
As a first, general comment, we note that the above equation has the feature that if $\kappa$ is a solution at time $t$, then $\kappa^*$ is a solution at time $-t$.  To see this, note that taking the complex conjugate of the entire equation simply sends 
$t \rightarrow -t $ and
$A,B,C \rightarrow A^*, B^*, C^*$. 
There is no effect on the RHS since $\alpha_I$ is always either pure real or pure imaginary.\footnote{We have used the fact that the $m_{\pm}$ are just products of $\Gamma$ functions whose arguments are linear combinations of $A,B$ and $C$, and $\Gamma(x^*) = (\Gamma(x))^*$, so taking the complex conjugate of $A,B,C$ has the effect $m_{\pm} \rightarrow m_{\pm}^*$.}  By inspection of (\ref{eq:RedefinitionofParameters}), conjugating $A,B,C$ just conjugates $\kappa$, and has no effect on real $\eta_L, \eta_H$.\footnote{In fact, this conclusion is more general than the small $|1-z|$ regime.  Starting from the monodromy differential equation (\ref{eq:MonodromyEquation}),(\ref{eq:MonodromyStressTensor}), the complex conjugate of any solution $\psi(y,z)$ is a solution to the same equation with $\kappa, y,z \rightarrow \kappa^* , y^*, z^*$.  Changing $y\rightarrow y^*$ simply changes the sign of the exponent of the eigenvalues of the monodromy matrix, which does not affect its overall trace.  Thus if $\kappa$ is a solution at $t$ then $\kappa^*$ is a solution at $-t$, regardless of the magnitude of $|1-z|$.}

\begin{figure}[t!]
\begin{center}
\includegraphics[width=\textwidth]{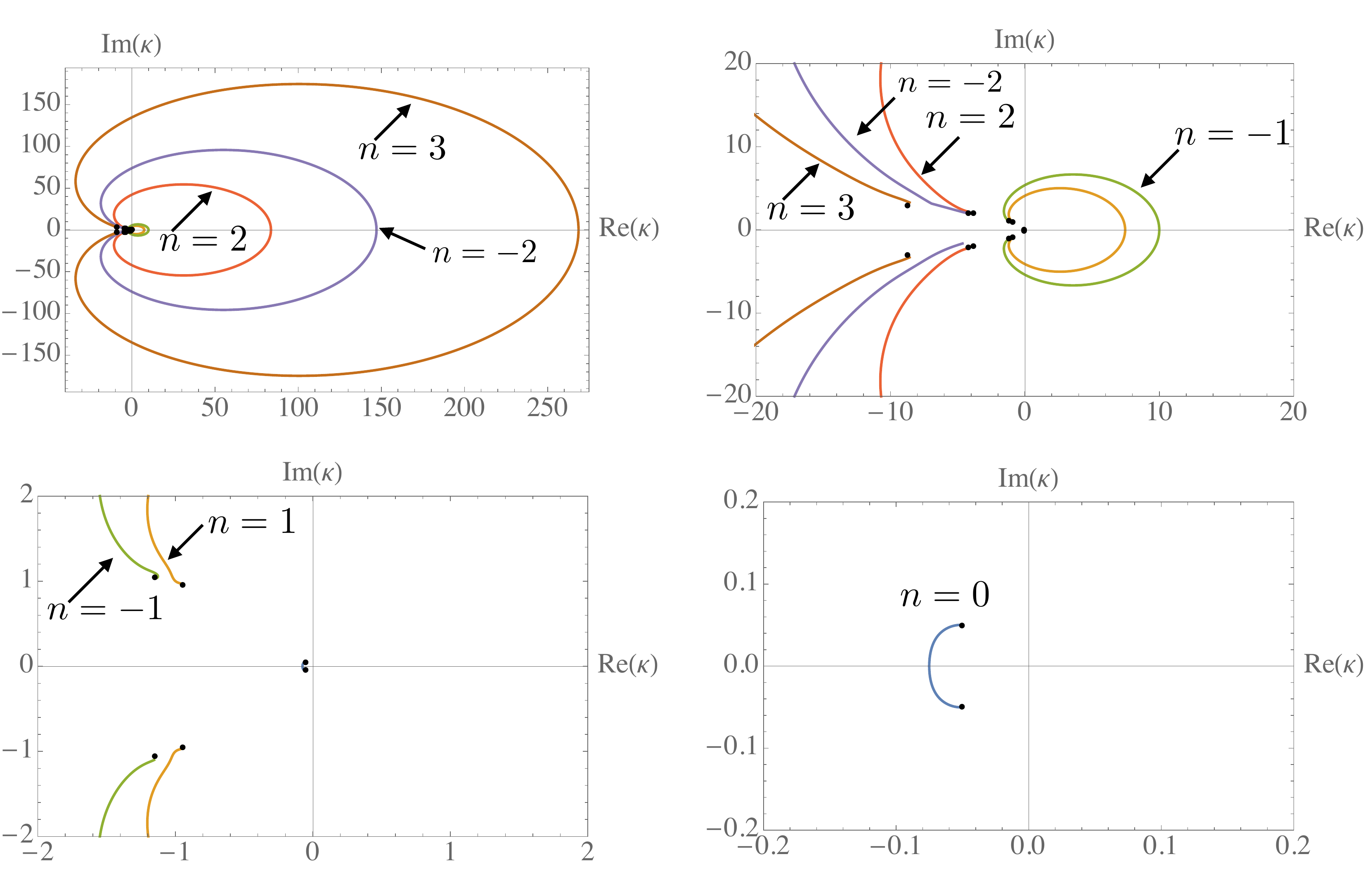}
\caption{Plots of solutions for $\kappa$ as we interpolate between early and late time behaviors in the decaying (\ref{eq:KappaSolns}) class.  Recall that  Im$(\kappa)$ (the vertical axes) determines the exponential decay or growth rate of the Virasoro blocks.  The upper left plot shows the full region covering all solutions, and the remaining three plots show scaled up regions of the first plot for better visibility.  Parameters are chosen to be $\alpha_H = i, \alpha_L = 0.99, |1-z|= 0.01 $.   Dots indicate the analytic solutions $\kappa = n(1-n) - \frac{1}{2} + (\frac{1}{2} -n) (\alpha_L \pm \alpha_H) \mp \frac{\alpha_L \alpha_H}{2}$ for $n =-2, -1, \dots 3$; each path interpolates between two points with the same $n$ and opposite choice of sign for $\pm \alpha_H$.  All of these saddles decay exponentially at late times; and all but the leading $n=0$ saddles decay even faster at most intermediate times. }
\label{fig:kappatransition}
\end{center}
\end{figure}

\begin{figure}[t!]
\begin{center}
\includegraphics[width=0.32\textwidth]{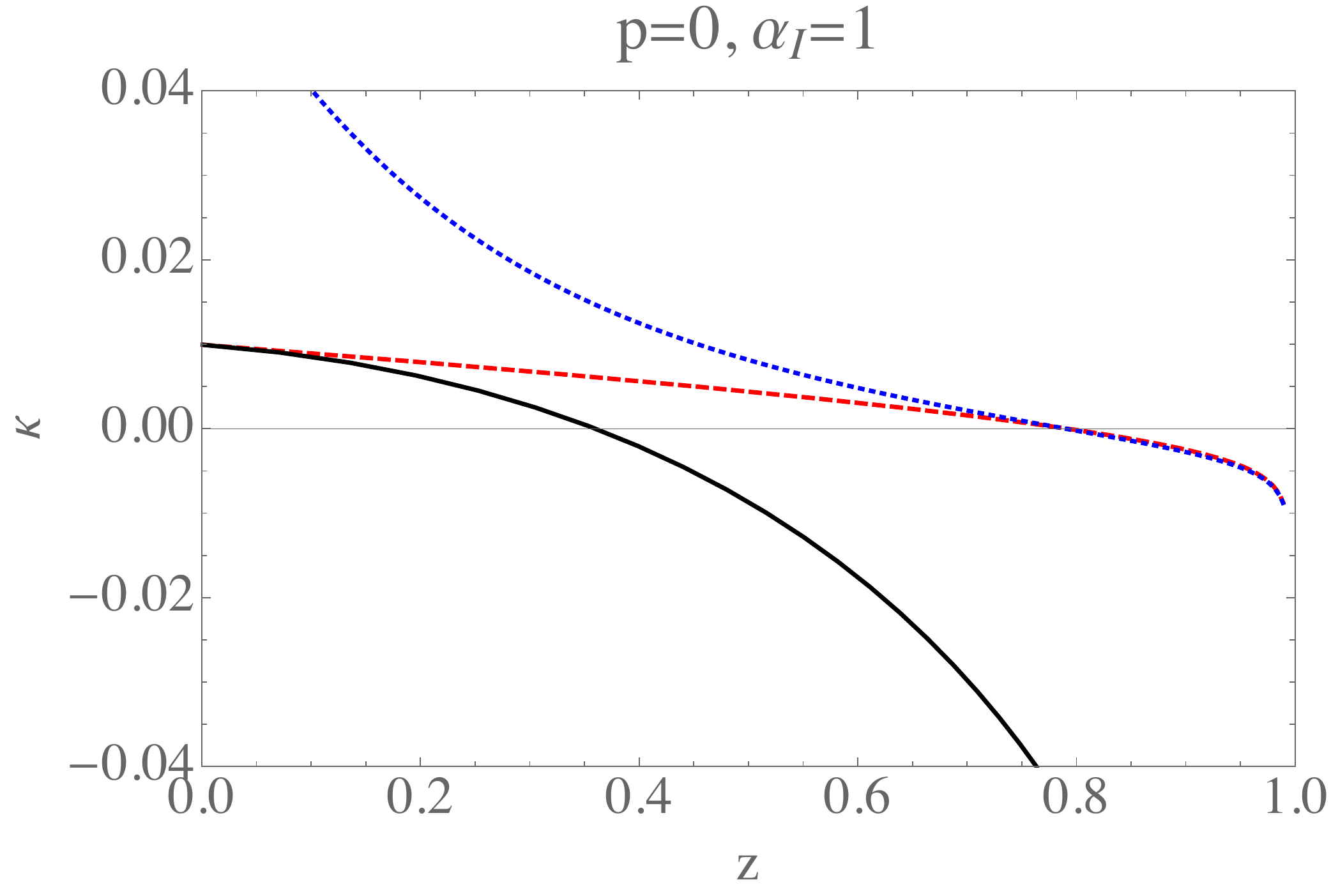}
\includegraphics[width=0.32\textwidth]{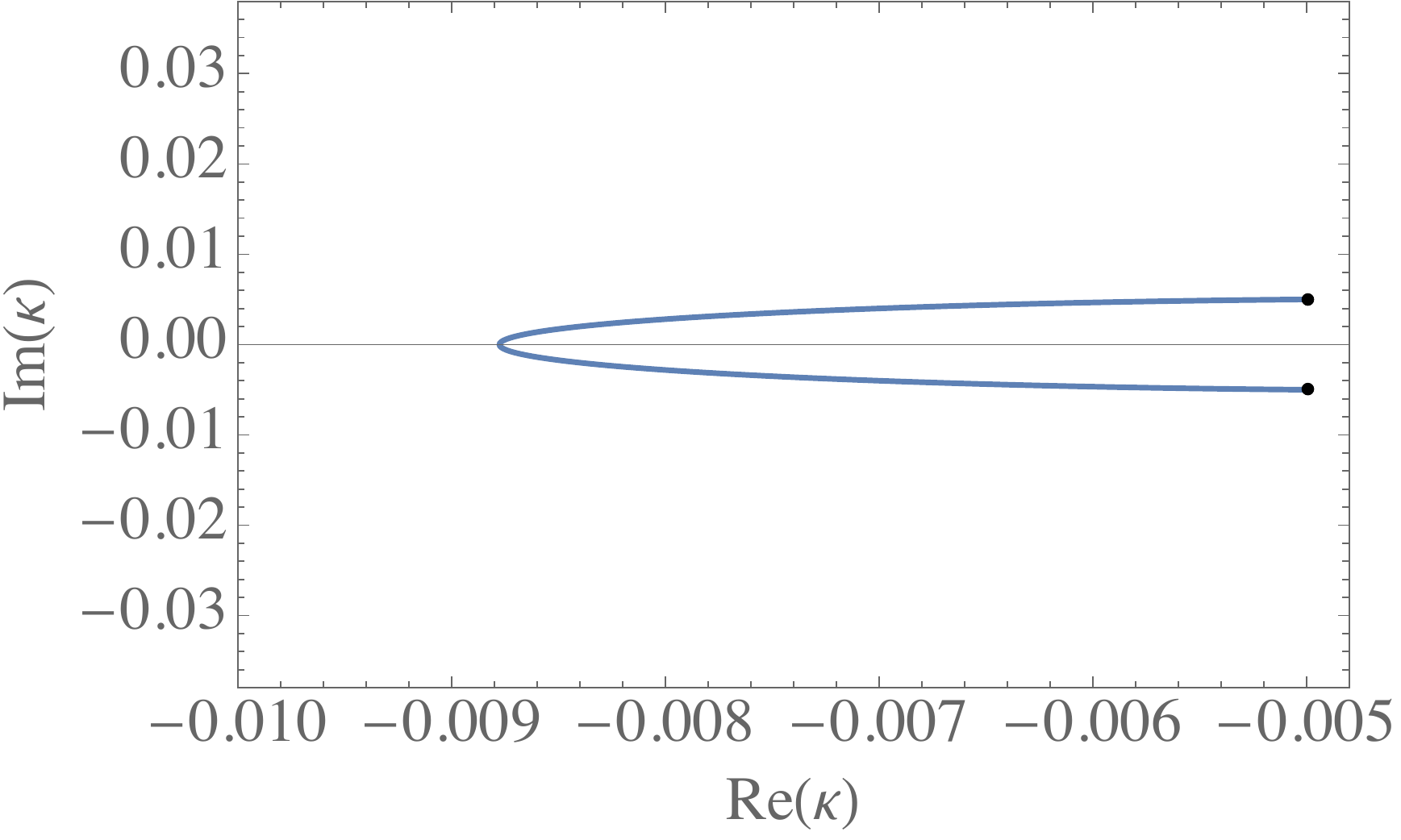}
\includegraphics[width=0.32\textwidth]{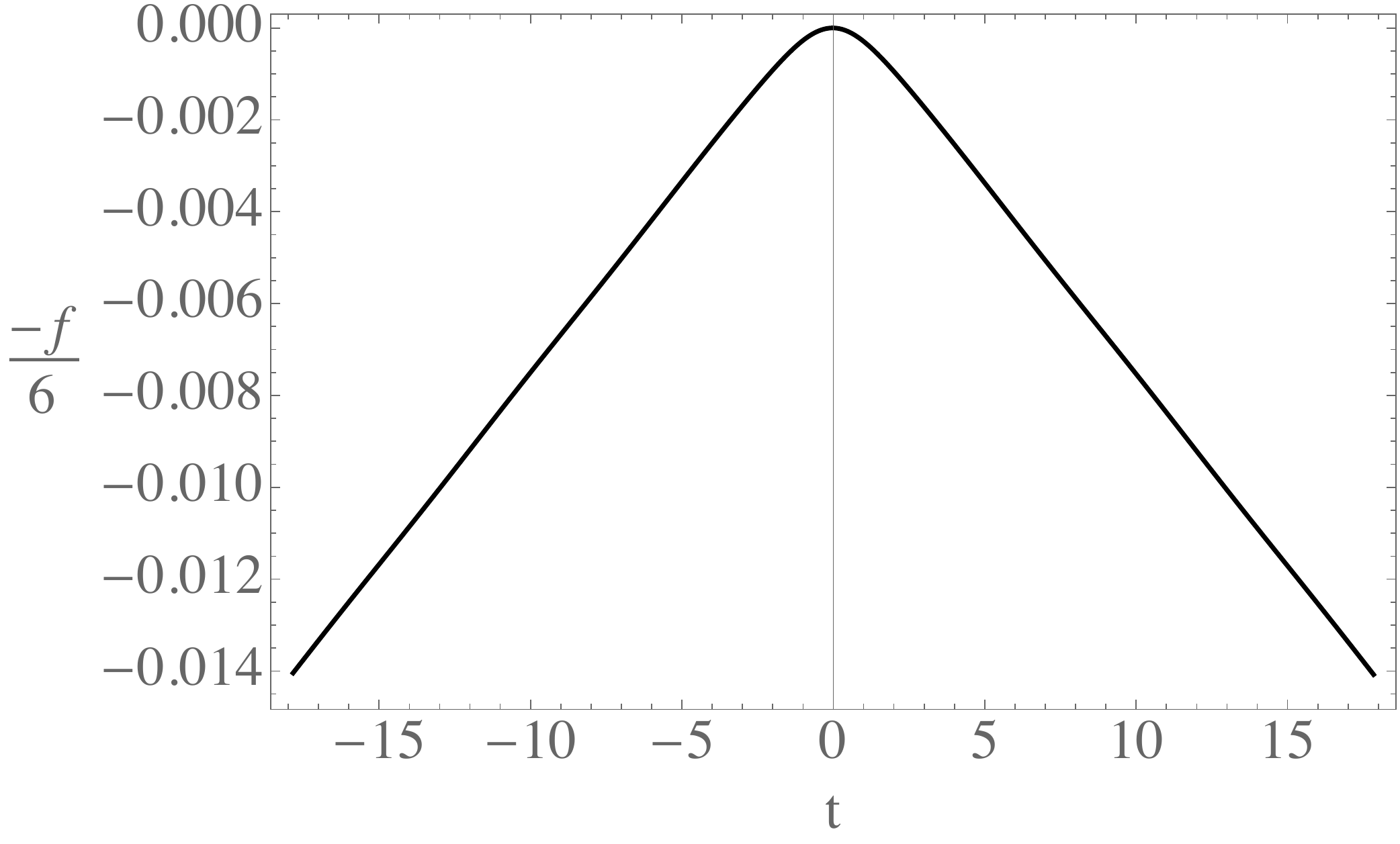}
\includegraphics[width=0.32\textwidth]{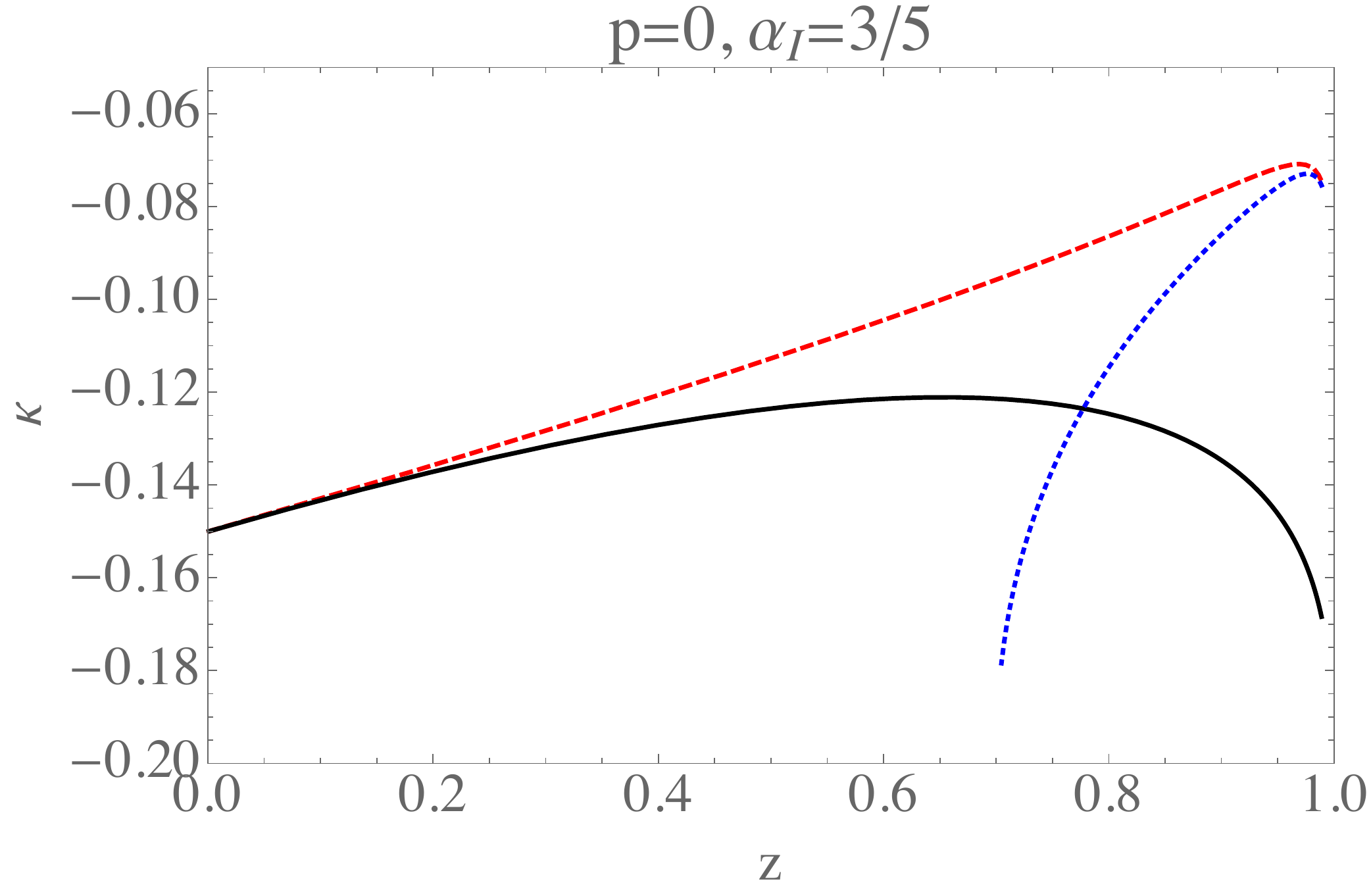}
\includegraphics[width=0.32\textwidth]{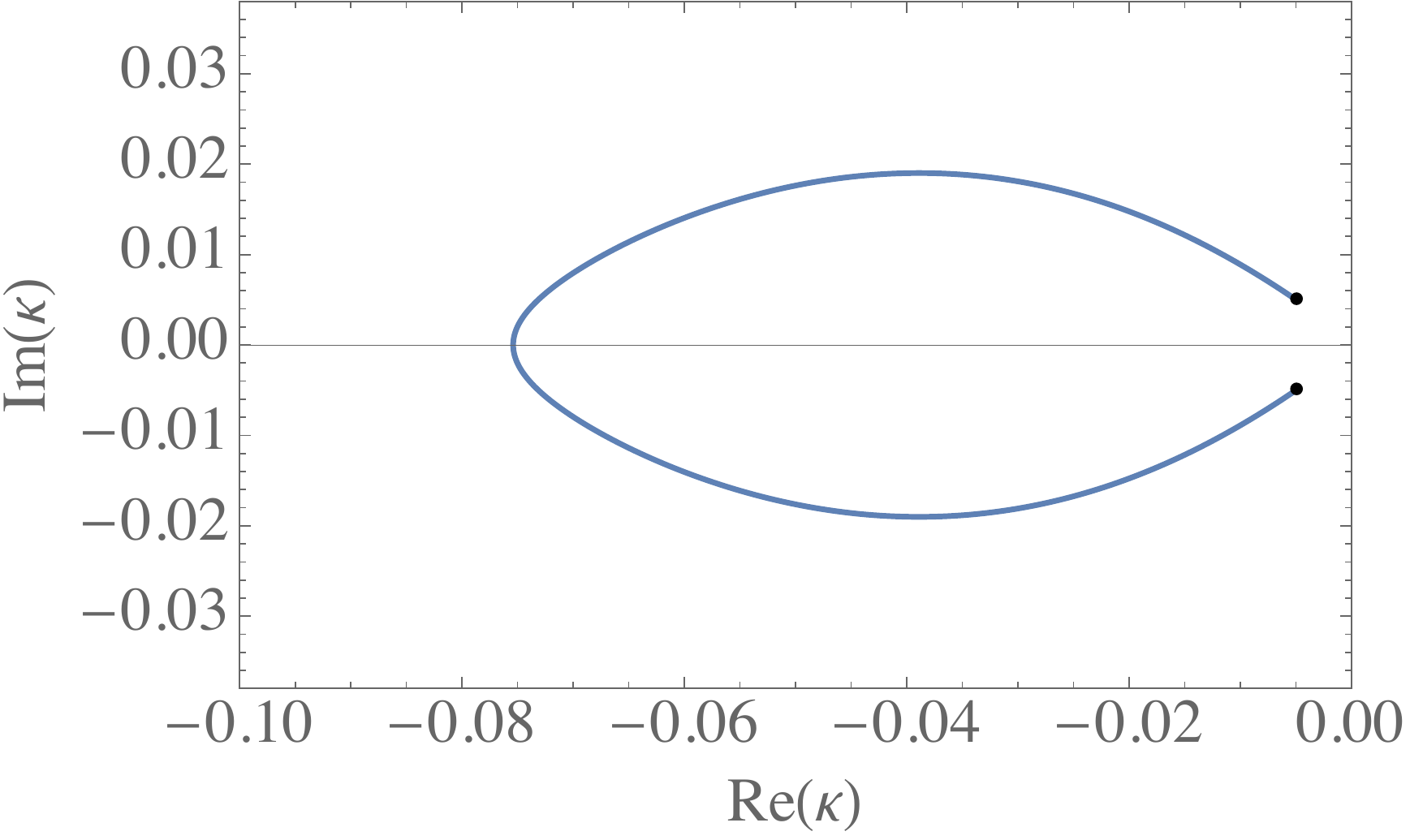}
\includegraphics[width=0.32\textwidth]{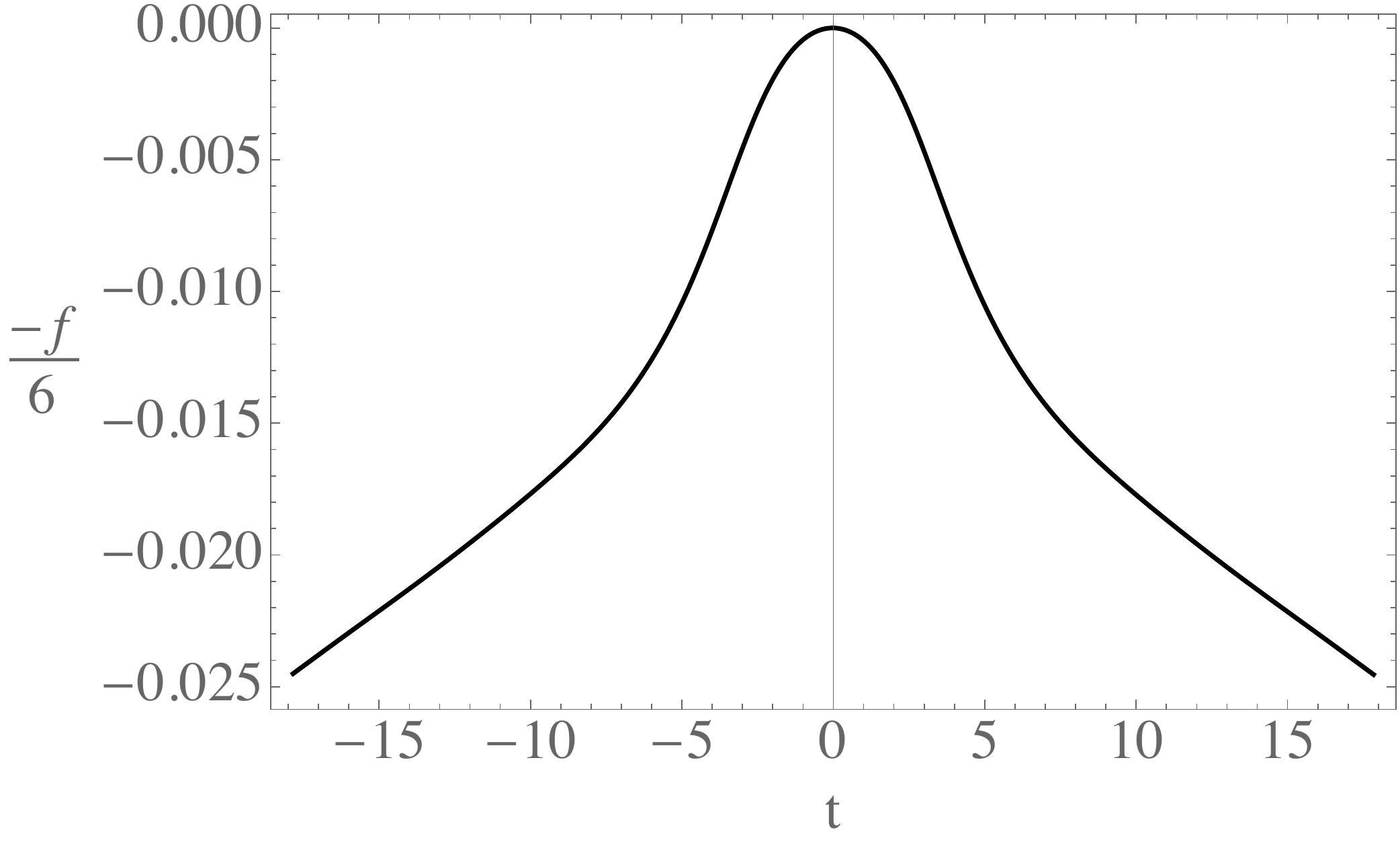}
\includegraphics[width=0.32\textwidth]{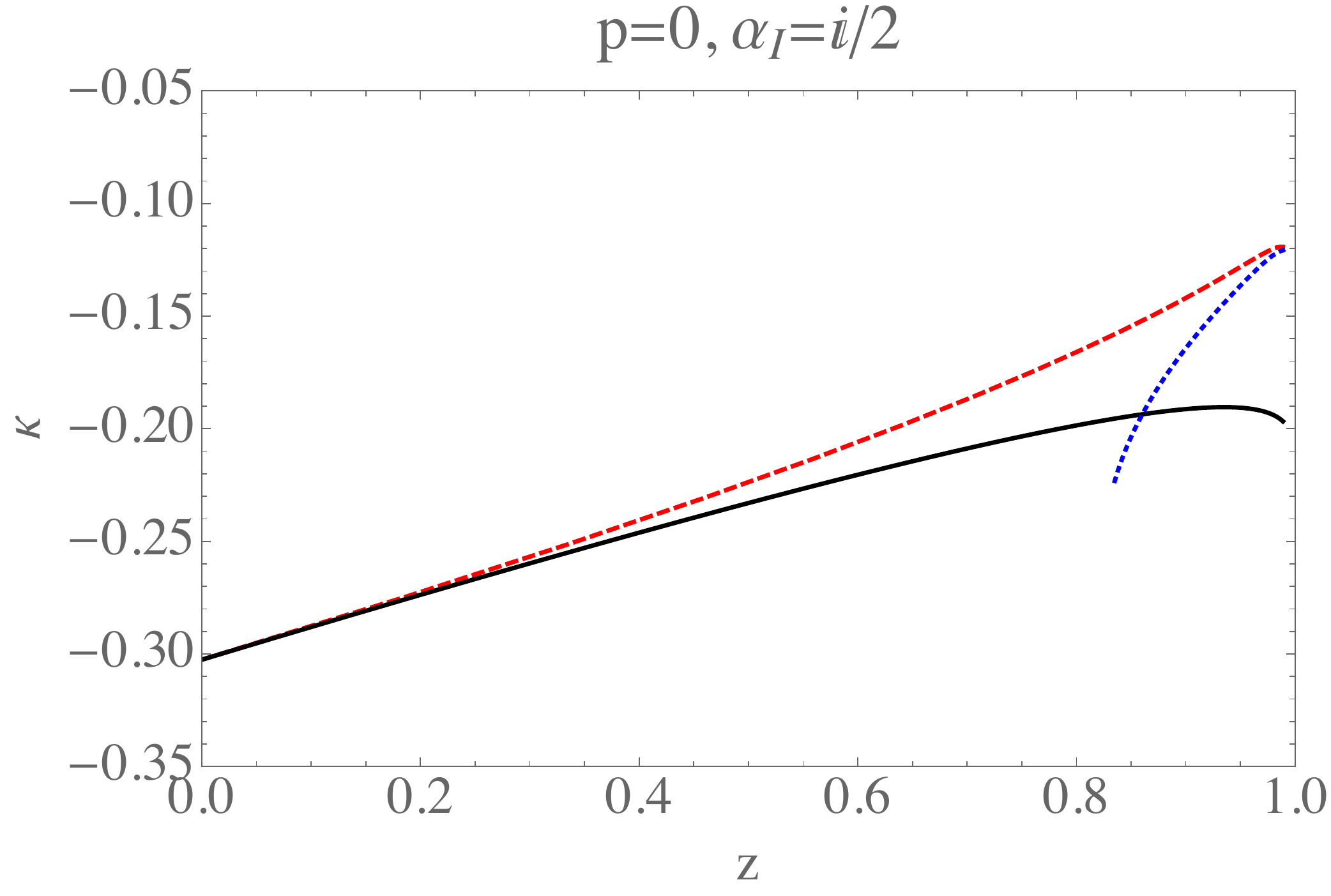}
\includegraphics[width=0.32\textwidth]{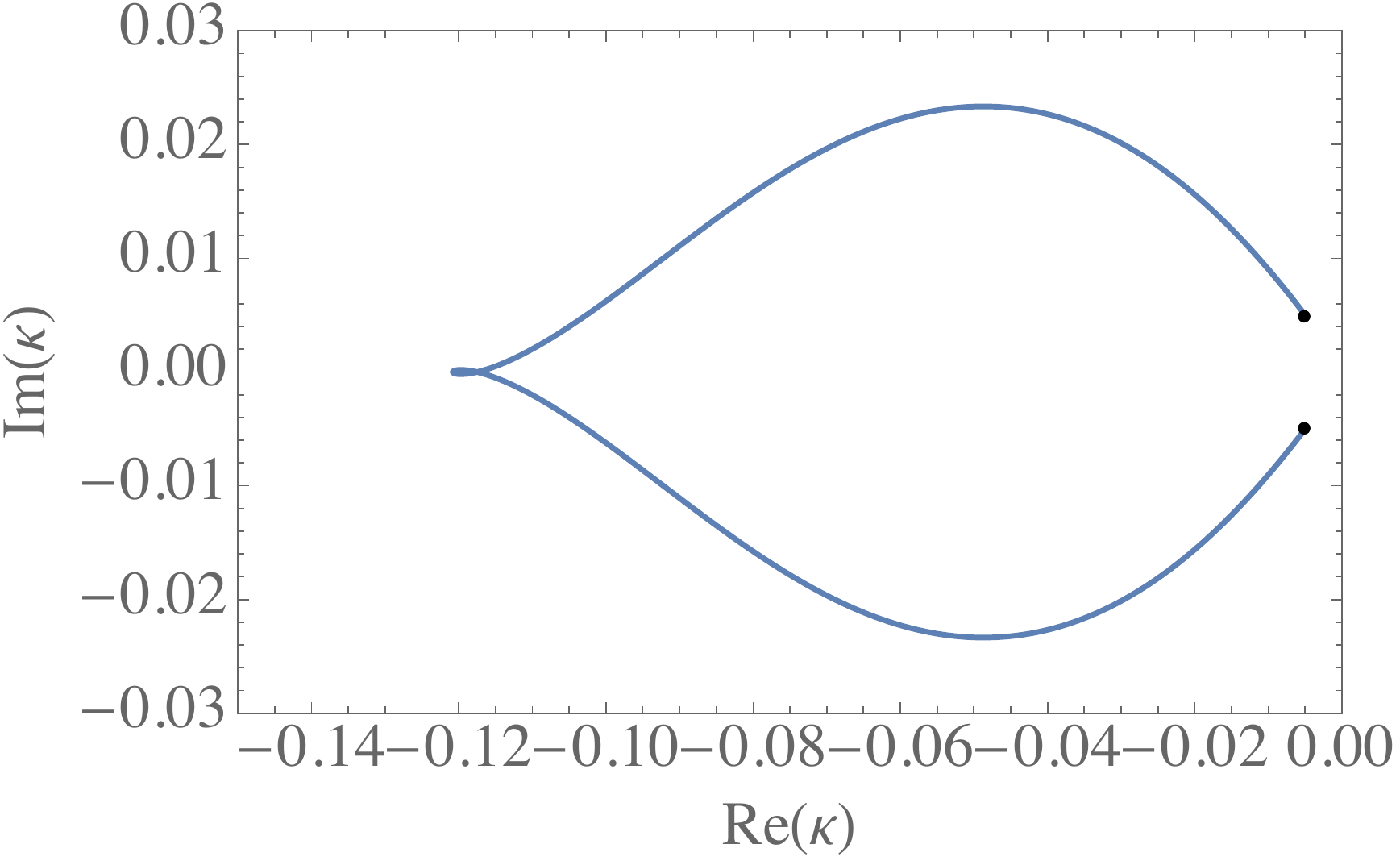}
\includegraphics[width=0.32\textwidth]{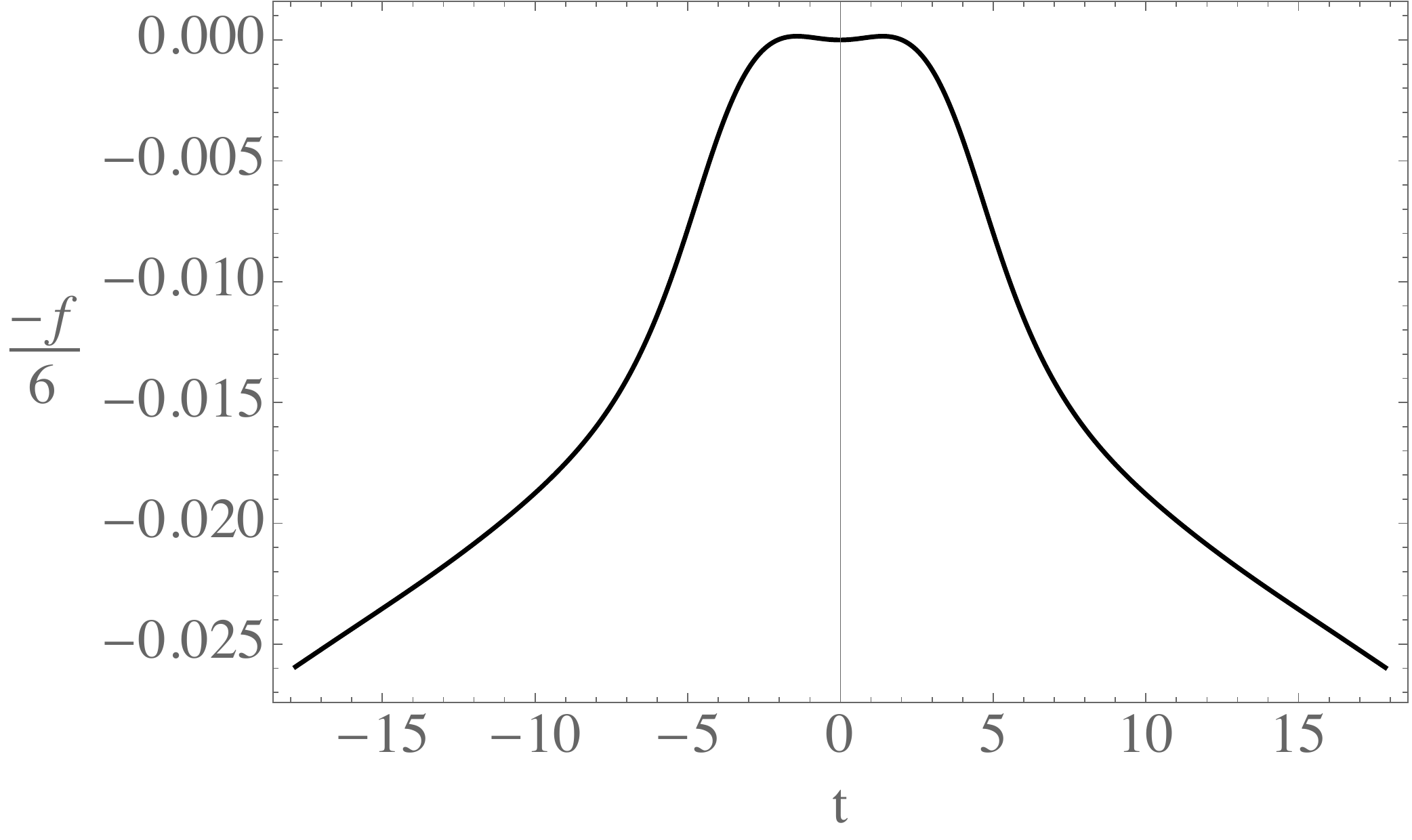}
\includegraphics[width=0.32\textwidth]{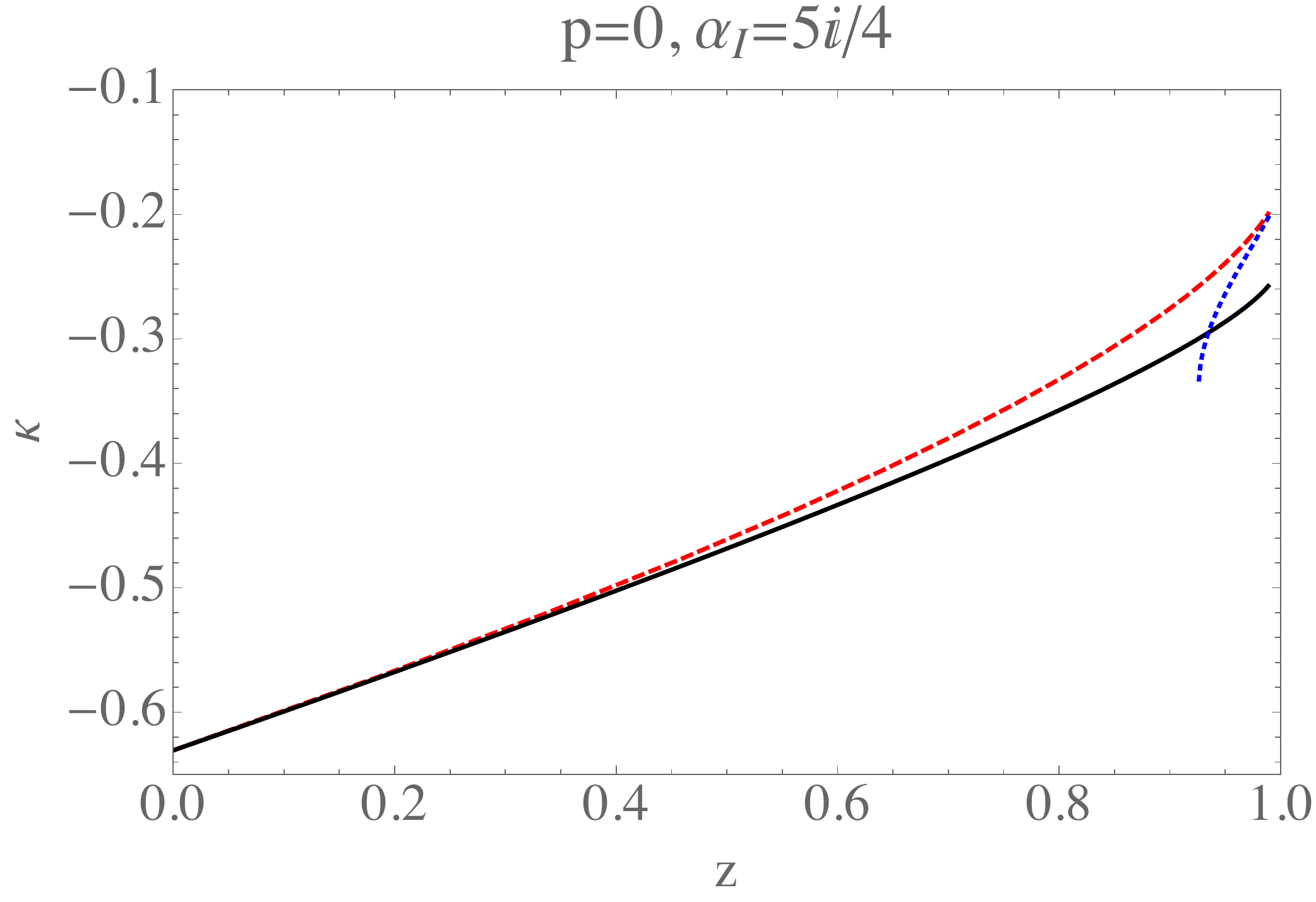}
\includegraphics[width=0.32\textwidth]{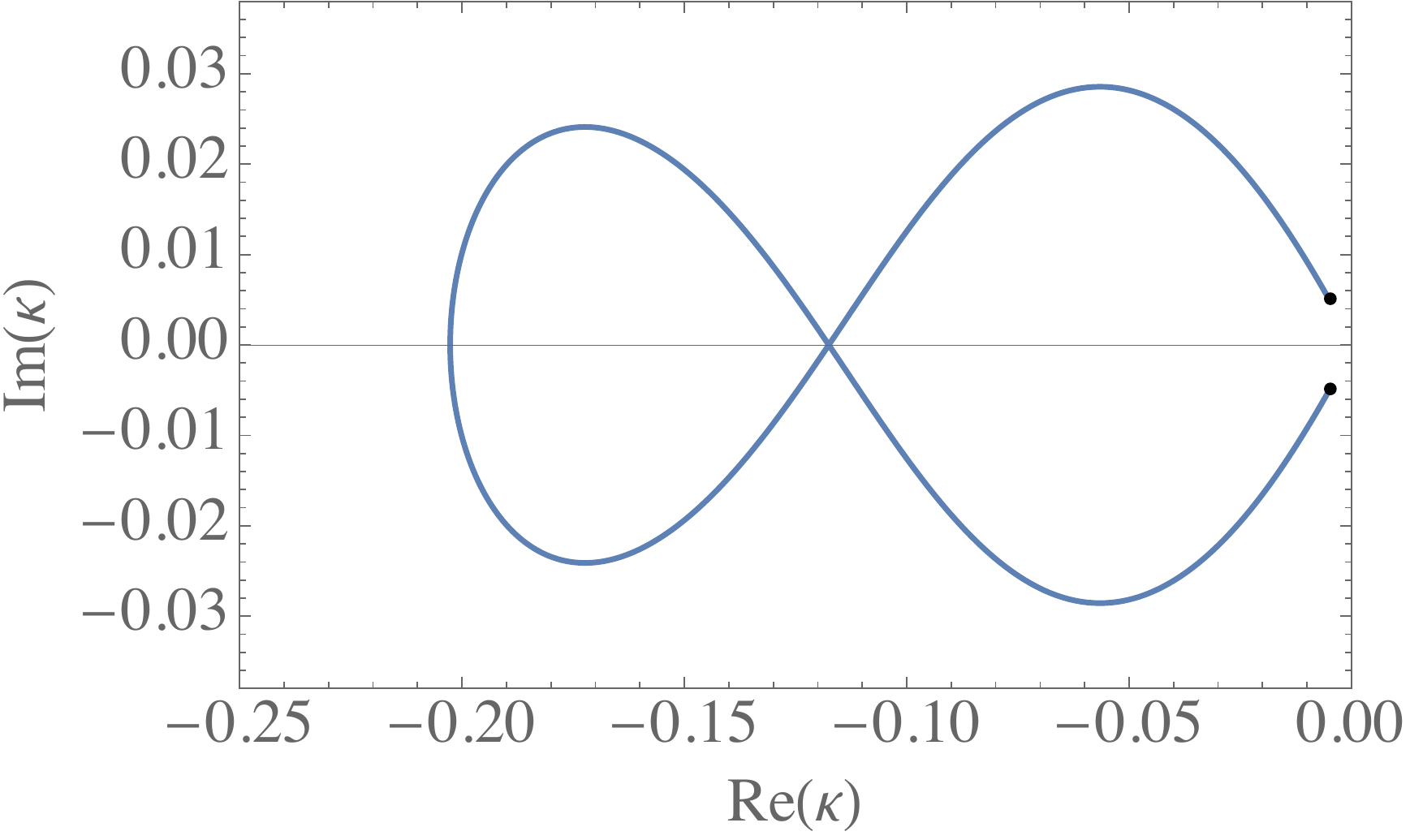}
\includegraphics[width=0.32\textwidth]{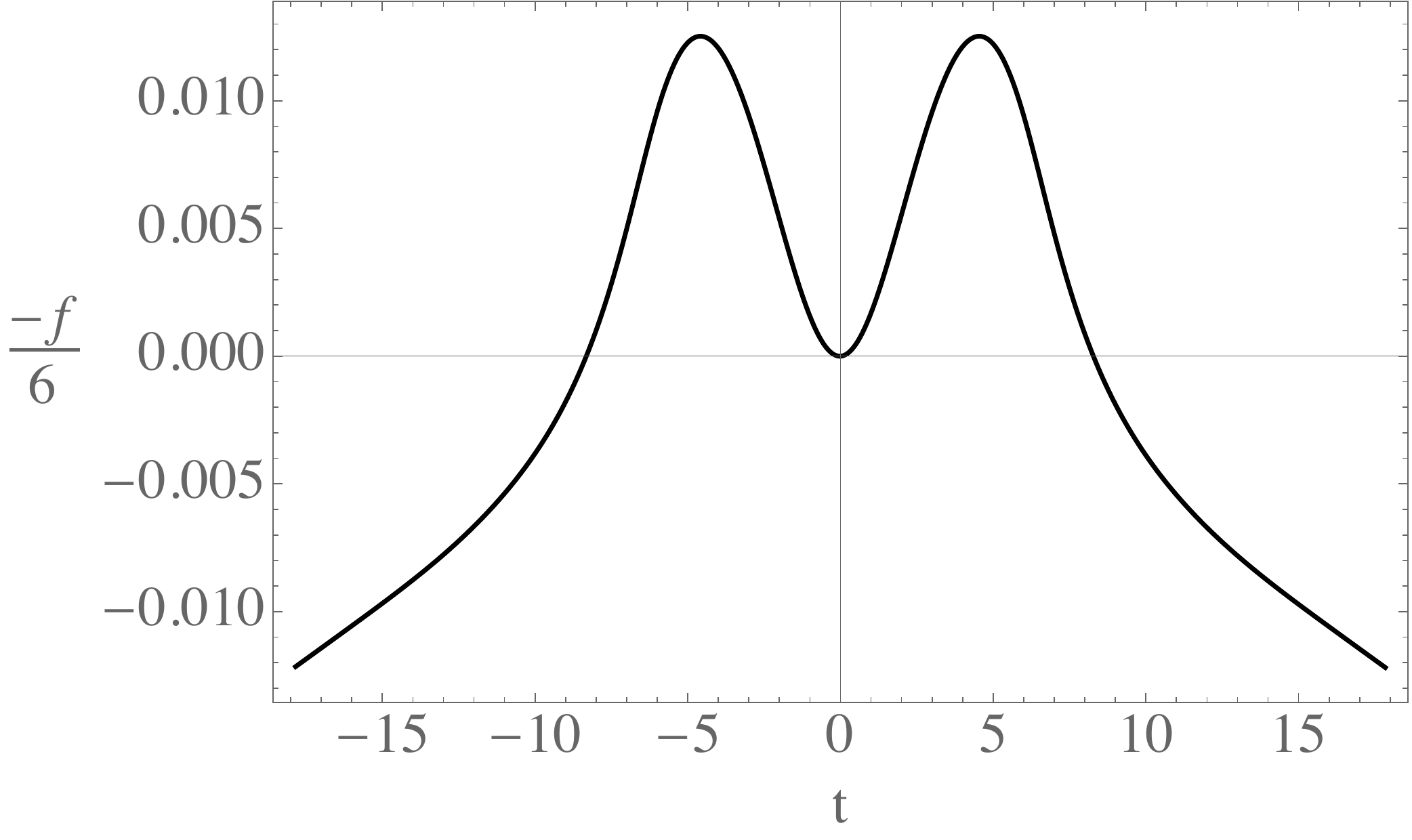}
\caption{Leading $(p=0)$  saddle for the $\alpha_I = 1, 3/5, i/2, 5i/4$ block (from top to bottom row, respectively). {\it Left:} Interpolation between $z\sim 0$ and $z \sim 1$.  Legend for the curves is as in fig.  \ref{fig:traceMcomparisons}: ({\it black, solid}) is the large $\kappa$/small $z$ approximation, ({\it blue, dotted}) is the small $|1-z|$ approximation ($\kappa$ becomes complex where the curve ends), and ({\it red, dashed}) is the exact numeric result.   {\it Middle: } Trajectory of the $p=0$ saddle at fixed $|1-z|= 0.011$ from $t=-\infty$ (where Im$(\kappa)>0$) to $t\rightarrow \infty$ (where Im$(\kappa)<0$).  At $t=0$, it crosses the real axis, where the value of $\kappa$ matches the value at $z \approx 1$ in the left plot. {\it Right}: Plot of the term $-\frac{f}{6}= \frac{1}{c} \log({\cal V})$ in the exponent of the block at fixed $|1-z|=0.11$ as a function of $t$, from integrating $\kappa$; the integration constant has been chosen so $f(0)=0$.  Parameters $\alpha_L =0.99, \alpha_H = i$ are chosen as in fig. \ref{fig:traceMcomparisons}.  Thus we see that some saddles grow before their ultimate late-time decay.}
\label{fig:FollowLeadingNonVac}
\end{center}
\end{figure}

In the case of the vacuum block, $\alpha_I = 1$ so the RHS above vanishes. We begin with this case. 
 For the choice of parameters $\alpha_L=0.99$, the initial value of the leading saddle at $z=0$ is $\kappa = \frac{1-\alpha_L^2}{2}$.  As discussed in the previous subsection, one can identify which saddles at $z \sim 1$ correspond to which leading saddles at $z =0$ by following the approximations (\ref{eq:TraceMZam}) and (\ref{eq:TraceM}), with the exact solution transitioning from one to the other approximation.  This is shown explicitly in fig. \ref{fig:FollowLeadingVac} for the $p=0$ and $p=1$ saddles.  From there, one can follow the saddles to late times using (\ref{eq:TraceM}).

In fig. \ref{fig:kappatransition}, we show the trajectory of several saddles that end on one of the late-time solutions with non-vanishing imaginary part.  These are the solutions from equation (\ref{eq:KappaSolns}), and we have indicated which values of $n$ in (\ref{eq:KappaSolns}) correspond to which curves.  The $t=0$ point of each curves always lies on the real axis.  By comparison with fig. \ref{fig:FollowLeadingVac}, one can read off that the $n=0$ curve at $t=0$ matches the $p=0$ curve at $z\sim 0$, i.e. it is the leading OPE saddle.

A similar analysis can be applied to non-vacuum blocks.  In fig. \ref{fig:FollowLeadingNonVac}, we show the leading saddle for the non-vacuum block with $\alpha_I = i/2$ and $\alpha_I = 5i/4$.  In these cases, the leading saddle $(p=0)$ still asymptotically approaches the $n=0$ solutions from (\ref{eq:KappaSolns}) at early and late times, so it has negative (positive) imaginary part at $t \rightarrow \infty$ ($-\infty$). Thus its contribution to the Virasoro block decays asymptotically to 0 at $t\rightarrow \pm \infty$.  However, it has a ``figure 8'' pattern where the imaginary part of $\kappa$ changes sign at intermediate $t$, and therefore the saddle actually grows for a period of time before decaying.  This is visible in the far right panels of fig. \ref{fig:FollowLeadingNonVac}, where we integrate $\partial_z f = \kappa(z)/(z(1-z))$ to obtain the exponent $-\frac{f}{6} = \frac{1}{c} \log ({\cal V})$.   Note that the {\it exponent} is $c$ times the value shown, so the growth and subsequent decay are extremely rapid.

\begin{figure}[t!]
\begin{center}
\includegraphics[width=0.31\textwidth]{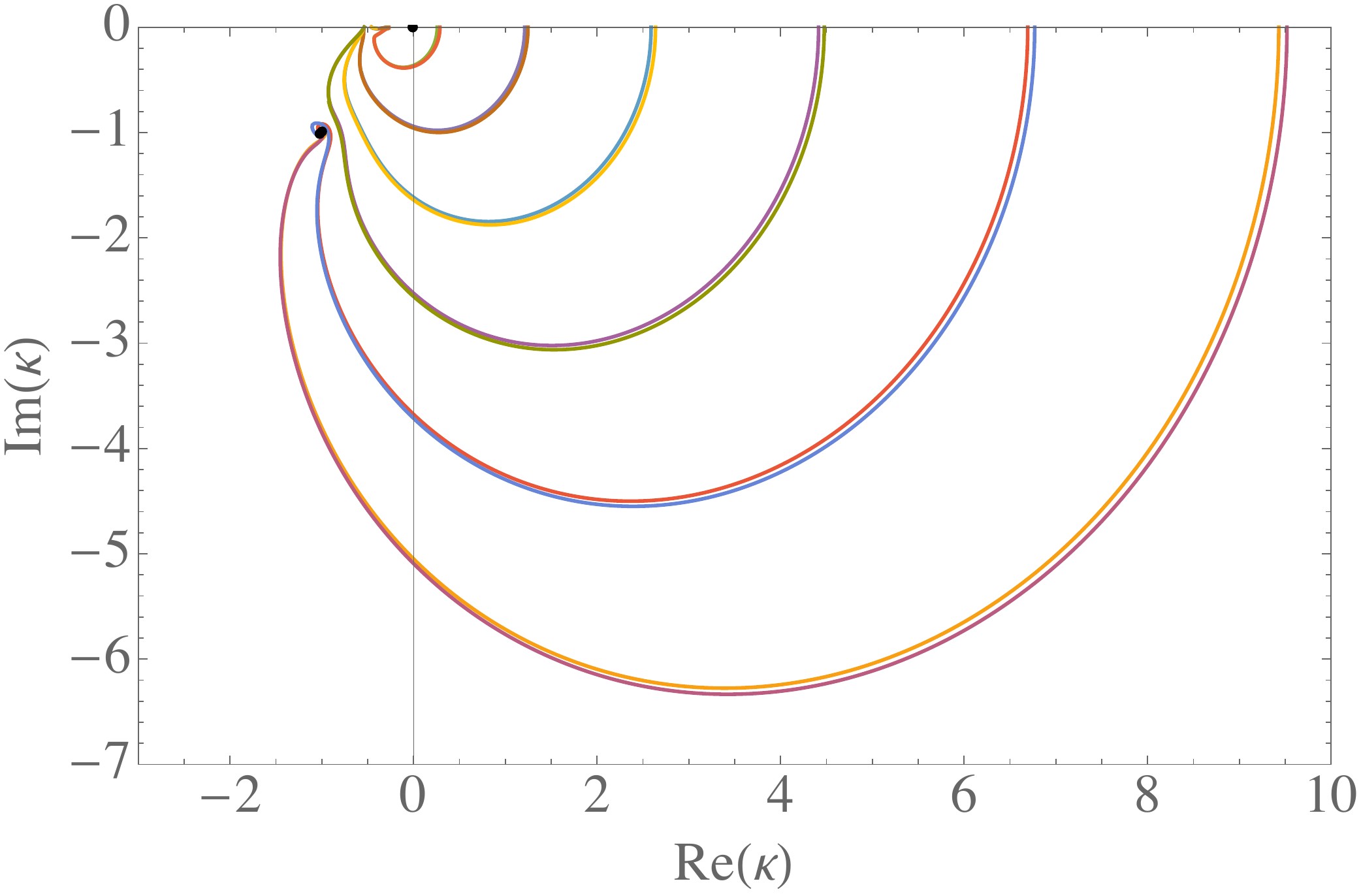}
\includegraphics[width=0.32\textwidth]{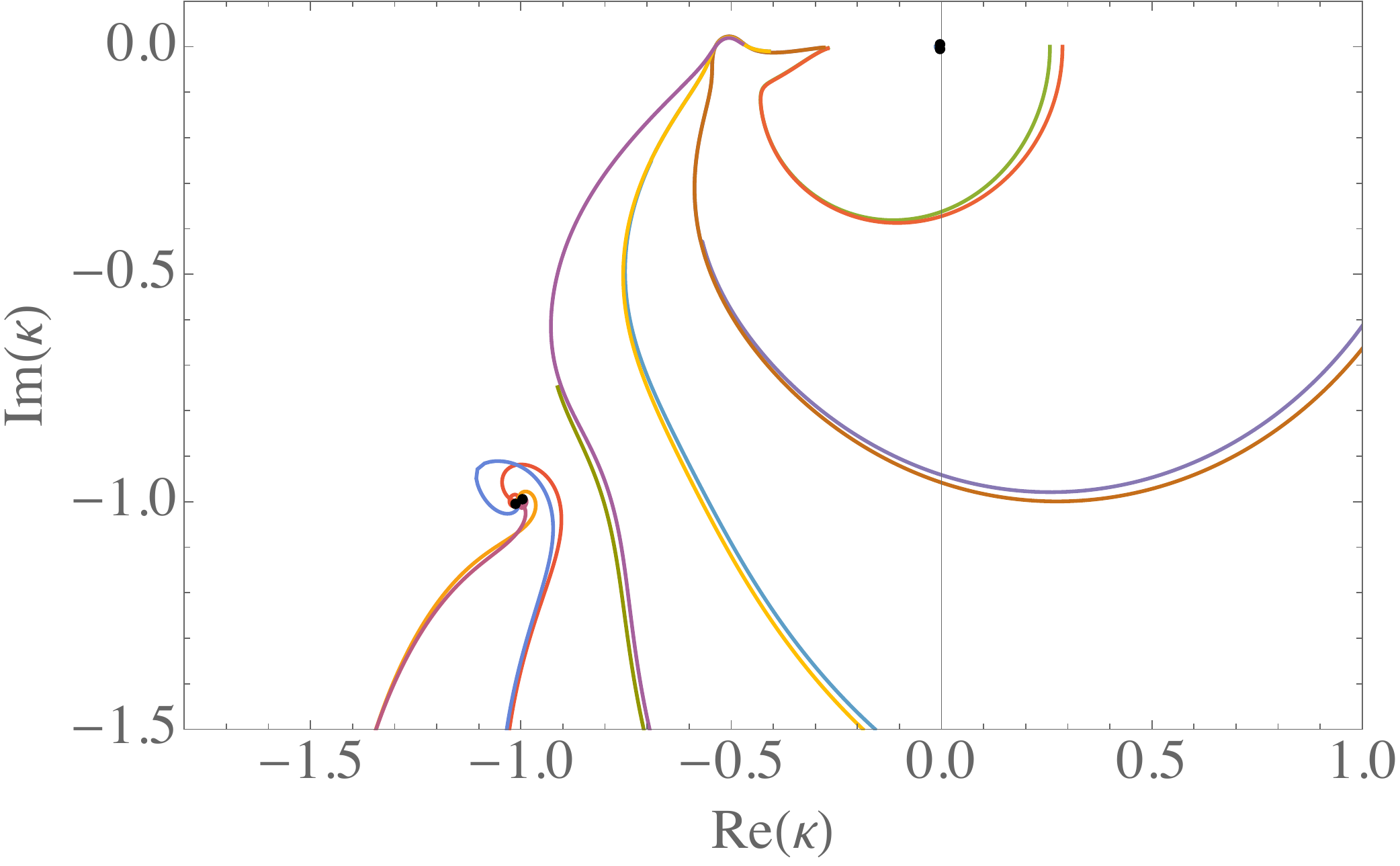}
\includegraphics[width=0.34\textwidth]{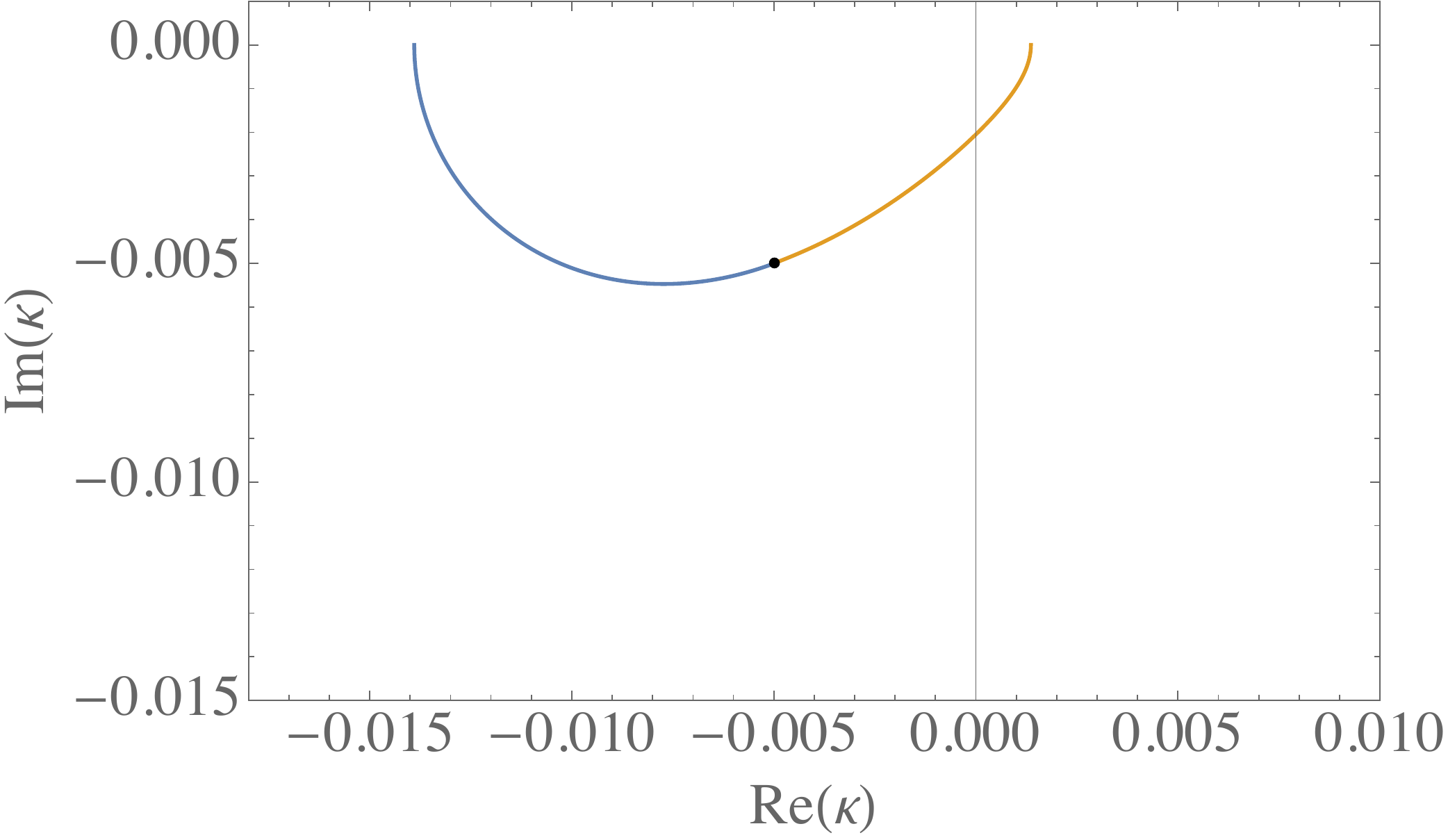}
\caption{The trajectory of a host of saddles as $t$ increases from $0$ to $\infty$ at fixed $|1-z| = 0.022$, shown at various levels of magnification.  Black dots indicate the asymptotic positions of decaying saddles, however most saddles are seen to be oscillating saddles in these plots. Parameters are $\alpha_L = 0.99, \alpha_H =i, \alpha_I =0.97, |1-z| = 0.022$.}
\label{fig:lotsofsaddles}
\end{center}
\end{figure}

Going beyond the leading ($p=0$) saddle, some saddles are in the decaying class (\ref{eq:KappaSolns}) that have non-zero imaginary parts asymptotically, and some are from the oscillating class (\ref{eq:newsolutions}) that have zero imaginary parts asymptotically.  In fig. \ref{fig:lotsofsaddles}, we show several trajectories, starting from $t=0$ and following $\kappa$ to $t \rightarrow \infty$.  While there may exist a simple rule for which values of $p$ map to saddles in the decaying vs oscillating classes, we have not found such a rule and the most we can say is that for generic values of $\alpha_I, \alpha_L,$ and $\alpha_H$, the two classes of solutions are interspersed with each other as one looks at greater and greater $|p|$.

\section{Semiclassical Virasoro Blocks for Degenerate States} 
\label{sec:DegenerateStates}

In this section we will discuss a different method of calculation using degenerate states and operators.  By definition, these operators are annihilated by a polynomial in the Virasoro generators $L_{-n}$, which means that their correlators satisfy differential equations.  Expanding these differential equations in the semiclassical limit $h_i/c$ fixed as $c \to \infty$ produces an algebraic equation for the degenerate operators' $\kappa(z) = \frac{6}{c} z(z-1) \partial_z \log \CV$.  

For any fixed degenerate operator $\CO_{r,1}$, the resulting algebraic equations for $\kappa$ are efficient to study both numerically and analytically, and so give us the solutions for the semiclassical saddles for all values of $z$.  In particular, at any $z$, the problem of finding the behavior of the saddles is reduced to finding the eigenvalues of a finite-dimensional matrix, which is numerically very efficient.  

In the limit that $z \to 1$, we can solve these algebraic equations for every $h_{r,1}$ degenerate state.  For degenerate operators this limit is much simpler than for general operators, because the $z \to 1$ limit no longer depends on the phase of $1-z$. Rather, the $z \to 1$ limit corresponds to a crossed channel OPE limit, and the semiclassical solutions simply approach values corresponding to the allowed dimensions of operators in this crossed channel.  We will see that we reproduce the results of the large $c$ limit of such dimensions derived in \cite{Turiaci:2016cvo,Fitzpatrick:2016ive}.  

By analytic continuation of these results, we obtain an alternate derivation for the infinite class of solutions derived in section \ref{sec:ABCneq0}.  The algebraic method can also be used it to provide a partial derivation of the monodromy method itself, as we explain in section \ref{sec:MonodromyMethodfromAlgebraicMethod}.

\subsection{An Algebraic Description of Semiclassical Degenerate Correlators}  
\label{sec:AlgebraicMethod}

Bauer, Di Francesco, Itzykson, and Zuber have developed a systematic method \cite{Bauer:1991qm, Bauer:1991ai} (for a review see \cite{DiFrancesco:1997nk} section 8.2 and exercise 8.8) for obtaining the combination of Virasoro generators that annihilate degenerate states.  
When studying degenerate states, it is convenient to the write the central charge in terms of a paramter $b$ via
\be
c = 1 + 6 \left(b + \frac{1}{b} \right)^2 . 
\ee
We will take the limit of large $c$ via the limit of large $b$.  Using this notation, the degenerate states have dimensions 
\be
\label{eq:DegenerateDimensions}
h_{r,s} = \frac{b^2}{4} (1-r^2) + \frac{1}{4 b^2}(1-s^2) + \frac{1}{2}(1-rs),
\ee  
parameterized by the positive integers $r,s$;  we will be focusing on states with dimension $h_{r,1}$.  

Now let us define the null state equations.  Let $D_{r,1}$ be the following matrix:\footnote{This formula differs from the analogous one in \cite{Bauer:1991qm, Bauer:1991ai,DiFrancesco:1997nk,Fitzpatrick:2016ive} by some factors of $b$; the difference is equivalent to rescaling the representations (\ref{eq:degenRep}) for $L_\pm$ by factors of $b$ so that the algebra is unchanged, and rescaling $D_{r,1}$ itself by an overall power of $b$.  We have also taken $b \rightarrow b^{-1}$, which is a  choice of convention since it amounts to taking the other branch of $c=1+6(b+1/b)^2$.} 
\be 
D_{r,1} = -J_-  +  \frac{1}{b^2} \sum_{m=0}^\infty \left(J_+ \right)^m L_{-m-1},  
\ee
where $J_\pm$ are matrix generators of the spin $(r-1)/2$ representation of $SU(2)$:
\be
(J_0)_{ij} &=& \frac{1}{2} (r-2i+ 1)\delta_{ij}, \nn\\
(J_-)_{ij} &=& \left\{ \begin{array}{cc}  \delta_{i,j+1} & (j=1,2, \dots, r-1) \\ 0 & \textrm{else}\end{array} \right. ,\qquad \qquad  \qquad \qquad \begin{array}{c} \left[J_+, J_- \right] = 2 J_0 , \\ \left[J_0, J_\pm \right] = \pm J_\pm . \end{array}  \nn\\
(J_+)_{ij} &=& \left\{ \begin{array}{cc} i(r-i)\delta_{i+1,j} & (i=1, 2, \dots, r-1), \\ 0 & \textrm{else} \end{array} \right.  . 
\label{eq:degenRep}
\ee
Note that  $J_-$ and $J_+$ are nilpotent. 
The degenerate state equation is obtained by  eliminating $f_1, \dots, f_{s-1}$ from the equations
\be
D_{r,1} \left( \begin{array}{c} f_1 \\ f_2 \\ \vdots \\ f_r \end{array} \right) &=& \left( \begin{array}{c} 0 \\ 0 \\ \vdots \\ 0 \end{array} \right) .
\ee
Formally, this can be re-written as
\be
\label{eq:BasicDetFormula} 
0 &=& \Delta_{r,1}(b) |h_{r,1}\>,  \nn\\
\Delta_{r,1}(b)  &\equiv& \det \left[ -J_- + \frac{1}{b^2} \sum_{m=1}^\infty\left( J_+ \right)^m L_{-m-1} \right].
\ee
The  factor of $\frac{1}{b^2}$ compensates for the single power of $b^2$ or $h$ that will be obtained when $L_{-m-1}$ acts on a semiclassical Virasoro block.  The simplest example of this formalism is the case $r = 2$, where we obtain
\be
\Delta_{2,1}(b) = \det \left[ \begin{array}{cc}
\frac{1}{b^2} L_{-1} & \frac{1}{b^2}   L_{-2}\\
 -1 &\frac{1}{b^2}  L_{-1}   \end{array} \right]  \propto L_{-1}^2 + b^2 L_{-2} .
\ee
It is easy to check that states with dimension $h_{2,1}= - \frac{3b^2 }{4}  - \frac{1}{2}$ are annihilated by this operator.

We can obtain a differential operator that annihilates the four-point function by studying
\be
\label{eq:NullStateOperatorRelation}
0 &=& \< h_{r,1}|\left( \Delta_{r,1} \right)^\dagger \CO_{r,1}(0) \CO(x) \CO(y)\>
\ee 
and commuting all $L_{n}$s to the right.   The $L_m$ act on an operator $\CO(z)$ via 
\be
[ L_m, \CO(z) ] = z^m \left( h(m  +1) + z \partial_z \right) \CO(z) .
\ee
The $L_n$ in $\Delta_{r,1}^\dag$ all annihilate $\CO_{r,1}(0)$ because it is a primary and all $n > 0$.
So within the correlator, we need only act the Virasoro generators on $\CO(x)$ and $\CO(y)$, which we take to have dimension $h$. 

We would like to obtain the resulting differential equation in the semiclassical limit. We will approximate the correlator as itself as 
\be
 \< h_{r,1}|  \CO_{r,1}(0) \CO(x) \CO(y)\> = \frac{e^{-b^2 f(1 - \frac{x}{y})} }{(x-y)^{2h}} ,
 \label{eq:hOOO}
\ee
identifying $z =  1 - \frac{x}{y}$.  We wish to keep only the leading results at large $b^2 \propto c \propto h_L$. 
 Raising $\< h_{r,1}| $ with an $L_{m}$ acts on the correlator (\ref{eq:hOOO}) as the differential operators
\be
L_{m} = x^{m+1} \partial_x + y^{m+1} \partial_y + h_L (m+1) (x^m + y^m).
\ee
We are taking $h_L \propto b^2$, so we can ignore actions of the derivatives in $L_m$ that do not produce factors of $b^2$ or $h_L$.
This means that we can ignore  actions of the derivatives in $L_m$ on kinematic factors such as $x^n$ and $y^n$ within $L_n$, and so effectively
\be
[ L_m, L_n ] \sim 0
\ee
when acting on the correlator in equation (\ref{eq:NullStateOperatorRelation}).  Using these simplifications, we find that when the $L_m$ act on the correlator, 
\be
&&\left( \frac{e^{-b^2 f(1 - \frac{x}{y})} }{(x-y)^{2h}} \right)^{-1} L_m \cdot \left( \frac{e^{-b^2 f(1 - \frac{x}{y})} }{(x-y)^{2h}}  \right)
\\
&& = \left( \frac{h_L \left((m-1) \left(x^{m+1}-y^{m+1}\right)+(m+1) \left(x y^m-y x^m\right)\right) }{x-y}+\frac{b^2 x \left(x^m-y^m\right) f'}{y}  \right)  .
\nn
\ee
Using equation (\ref{eq:BasicDetFormula}) and taking $y \to 1$ and $x \to 1-z$ leads to the determinant formula
\be
\label{eq:DeterminantFormulawithSum}
0 &=& \det \left[ J_- + \frac{\kappa(z)}{(1-J_+)(1-(1-z)J_+)} - \frac{h_L}{b^2} \frac{J_+ z^2}{(1-J_+)^2 (1-(1-z)J_+)^2} \right],
\ee
where we have replaced $\kappa(z) = z(z-1) f'(z)$ and performed the sum over $m$ in equation (\ref{eq:DeterminantFormulawithSum}) in closed form to simplify the result.  The factors of $\frac{1}{b^2}$ have canceled against factors of $b^2$ and $h$ to produce a result that is fixed in the semiclassical limit.  Crucially for the following, the elements of the above matrix are all now just numbers, and so the determinant is no longer defined formally but is instead just has its standard meaning.  This fact allows us to multiply the matrix inside the determinant by any invertible matrix, since doing so does not change the condition that the determinant vanishes.  In particular, we can multiply by $(1-J_+)(1-(1-z)J_+)$, yielding
\be 
\label{eq:MasterDeterminantFormula}
0 &=& \det \Big[ H(z) +\kappa(z)  \Big] , \nn\\
H(z) &=&(1-J_+)  (1- (1-z) J_+) J_- - \frac{h_L}{b^2}  \frac{ J_+ z^2 }{(1-J_+)(1-(1-z) J_+)} .
\ee
Equation (\ref{eq:MasterDeterminantFormula}) is exactly the condition that $-\kappa(z)$ is the set of eigenvalues of  of $H(z)$!  This equation also provides an $r^{\mathrm{th}}$ order algebraic equation for $\kappa(z)$ that can be solved in closed form for the first several values of $r$.

As a check, let us consider what happens in the limit $z \to 1$.  In this case, we know all solutions for $\kappa$ should reduce to operators allowed by the fusion rules of the degenerate operator.  The dimensions of allowed operators are most easily seen in the Coulomb gas formalism for the shift of the weights of operators when they fuse with a degenerate operator at large central charge.\footnote{See e.g. \cite{Turiaci:2016cvo,Fitzpatrick:2016ive} for similar observations in the large $c$ limit.}  That is, a degenerate operator $\CO_{r,1}$ can fuse with a general operator $\CO_L$ to make a new operator $\CO'$ with weight $h'$ satisfying
\be
h' = h_L - (r-2n-1) \frac{b}{2} \left(\pm \sqrt{Q^2 -4 h_L} + (r-2n-1) \frac{b}{2}\right),
\label{eq:coulombshift}
\ee
where $Q = (b+1/b)$ and the integers 
\be
n =0, \dots, \lfloor \frac{r-1}{2} \rfloor.
\label{eq:allowedCoulombN}
\ee  Equation (\ref{eq:coulombshift}) follows immediately from the fact that fusing with $\CO_{r,1}$ can shift the Coulomb gas charge of $\CO_L$ by $-(r-2n-1)\frac{b}{2}$ for any $n$ in (\ref{eq:allowedCoulombN}).  The parameter $\kappa$ is related to $h'$ by 
\be
-\frac{c}{6} \kappa(1) = h' - h_L - h_{r,1},
\ee
since the OPE singularity at $z \sim 1$ is $\sim (1-z)^{h' - h_L - h_{r,1}}$.  Taking the limit $b\rightarrow \infty$, one finds that $\kappa$ reduces to the following tower of pairs of solutions:
\be
\kappa (1) = \frac{r-1 + 2n (r-1-n) \pm (r-2n-1) \sqrt{1-  \frac{4 h_L}{b^2}}}{2}.  
\label{eq:AlgebraicKappaof1}
\ee
For comparison, take (\ref{eq:MasterDeterminantFormula}) in the limit $z \to 1$, in which case it greatly simplifies to
\be
0 = \det \left[ (1-J_+)^2 J_- + (1-J_+) \kappa(1)  - \frac{h_L}{b^2}J_+ \right] .
\ee
This can be solved in closed form for all values of $r$, yielding the pairs of solutions in (\ref{eq:AlgebraicKappaof1}).

These solutions exactly match those of equation (\ref{eq:KappaSolns}) by identifying $6\eta_L = \frac{h_L}{b^2}$ or $\alpha_L = \sqrt{1 - \frac{4 h_L}{b^2}}$, analytically continuing $r = \alpha_H$, and letting $n$ range over all integers.  The ambiguity of sign just corresponds to the choice of sign of square roots in $\alpha_L$ and $\alpha_H$.  One can also see that these results accord with the derivation of BTZ quasi-normal modes given in \cite{Turiaci:2016cvo,Fitzpatrick:2016ive}.
So in the semiclassical limit our algebraic method matches our results from the monodromy method.  Note however that we have not obtained the other infinite class of semiclassical solutions discussed in section \ref{sec:ABCto0}.

\begin{figure}[t!]
\begin{center}
\includegraphics[width=0.45\textwidth]{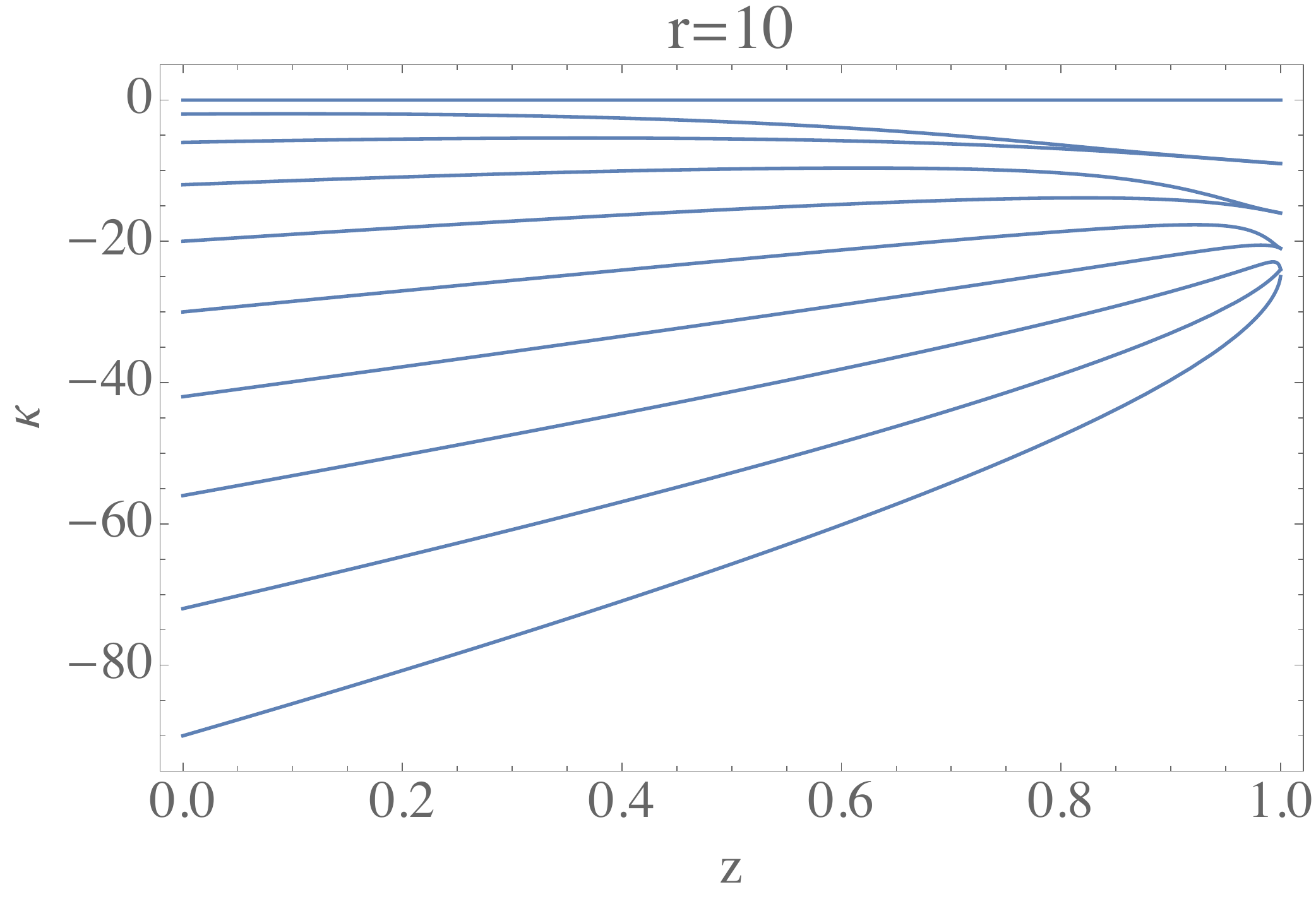}
\includegraphics[width=0.45\textwidth]{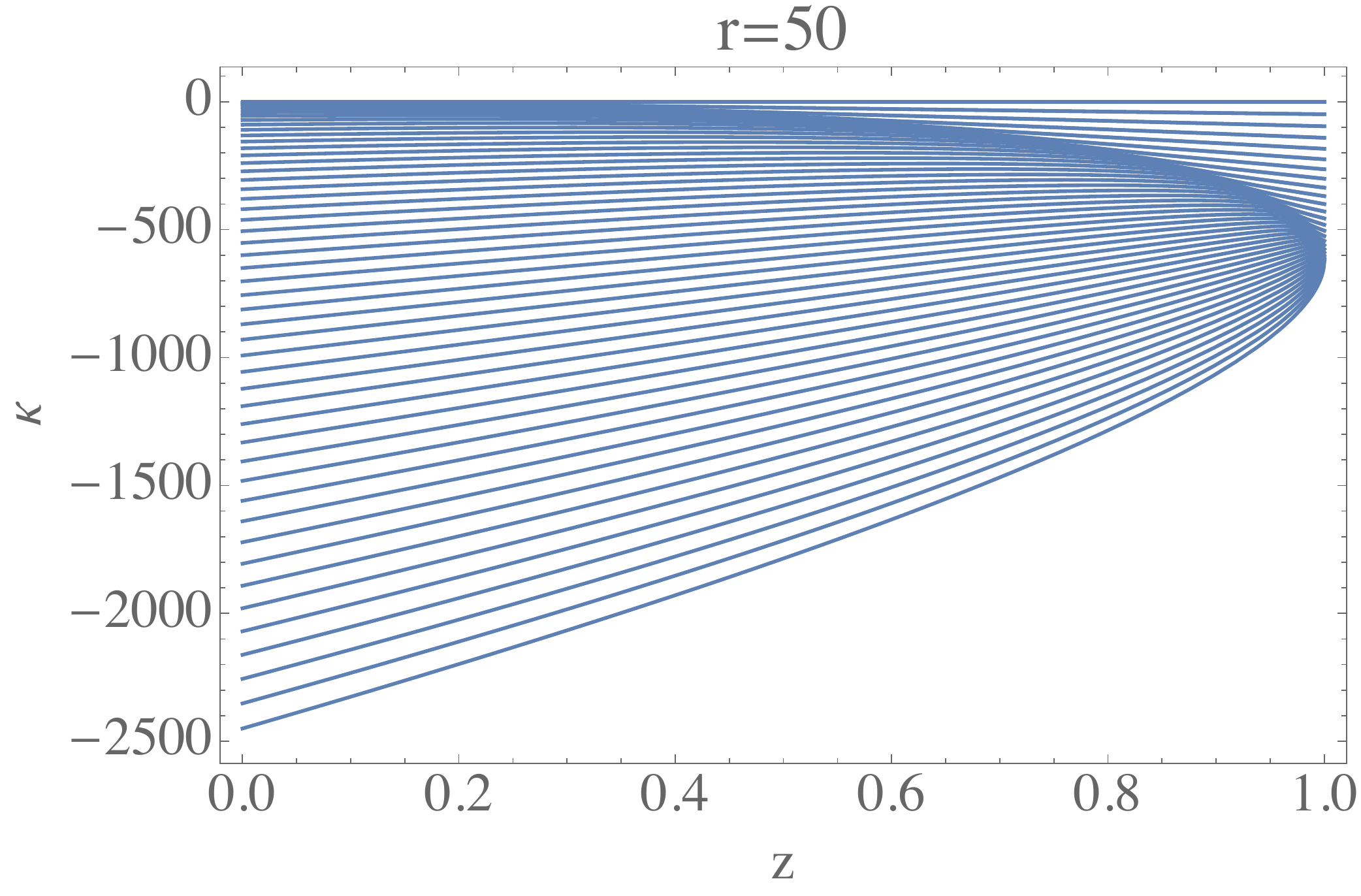}
\caption{Plot of values of $\kappa(z)$ with $h_L=0$ as a function of $z$ for $r=10$ (top) and $r=50$ (bottom).}
\label{fig:KappaDegNoHL}
\end{center}
\end{figure}

The formula (\ref{eq:MasterDeterminantFormula}) is quite useful since it allows us to find the instantons at general $z$ for $\CO_H$ degenerate by computing the eigenvalues of a matrix.  Such computations are fairly efficient, and so we can quickly get a sense of how the instantons behave for a large range of $r$ and $h_L$ values.    For instance,  in fig. \ref{fig:KappaDegNoHL}, we plot the solutions for $\kappa(z)$ at $r=10$ and $r=50$ as a function of $z$.  One can see the values corresponding to the OPE behavior at $z\sim 0$, and the pairwise merging of eigenvalues at $z \sim 1$. 

 \begin{figure}[th!]
\begin{center}
\includegraphics[width=0.8\textwidth]{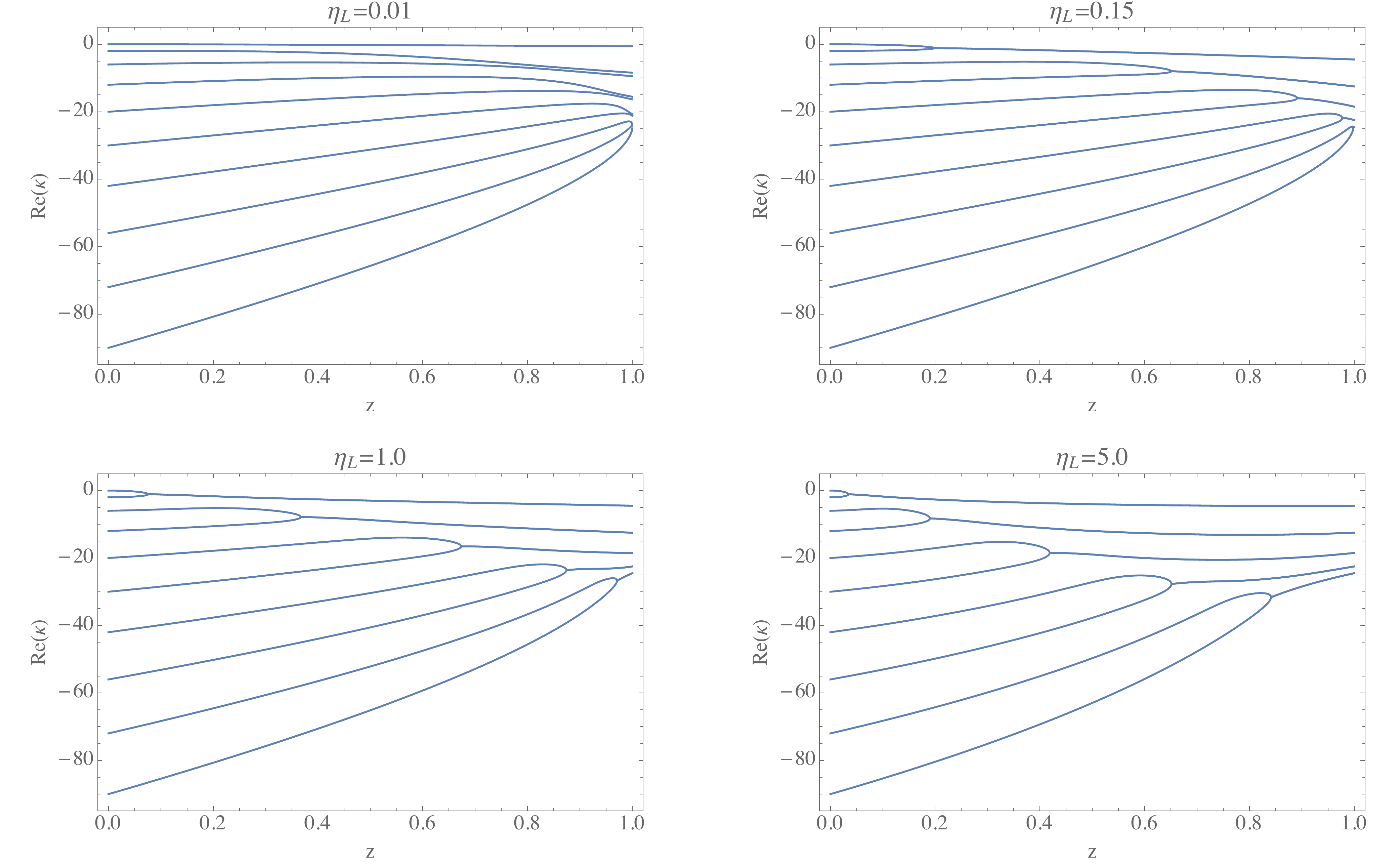}
\caption{Plot of values of Re$(\kappa)$ as a function of $z$ for $r=10$ with various values of $\eta_L = h_L/c$.  We see that the (real parts of the)  solutions merge in pairs even before they reach $z=1$ when $\eta_L > 0$. }
\label{fig:KappaDegHL}
\end{center}
\end{figure}

 \begin{figure}[th!]
\begin{center}
\includegraphics[width=0.35\textwidth]{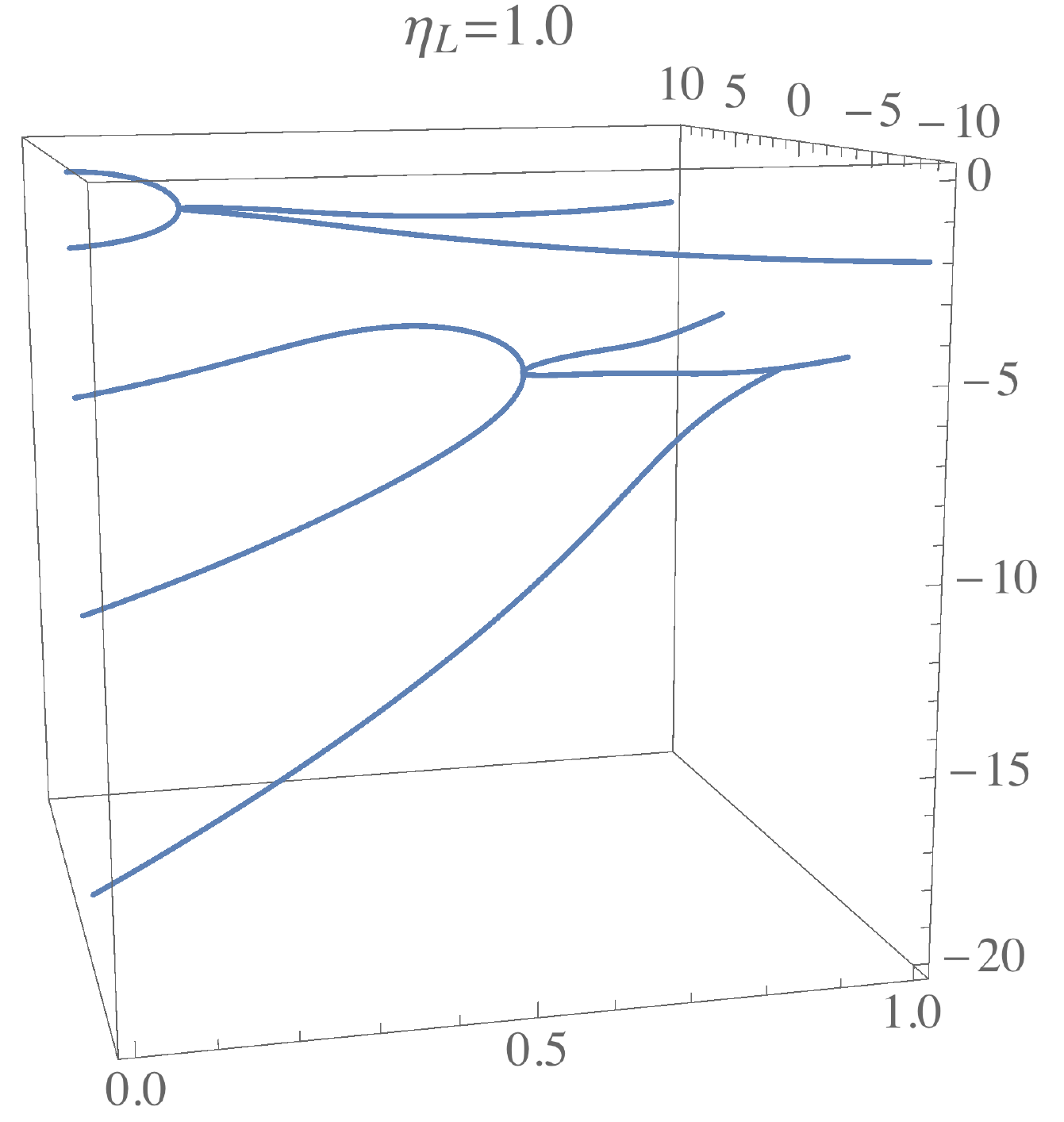}
\caption{Plot of values of both Re$(\kappa)$ and Im$(\kappa)$ as a function of $z$ for $r=5$ and $\eta_L = 1.0$. After the  solutions merge in pairs, they move off into the complex plane. }
\label{fig:KappaCpxDegHL}
\end{center}
\end{figure} 

Once we turn on a non-zero $h_L$, the behavior becomes much more interesting.  For $0 < h_L < c/24$, the solutions repel each other slightly and no longer merge at $z=1$. However, taking $h_L > c/24$, the solutions start to merge at $z < 1$, at which point they develop complex parts.  This is shown in fig. \ref{fig:KappaDegHL} and \ref{fig:KappaCpxDegHL}.  Perhaps surprisingly, the degenerate operators have only ``decaying'' saddles, i.e. saddles in the class (\ref{eq:KappaSolns}) that pick up non-zero imaginary parts at $z \sim 1$ (when $h_L > c/24$).  Conceivably, this could be an indication that the generation of non-decaying saddles under analytic continuation in $z$ is more subtle than that of the decaying solutions, or it could be an indication that instantons for conformal blocks of degenerate operators simply do not contain key information about the generic case.

\subsection{The Monodromy Method from the Algebraic Method}
\label{sec:MonodromyMethodfromAlgebraicMethod}

We would like to `analytically continue' results from the methods of the previous section section in order to re-derive the monodromy method for the vacuum conformal block.  However, the methods of section \ref{sec:AlgebraicMethod} involved deriving differential equations of $r^{\mathrm{th}}$ order from an $r \times r$ matrix appropriate for $h_{r,1}$ degenerate states.  Throughout it was crucial that $r$ be an integer -- so how can we analytically continue the dimensions of a matrix?  This is easy if we interpret the matrix as a product of lie algebra generators, and then generalize to an infinite dimensional representation of the group.  So our first step will be to reformulate the method using a representation where $J_{\pm}, J_0$ act as differential operators in the case where $r$ is an integer.

Our starting point will be equation (\ref{eq:MasterDeterminantFormula}), which states that a certain matrix, written in terms of the $su(2)$ matrices $J_+$ and $J_-$, must have a zero eigenvalue.  For integer $r$ we used the spin $\frac{r-1}{2}$ representation.  In order to treat general values of $r$, we will use the following $su(2)$ representation 
\be
J_- &=& i y \partial_y^2   + i (1 - r) \partial_y
\nn \\
J_+ &=& i y  
\nn \\
J_0 &=& y \partial_y + \frac{1 - r}{2} 
\ee
where we chose a particularly simple $J_+$ because it appears most frequently in equation (\ref{eq:MasterDeterminantFormula}).  The space of vectors of the theory is by definition spanned by the eigenvectors of $J_0$ with eigenvalues $\left\{ \frac{1}{2} (r-2i+1) \right\}_{i=1, \dots, r}$.  We have chosen the above representation so that this exactly corresponds to the case where the vector space that $J_\pm, J_0$ act on is polynomials of degree $r$.  Naively, in this representation, the statement that the determinant of $D_{r,1}$ vanishes is just the condition that `matrix' $H(z)+\kappa(z)$ of equation (\ref{eq:MasterDeterminantFormula}) has a zero eigenvector.  However, because the elements of $H(z)$ are no longer pure numbers, we must be more careful and return to the ``formal'' definition of the determinant and consider how to process it correctly.  So let us analyze eq. (\ref{eq:DeterminantFormulawithSum}) in this context.  

The formal definition of the determinant in this expression is that $D_{r,1}$ acting on the vector $\{ f_i\}_{i=1, \dots, r}$ gives the proper differential equation after eliminating $f_1, \dots, f_{r-1}$.  In our new representation, this just means that the condition for $\kappa$ is that $D_{r,1}$ acting on the polynomial  
\be 
\chi(y,z) &\equiv& \sum_{j=0}^{r-1} y^{r-j} f_j(z) 
\label{eq:FiniteOrderPoly}
\ee
vanishes.  More properly, since $J_+ = i y$ is nilpotent, $\chi(y,z)$ should be thought of as an element of the quotient ring $\mathbb{C}[y]/(y^r)$.  Now we see that although multiplying by functions of $J_+ = i y$ inside the determinant does not change the condition that the matrix has a zero eigenvalue, it {\it does} change whether or not the functions that it acts on are degree-$r$ polynomials or some other space of functions.  
 
The point of this discussion is that it tells us only a certain space of functions are acceptable as zero eigenfunctions of the matrix $H(z) +\kappa(z)$. Uplifting from the space $\mathbb{C}[y]/(y^r)$ to $\mathbb{C}[y]$ is necessary in order to connect with the usual monodromy method, but is not obviously straightforward.  When $\eta_L=0$, one of the solutions to the monodromy matrix is in fact the finite order polynomial $\chi$ in (\ref{eq:FiniteOrderPoly}) (after factoring out some simple prefactors, see below), so the uplift is trivial.  In any case, though, once we move away from integer $r$ it is necessary to justify what infinite dimensional space of functions one should allow and we do not have a sharp argument using the present method that it should be the space of functions with the proper monodromy.  A somewhat {\it ad hoc} generalized criterion would be that the series expansion in $y$ should converge in a particular region; this is clearly satisfied by finite order polynomials, and any cycles in the region of convergence would have trivial monodromy as a necessary consequence.  

Demanding that the operator $H(z)+\kappa(z)$ has a zero eigenvector is equivalent to  the differential equation
\be
\psi''(y) +  \left( \frac{6 \eta_L (1-z)^2}{ (y+1)^2 (y z+1)^2}  + \frac{(1-r^2)}{4
   y^2}+\frac{\kappa(z) }{y\left(y+1\right) (y z+1)} \right) \psi(y) = 0
\ee
for the function $\psi(y) \equiv y^{\frac{1-r}{2}} \chi(y,z)$, where we recall that $\eta_H = \frac{h_H}{c} = \frac{h_H}{6 b^2}$, and we sent $y \to i y$ and $z \to 1 -z$ to simplify the result.  To analytically continue, we need only identify $r = \alpha_H = \sqrt{1  - \frac{24 h_H}{c}}$, in which case we find 
\be
\psi''(y) +  \left( \frac{6 \eta_L (1-z)^2}{ (y+1)^2 (y z+1)^2}  + \frac{6 \eta_H}{y^2}+\frac{\kappa(z) }{y\left(y+1\right) (y z+1)} \right) \psi(y) = 0.
\ee
Finally, let us send $z$ back via $z \to 1-z$ and take $y \to \frac{y-1}{1-z}$:
\be
\label{eq:NewlyDerivedMonodromy}
\psi''(y) +  \left( \frac{6\eta_L z^2}{y^2 (y-z)^2}+\frac{6 \eta_H}{(y-1)^2}+\frac{\kappa(z) }{(y-1) y (y-z)} \right) \psi(y) = 0.
\ee
This coincides exactly with the differential equation associated with the monodromy method, equations (\ref{eq:MonodromyEquation}) and (\ref{eq:MonodromyStressTensor}), once we shift our definition of by $\kappa \to \kappa + 12 \eta_L (z-1)$.  This last step can be explained by recalling that in section \ref{sec:AlgebraicMethod} we wrote $\CV = z^{-2h_L} e^{-b^2 f(z)}$, whereas in the monodromy method of section \ref{sec:LateTimeMonodromy} we absorb the $z^{-2h_L}$ factor into the definition of $f(z)$.  The factor of $y^{\frac{1-r}{2}}$ between $\psi(y)$ and $\chi(y,z)$ introduces a non-trivial monodromy in $\psi$ around certain cycles, but tracking our changes of variables one sees that this factor becomes $\propto \left( 1-y \right)^{\frac{1-r}{2}}$, so that no monodromy is introduced by this factor on cycles where $y$ circles $0$ and $z$ but not $1$. 

We have now derived the differential equation of the monodromy method by analytically continuing results relevant to degenerate external states.  This derivation started from the case of degenerate external operators, i.e. $\alpha_H = r$ was an integer, so $\eta_H$ was restricted to specific values.  However, in the final formulation, the parameter $\eta_H$ appears only as an argument in the differential equation (\ref{eq:NewlyDerivedMonodromy}), which is trivial to analytically continue by letting $\eta_H$ take arbitrary complex values. 
More significantly, the complete formulation of the monodromy method also requires the statement that the space of allowed functions generalizes to the space of functions with trivial monodromy around certain cycles.  We have made some heuristic comments to try to motivate this last step, but we do not have a proof along these lines that this is the correct analytic continuation. Perhaps the best that can be said is that such a generalization seems natural, and in any case is the one that reproduces the standard monodromy method.

\section{Discussion}
\label{sec:Discussion}

Even with the AdS/CFT correspondence in hand, it has been difficult to resolve the most conceptually fascinating conundrums of black hole physics -- while AdS/CFT may be an exact description of quantum gravity in principle, it has yet to become one in practice.  Some of the difficulty arises from limitations in our understanding of the AdS/CFT dictionary, in particular how or even if one can generally reconstruct bulk observables in terms of CFT dynamics.  However, another major obstacle is that we often do not know how to explicitly compute key observables of the boundary CFT.  Such quantities include boundary CFT correlators related to ``easier'' versions of the information loss problem \cite{Maldacena:2001kr,Fitzpatrick:2016ive}.  

Recently, there has been remarkable progress towards an understanding of the robust features of AdS$_3$/CFT$_2$ using methods closely related to the conformal bootstrap  \cite{HartmanLargeC,Headrick,Hellerman:2009bu,Fitzpatrick:2016ive,Lewkowycz:2016ukf,Fitzpatrick:2015foa,Fitzpatrick:2014vua,Fitzpatrick:2015dlt,Fitzpatrick:2016thx,Fitzpatrick:2015zha, Anous:2016kss, Chen:2016cms,Alkalaev:2015wia, Perlmutter:2016pkf,HartmanExcitedStates,Jackson:2014nla, TakayanagiExcitedStates,Chen:2013kpa,Keller:2014xba,Chang:2015qfa,Chang:2016ftb,Roberts:2014ifa,Benjamin:2016fhe,Caputa:2015waa,Galliani:2016cai,Chen:2016dfb,Kraus:2016nwo,Friedan:2013cba}.  An organizing principle underlies many of these results:  black hole physics emerges directly from the structure of the Virasoro conformal blocks at large central charge, and is largely independent of the precise details of the CFT spectrum and OPE coefficients.  This suggests that it may be possible to understand information loss and unitarity restoration without  solving any specific holographic CFT. To understand AdS$_3$ quantum gravity we appear to need all of the foundational principles of conformal symmetry, modular invariance, locality, and quantum mechanics, but not so much more. 

As a starting point, it was important to understand how to reproduce semiclassical results on the geometry and thermodynamics of strongly-coupled gravitational backgrounds directly from CFT \cite{Hellerman:2009bu, Fitzpatrick:2015zha,Fitzpatrick:2014vua,Hartman:2014oaa,Anous:2016kss, Kraus:2016nwo, HartmanLargeC, Roberts:2014ifa}.  In other words, it was crucial to see how unitary CFTs mimic the information-destroying effects of black holes, pointing the way towards an understanding of what is missing from the semiclassical gravity description.    We have   \cite{Fitzpatrick:2016ive, Chen:2016cms} explicitly computed some of the `$e^{-c}$' effects responsible for the restoration of unitarity.   However, there remain important kinematical regimes, such as late Lorentzian times, where tractable and sufficiently accurate approximations to the Virasoro blocks are not yet known.

In this paper we have focused on identifying and classifying non-perturbative effects in AdS$_3$ gravity through the study of the remarkable semiclassical `saddles' of Virasoro conformal blocks.  The motivation for this investigation was twofold.

 First, black hole information loss manifests as unitarity violation in semiclassical correlation functions, and this violation is present at the level of the individual Virasoro blocks.  Studying the large $c$ saddles teaches us about the nature of the exponentially small corrections involved in the restoration of unitarity.  The most important lesson from the present analysis is that the leading semiclassical contribution to Virasoro blocks with $h_H > \frac{c}{24} > h_L$ decays exponentially\footnote{Note the order of limits: a semiclassical block with finite $h_I$ will decay exponentially at sufficiently late times.  An infinite sum over blocks including $h_I \to \infty$ might not decay; this deserves further study. }   at late times, and at a universal rate independent of internal operator dimensions, as we illustrated with figure \ref{fig:MoneyPlot}.  This feature was observed previously for conformal blocks of light external and intermediate states  \cite{Fitzpatrick:2014vua}, but it was far from obvious that it would persist for all semiclassical blocks.
 
 We have also identified other saddles that either decay or approach a constant magnitude at late times.  The latter may play an important role in resolving information loss, but as we explained in section \ref{sec:BlockExpansions}, their mere existence is not sufficient by itself.  It would be very interesting to understand how the semiclassical saddles that we have uncovered here relate to the non-perturbative resolution of forbidden singularities, and to the `master equation' \cite{Fitzpatrick:2016ive} that seems to be a first step towards a determination of the true late-time behavior of the vacuum Virasoro block.

Second, we hope that our analysis may provide key insights towards a complete  path integral description of the Virasoro blocks \cite{Witten:1988hf, Verlinde:1989ua, Witten:1988hc, Elitzur:1989nr, Witten:2010cx, Gaiotto:2011nm}.  Ideally, one could start directly from an action, possibly the Chern-Simons  action for holomorphic `gravitons', together with a coherent, self-contained set of rules for when particular saddles contribute.  

Knowledge of the saddles likely gives us a very precise indication of what semiclassical effects will look like in a gravitational context. The simple reason for this optimism is that the saddles directly indicate the corresponding background value for the boundary stress tensor, and we can attempt to extend the stress tensor to the bulk metric.  We expect our metrics should be vacuum metrics since in the semiclassical limit sources are localized on geodesics.  A general vacuum metric can be written
\be
ds^2 &=& \frac{L(y)}{2} dy^2 + \frac{\bar{L}(\bar{y})}{2} d\bar{y}^2 + \left( \frac{1}{u^2} + \frac{u^2}{4} L(y) \bar{L}(\bar{y}) \right) dy d\bar{y} + \frac{du^2}{u^2},
\ee
where $L(y)$ and the boundary stress tensor $T(y;z)$ are related by 
\be
T(y; z) &=& - \frac{c}{12} L(y)
\ee
and $T(y;z)$ is given in terms of $\kappa$ by (\ref{eq:MonodromyStressTensor}).  Restoring the explicit dependence on the positions $x_i$ of the $\CO_H$ and $\CO_L$ operators, $T(y;z)$ can be written\footnote{More precisely, this is the contribution to $\frac{\< \CO_H(x_1) \CO_H(x_2) \CO_L(x_3) \CO_L(x_4) T(y)\>}{\< \CO_H(x_1) \CO_H(x_2) \CO_L(x_3) \CO_L(x_4)\>}$ from the saddle  $\kappa(z)$.}
\begin{equation}
 T(y;x_i) = \left[ \frac{h_H x_{12}^2}{x_{15}^2 x_{25}^2}+ \frac{h_L x_{34}^2}{x_{35}^2 x_{45}^2} - \frac{c}{6} \frac{x_{13} x_{24} \left( \kappa(z )- 12 (1-z) \eta_L \right)_{z=\frac{x_{12} x_{34}}{x_{13} x_{24}}}}{x_{15} x_{25} x_{35} x_{45}}\right],
\end{equation}  
where $x_5 \equiv y$.  Interestingly, generic saddles therefore depend on the positions $x_3, x_4$ of the probe operators $\CO_L$ even in the limit $\eta_L \rightarrow 0$.

Knowledge of the $\CO(c)$ part, i.e. $\log \CV = -\frac{c}{6} f$, of the saddle exponents is of course not sufficient to determine their absolute contribution to the blocks, since there is an unknown $\CO(1)$ prefactor.  This prefactor vanishes for all but the leading saddle in the $z \sim 0$ OPE limit, and to ascertain the contribution from the subleading saddles we need to identify Stokes lines/walls.  This should be possible, since Stokes lines occur when leading and subleading saddles have equal imaginary parts.  
However, obtaining a sharp prediction for the prefactor of the subleading saddles is more challenging.  The prefactor is in principle known quasi-analytically for blocks with degenerate external operators from the crossing matrices \cite{Dotsenko:1984nm,Dotsenko:1984ad,Esterlis:2016psv}, but the existence of the non-decaying saddles which do not appear in degenerate conformal blocks suggests that perhaps there are limitations to analytic continuation from degenerate states. Alternatively, it may be possible to extract the prefactor from the results of \cite{Ponsot:1999uf,Teschner:2003en} for the braiding matrices of Virasoro conformal blocks.

\section*{Acknowledgments}

We would like to thank Hongbin Chen, Tom Hartman, Ami Katz, Zohar Komargodski, Daliang Li, Miguel Paulos, Jo\~ao Penedones, Eric Perlmutter, Martin Schmaltz, Julian Sonner, Douglas Stanford, Mithat \"Unsal, Matt Walters, Huajia Wang, Junpu Wang, and Sasha Zhiboedov for useful discussions, and the GGI for hospitality while parts of this work were completed.  ALF is supported by the US Department of Energy Office of Science under Award Number DE-SC-0010025.  JK  has been supported in part by NSF grants PHY-1316665 and PHY-1454083, and by a Sloan Foundation fellowship.  ALF and JK are supported by a Simons Collaboration Grant on the Non-Perturbative Bootstrap.

\addtocontents{toc}{\protect\setcounter{tocdepth}{1}}

\appendix

\section{Comments on the Generality of Our Results}
\label{app:Generality}

In this paper and other recent work \cite{Fitzpatrick:2014vua, Fitzpatrick:2015zha, Fitzpatrick:2015foa, Fitzpatrick:2015dlt, Fitzpatrick:2016ive, Chen:2016cms} we have been developing an understanding of black hole thermodynamics, information loss, and its resolution in AdS$_3$/CFT$_2$.    This means that we are interested in `irrational' or non-integrable CFT$_2$ at large central charge, and specifically theories that may have a dual interpretation involving finite temperature black holes in AdS$_3$.  

Our results are extremely general because they follow entirely from the structure of the Virasoro algebra.    Roughly speaking, our results pertain to any theory where the Virasoro algebra is not embedded in a much larger algebra.  This means that our results are not relevant for integrable theories or free theories, but will apply to virtually all other large $c$ theories.  In this appendix we will discuss the generality of our results more precisely, and  explain why they do not apply to integrable theories.

\subsection{Forbidden Singularities are Forbidden From All Virasoro Blocks}
\label{app:ForbiddenfromVirasoro}

In recent work \cite{Fitzpatrick:2016ive} we discussed two signatures of information loss in heavy-light 4-pt correlators:  late Lorentzian time behavior, which has been the major focus in this paper, and forbidden singularities in the Euclidean region.   Let us now focus on the latter; for a more extensive discussion see \cite{Fitzpatrick:2016ive}.  

Forbidden singularities arise because the CFT 4-pt function approximates the 2-pt function of a light probe field in a BTZ black hole background.  These correlators are thermal, which means that they are periodic in Euclidean time.  In particular, this means that the OPE singularity has an infinite sequence of image singularities under $t_E \to t_E + n \beta$ for any integer $n$, where $t_E$ is the Euclidean time and $\beta = \frac{1}{T_H}$.  However, from the point of view of the CFT correlator, these are not OPE singularities, rather they are extraneous, forbidden singularities in the Euclidean plane at $z = 1 - e^{n \beta}$.  

It should be noted that these singularities are allowed in a true thermal 2-pt function (ie in the canonical ensemble), but are forbidden from 4-pt functions (which can be viewed as 2-pt functions in a pure state background) and also from the microcanonical ensemble.  The singularities  can only emerge after summing over an infinite number of external states.

We outlined an argument \cite{Fitzpatrick:2016ive} that forbidden singularities are forbidden not just from full CFT correlators, but also from individual Virasoro blocks.  To clarify the situation, here we provide a more complete proof.  As we previously noted  \cite{Fitzpatrick:2016ive}, all of the hard work was done by Maldacena, Simmons-Duffin, and Zhiboedov (MSZ) in section 7 and appendix D of \cite{Maldacena:2015iua}, so we recommend that interested readers consult that reference for details.  We will simply summarize their results.  

MSZ proved that individual Virasoro blocks for the correlator $\< \CO_1(\infty) \CO_2(1) \CO_1(z) \CO_2(0) \>$ in the $\CO_1(z) \CO_2(0)$ OPE channel can only have OPE singularities (in fact their proof is stated for $\CO_1 = \CO_2$ but the argument easily generalizes).  In particular, the Virasoro blocks have an expansion
\be
\CV^{12,12}_h(z) = \Lambda(z)  \sum_n a_n q^n
\ee
where $\Lambda(z)$ is a universal prefactor with only OPE singularities, the $a_n > 0$ for all $n$, and 
\be
q(z) = e^{-\pi \frac{K(1-z)}{K(z)}}
\ee
where $K(z)$ is an elliptic integral of the first kind. 

The full CFT correlator must be finite for $|q| < 1$, and this condition obtains for all $z$ away from OPE singularities, including after analytically continuing $z$ arbitrarily far from the original Euclidean sheet of the complex plane.  Because the coefficients of the individual Virasoro blocks as well as the coefficients $a_n > 0$, this necessarily implies that the Virasoro blocks themelves must converge for $|q| < 1$.  Thus these blocks cannot have any non-OPE singularities.   Note that all of these results assume \emph{finite} values of $h_i, h, c$, so they do not contradict our semiclassical computations, where we took $c \to \infty$ with $h/c$ fixed.

However, this is not quite what we wanted, because we are studying the Virasoro blocks appropriate for the OPE limit $\CO_1(z) \CO_1(1)$ rather than the $\CO_1(z) \CO_2(0)$ channel.  The relevant block will still have a $q$ expansion (transforming now to $z \to 1-z$ for convenience) 
\be
\label{eq:V1122Expansion}
\CV^{11,22}_h(z) = \Lambda(z)  \sum_n b_n q^n
\ee
for some $b_n$ that are not necessarily positive.  We would like to bound these coefficients to prove convergence for $|q| < 1$.  For this purpose let us consider the $2 \times 2$ matrix of Virasoro conformal blocks
\be
 \left( \begin{array}{cc}
\CV^{11,11} & \CV^{22,11} \\
\CV^{11,22} & \CV^{22,22}  \end{array} \right) = \left\< \left( \begin{array}{c}
\CO_1(\infty) \CO_1(1) \\
\CO_2(\infty) \CO_2(1)  \end{array} \right) {\bf P}_h \left( \CO_1(z) \CO_1(0), \CO_2(z) \CO_2(0) \right)  \right\>
\ee
where ${\bf P}_h$ is a formal projector onto the primary state $h$ and all of its Virasoro descendants.  This matrix must be positive definite.  Equivalently, the linear combination of correlators
\be
\left\< \left( \CO_1(\infty) \CO_1(1) + \lambda \CO_2(\infty) \CO_2(1)  \right)  \left( \CO_1(z) \CO_1(0) + \lambda \CO_2(z) \CO_2(0) \right) \right\>
\ee
can be interpreted as a (somewhat unusual) inner product of normalizable states in  MSZ's pillow metric \cite{Maldacena:2015iua}.  This fact bounds the absolute value of $|\CV^{11,22}_h(z)|$ in terms of $\CV^{12,12}_h(z)$, proving convergence of equation (\ref{eq:V1122Expansion}) for all $|q| < 1$.  Thus the heavy-light conformal blocks can never have unitarity-violating forbidden singularities at finite values of operator dimensions $h$ and the central charge $c$.

\subsection{Can Forbidden Singularities Cancel Between Semiclassical Blocks?}
\label{app:LightconeOPE}

In section \ref{app:ForbiddenfromVirasoro} we proved that the forbidden singularities that arise in the semiclassical approximation to the Virasoro blocks cannot be present in the exact Virasoro blocks.  This is sufficient to demonstrate that a major problem associated with black hole physics must be resolved block-by-block.

But  one can ask if it is nevertheless possible for forbidden singularities to cancel between the distinct semiclassical Virasoro blocks that make up a full CFT correlator in specific CFTs.  In other words, perhaps blocks $\CV_i$ all have forbidden singularities at the semiclassical level, but these singularities cancel in the linear combination $\sum_i \lambda_i \CV_i$ that appears in correlators.  We will see below that in integrable theories this can occur \cite{Galliani:2016cai,Caputa:2016tgt}, but we do not believe it occurs in the irrational CFT$_2$ of interest to the study of quantum gravity.

In any case, one can immediately argue that in large $c$ CFT$_2$ where the stress tensor is the only conserved current, \emph{there are forbidden singularities that can never cancel in this way}.  If we study semiclassical heavy-light correlators in the limit $\bar z \to 0$, which is the lightcone OPE limit \cite{Fitzpatrick:2012yx, KomargodskiZhiboedov}, the vacuum Virasoro block dominates over all other contributions, because the only states that can contribute to this limit have $\bar h = 0$, and are therefore conserved currents.  But in the heavy-light semiclassical limit the vacuum Virasoro block has forbidden singularities \cite{Fitzpatrick:2016ive}, so no other semiclassical Virasoro block can cancel them.

Note that this argument does not require any assumption about the sparseness of the light operator spectrum, which is relevant to other situations where the vacuum Virasoro block takes center stage \cite{HartmanLargeC, Hartman:2014oaa, Anous:2016kss}. The argument depends only on the behavior of the Virasoro blocks in large $c$ heavy-light limit, and the assumption that $T(z)$ is the only conserved current in the CFT.

\subsection{Theories with Larger Symmetry Algebras}

It is interesting to understand how the result from appendix \ref{app:LightconeOPE}  can break down in theories where the Virasoro algebra is contained within a larger symmetry algebra.  In such theories, it may be possible for the forbidden singularities to cancel among distinct Virasoro blocks \cite{Galliani:2016cai,Caputa:2016tgt}.  Examples include WZW theories and rational CFTs with $\mathcal{W}$-algebra symmetries, but free CFTs with a large number of fundamental fields provide a more straightforward test-case.  The authors of \cite{Galliani:2016cai} gave interesting examples involving orbifolds of free theories relevant to the free limit of the D1-D5 system.

Here we will study an extremely simple example:  a CFT$_2$ with $c \gg 1$ and a $U(1)$ symmetry,\footnote{We thank Simeon Hellerman for discussions of this example at the 2015 Bootstrap workshop.} for which the conformal blocks are already known (see section 3.5 of \cite{Fitzpatrick:2015zha} for the derivation and relevant discussion).  In such a theory, the stress tensor $T$ can be written as
\be
T = T^{\mathrm{sug}} + T^{(0)}
\ee
where $T^{\mathrm{sug}}$ is the Sugawara stress tensor of the $U(1)$ current algebra, which has $c=1$.  We can break up the Virasoro generators in the same way, writing $L_n = L_n^{\mathrm{sug}} + L_n^{(0)}$, and so we can also write the scaling dimensions of operators and states as
\be
h = h^{(0)} + \frac{q^2}{2k}
\ee
where $k$ is the level and $q$ is the $U(1)$ charge of an operator.  

The theory may include states with large $\frac{q^2}{2k} \propto c$ but small or vanishing values of $h^{(0)}$.  Very naively, we might expect that heavy-light Virasoro blocks involving such states appear thermal when $\frac{q^2}{2k} > \frac{c}{24}$.  But this is impossible if $h^{(0)}$ is small.  In fact, the explicit form of the heavy-light $U(1)$ $+$ Virasoro blocks is \cite{Fitzpatrick:2015zha}
\be
\label{eq:FullChargedBlocks}
\CV_{T+J}(c, h_i, q_i, z) = \CV_T \left(c-1, h_i - \frac{q_i^2}{2k}, z \right) z^{-\frac{q_L^2}{k}} (1-z)^{\frac{q_H q_L}{k}}
\ee
Thus if $h_L^{(0)}  = h_H^{(0)} = 0$, then the blocks are simply $z^{-\frac{q_L^2}{k}} (1-z)^{\frac{q_H q_L}{k}}$, which clearly cannot be thermal.  We only see a Hawking temperature when $h_H^{(0)} > \frac{c}{24}$, and this condition is independent of $q_H$.  This is consistent with expectations from AdS$_3$ theories with a bulk $U(1)$ Chern-Simons gauge field \cite{Fitzpatrick:2015zha}.  

Since equation (\ref{eq:FullChargedBlocks}) represents a correlator in a theory with a symmetry containing Virasoro, it can be expanded in terms of pure Virasoro blocks (ignoring the $U(1)$ current algebra), just as correlators in theories with Virasoro symmetry can be expanded in the global or sl$(2)$ conformal blocks.  This suggests that in theories with a symmetry larger than Virasoro, it is possible to see cancellations between Virasoro blocks that eliminate forbidden singularities.  Note that the coefficients $c^{LL}_h c^{HH}_h$ of individual blocks in the channel we are considering do not need to be positive, since this quantity is not the square of a real number.  The Virasoro blocks with $h_I, h_L \ll c$ are \cite{Fitzpatrick:2015zha}
\be
\CV_h(z) = \frac{(1-z)^{-h_L}}{ \left[ \frac{2}{\alpha} \sinh \left( \frac{\alpha}{2} \log(1-z) \right) \right]^{2h_L} }  \left(\frac{1 - (1-z)^\alpha }{\alpha} \right)^{h_I} {}_2 F_1 \left( h_I , h_I, 2h_I, 1 - (1-z)^\alpha \right)
\ee
where $\alpha = \sqrt{1 - \frac{24 h_H}{c}}$ as usual, and we have normalized the block to begin with $z^{h - 2 h_L}$ in a series expanion in $z$.  

We can  explicitly decompose the $U(1)$ $+$ Virasoro vacuum block into pure Virasoro blocks.  Let us take the case $h_L^{(0)}  = h_H^{(0)} = 0$ and set $k=1$ and $Q \equiv q_H q_L$ for notational convenience:
\be
\frac{\CV_{T+J}(c, q_L, q_H, z)}{\CV_0(z) } &=& \frac{z^{-q_L^2} (1-z)^{Q}}{\CV_0(z) }
 \\
&=& 1-Q z+\frac{1}{12} \left(6 (Q-1)
   Q+\left(\alpha ^2-1\right) h_L\right) z^2  + \cdots
   \nn
\ee
where in this case $\alpha = \sqrt{1 - \frac{12 q_H^2}{c} }$.  This can be matched order-by-order to the sum
\be
\sum_{h_I=0}^\infty P_{h_I} \left(\frac{1 - (1-z)^\alpha }{\alpha} \right)^{h_I} {}_2 F_1 \left( h_I , h_I, 2h_I, 1 - (1-z)^\alpha \right)
\ee
where $h_I$ only take integer values, and $P_{h_I}$ are the Virasoro block coefficients, which are products of Virasoro primary OPE coefficients.  This matching is tedious but straightforward to any fixed order; the first few terms are
\be
P_0 &=& 1
\nn \\
P_1 &=& -Q
\nn \\
P_2 &=& \frac{1}{12} \left(\alpha ^2 h_L-h_L+6 Q^2\right)
\nn \\
P_3 &=& \frac{1}{12} \left(Q h_L -\alpha ^2 Q h_L - 2 Q^3\right)
\ee
The $P_1$ term is simply the coefficient of the exchange of the Virasoro block of the $U(1)$ current $J(z)$, while the higher terms involve a combination of $J_{-n}$ generators (which include the Sugawara stress tensor within their algebra).  Thus these coefficients could also be computed directly using matrix elements of the current algebra.  We can also re-write them entirely in terms of $q_L$ and $q_H$, in which case we find that at large $c$
\be
P_0= 1,P_1= -Q,P_2= \frac{Q^2}{2},P_3=
   -\frac{Q^3}{6},P_4= \frac{1}{120} \left(5
   Q^4+Q^2\right),P_5= -\frac{1}{840} Q^3 \left(7
   Q^2+5\right) \nn
\ee
Thus at large $c$, the $P_{h_I}$ depend only on $Q$, but incorporating the other terms will be important for an expansion in the Virasoro blocks with fixed $q_H^2/c$ at large $c$. 

In any case, the point of this exercise is that we can express a full block of the form of equation \ref{eq:FullChargedBlocks} in terms of the relevant pure Virasoro blocks, meaning that at least at a formal level, the forbidden singularities cancel in this sum.  It would be interesting to see this cancellation more explicitly.

%
%
%

\bibliographystyle{utphys}
\bibliography{VirasoroBib}

\end{document}